\documentclass[12pt]{article}
\usepackage[utf8]{inputenc}
\usepackage[T1]{fontenc}
\usepackage{amsmath}
\usepackage{graphicx,psfrag,epsf}
\usepackage{enumerate}
\usepackage{natbib}
\usepackage{amsthm}
\usepackage{amssymb}
\usepackage{blkarray}
\usepackage{bbm}
\usepackage{mathtools}
\usepackage{float}
\usepackage{adjustbox}
\usepackage{algorithm}
\usepackage{algpseudocode}
\usepackage{natbib}
\usepackage{pdflscape}
\usepackage{longtable}
\usepackage{booktabs}
\usepackage{caption}
\usepackage{subcaption}
\usepackage{color}
\usepackage{hyperref}
\usepackage{titletoc}
\hypersetup{colorlinks=true,
citecolor = blue,
filecolor=cyan}
\usepackage{cleveref}
\usepackage{pdfpages}

\newcommand{\expec}{\mathbb{E}}
\newcommand{\prob}{\mathbb{P}}

\theoremstyle{plain}

\newtheorem{lemma}{Lemma}[section]
\newtheorem{claim}{Claim}[section]
\newtheorem*{claim.no}{Claim}
\newtheorem{proposition}{Proposition}[section]

\theoremstyle{definition}
\newtheorem{assumption}{Assumption}[section]
\newtheorem{remark}{Remark}[section]

\newcommand{\blind}{0}

\usepackage[margin=1in]{geometry}

\usepackage{setspace}

\begin{document}

%\def\spacingset#1{\renewcommand{\baselinestretch}%
%{#1}\small\normalsize} \spacingset{1}

%%%%%%%%%%%%%%%%%%%%%%%%%%%%%%%%%%%%%%%%%%%%%%%%%%%%%%%%%%%%%%%%%%%%%%%%%%%%%%

\title{Homophily in preferences or meetings? Identifying and estimating an iterative network formation model\thanks{We would like to thank Angelo Mele, Áureo de Paula, Braz Camargo, Bruno Ferman, Eduardo Mendes, Ricardo Masini, and Sérgio Firpo for their helpful comments and suggestions. We are also grateful to seminar participants at PUC-Rio, the 41st SBE meeting, the 2020 Econometric Society World Congress, the Penn State Econometrics seminar, LAMES 2021, the 2021 SBE seminar series and FEA-RP 2022 Seminar Series. Luis Alvarez gratefully acknowledges financial support from Capes. Cristine Pinto and Vladimir Ponczek acknowledge financial support from CNPq.}}

\if0\blind
{
  \author{Luis Antonio Fantozzi Alvarez\thanks{ Department of Economics, University of São Paulo:  \href{mailto:luis.alvarez@usp.br}{luis.alvarez@usp.br}.} \and Cristine Campos de Xavier Pinto\thanks{ Insper Institute of Education and Research: \href{mailto: CristineCXP@insper.edu.br}{CristineCXP@insper.edu.br}. }  \and Vladimir Pinheiro Ponczek\thanks{São Paulo School of Economics, FGV: \href{mailto:vponczek@fgv.br}{vponczek@fgv.br}.}}
  \maketitle
} \fi

\if1\blind
{

 \maketitle
%   \medskip
 } \fi

\bigskip
\begin{abstract}

      %{\color{blue}Is homophily in social and economic networks driven by a taste for homogeneity (preferences) or by a higher probability of meeting individuals with similar attributes (opportunity)? This paper studies identification and estimation of an iterative network game that distinguishes between both mechanisms. We provide conditions under which preference- and meeting-related parameters are identified. We propose estimating model parameters using a Bayesian likelihood-free algorithm to overcome computational and dimensionality-related issues. Our approach enables us to assess the counterfactual effects when changing the meeting protocol between agents. As an application, we study the role of preferences and meetings in shaping classroom friendship networks in Brazil. In a network structure in which homophily due to preferences is stronger than homophily due to meeting opportunities, tracking students may improve welfare. Still, the relative benefit of this policy diminishes over the school year.}\vspace{2em}
\noindent Is homophily in social and economic networks driven by a taste for homogeneity (preferences) or by a higher probability of meeting individuals with similar attributes (opportunity)? This paper studies identification and estimation of an iterative network game that distinguishes between these two mechanisms. Our approach enables us to assess the counterfactual effects of changing the meeting protocol between agents. As an application, we study the role of preferences and meetings in shaping classroom friendship networks in Brazil. In a network structure in which homophily due to preferences is stronger than homophily due to meeting opportunities, tracking students may improve welfare. Still, the relative benefit of this policy diminishes over the school semester.\vspace{1em}

\end{abstract}

\newpage

\onehalfspacing

%\spacingset{1.45} % DON'T change the spacing!

%Sections
\section{Introduction}
\label{introduction}
Homophily, the observed tendency of agents with similar attributes to maintain relationships, is a salient feature in social and economic networks \citep{chandrasekhar2016econometrics,Jackson2010social}. Inasmuch as it may drive network formation, homophily can produce relevant effects in outcomes as diverse as smoking behavior \citep{Badev2017}, test scores \citep{Hsieh2016,Goldsmith2013}, the adoption of health innovations \citep{Centola2011}, and health outcomes \citep{Kadelka2021}. Social connections are also an important driver of economic mobility \citep{Chetty2022socioeconomic}, which suggests that a proper account of homophily may improve the design of policies that aim to reduce economic inequality \citep{Jackson2021,Chetty2022social}. It is therefore unsurprising that the appropriate modelling of homophily has been a focus in the recent push for estimable econometric models of network formation \citep{Goldsmith2013, Mele2017, Graham2016, Graham2017,Jackson2016}.

Homophily that is due to choice is distinct from homophily that is due to opportunity  \citep[p. 68]{Jackson2010social}. We shall label the former homophily in ``preferences''; and the latter homophily in ``meetings''. This distinction has an important role: public policy may be able to alter the meeting technology between agents (say, by desegregating environments), but it may be less successful in changing preferences. Thus, the effect of a public policy that aims at changing individual connections (e.g. a tracking policy or a policy that induces people to move to a different neighborhood) is expected to depend on the type of homophily that is prevalent in that network \citep{Chetty2022social}. Theoretical models that distinguish between these mechanisms are studied by \cite{Currarini2009} and \cite{Bramoulle2012}.\footnote{\cite{Currarini2010} provides estimates of a parametric version of the  \cite{Currarini2009} model using AddHealth data.} These models have some relevant limitations, though: first, they focus on steady-state or ``long-run'' behaviour, which may not be appropriate in settings where transitional dynamics may matter (as in our empirical application in \Cref{application}); second, they are either purely probabilistic\footnote{In contrast to strategic models of network formation. See \cite{Jackson2010social} and \cite{dePaula2016} for examples.} \citep{Bramoulle2012} or, in the case of \cite{Currarini2009}, allow for only a restrictive set of pay-offs from relationships.\footnote{Pay-offs of \cite{Currarini2009} depend only on the number of relationships with individuals of the same or different types. There is no role for indirect benefits.} These limitations render these models unfit for some empirical analyses.\footnote{In a static choice setting, \cite{Zeng2008} propose an ordered logistic model that accounts for both homophily in preferences and opportunities. In their model, however, the structure of homophily in opportunities must either be known \textit{a priori} or depend on a disjoint set of traits than preferences, which limits its applicability even in settings where staticity may be a reasonable assumption.} 

In this paper, we intend to fill the gap in the literature by analysing an estimable econometric model that accounts for both types of homophily. We study identification and estimation of a sequential network-formation algorithm originally developed by \cite{Mele2017} (see also \cite{Christakis2010} and \cite{Badev2017}), where agents meet sequentially in pairs in order to revise their relationship status. The model is well-grounded in the theoretical literature of strategic network formation \citep{Jackson2002} and allows specifications that account for both ``homophilies''. However, our approach differs from previous work in several aspects. Firstly, whereas \cite{Mele2017} discusses identification and estimation of utility parameters when either a single large or several networks drawn from the model's induced stationary distribution are observed, we consider identification and estimation of \emph{both} preference- and meeting-related parameters in a setting where many networks are observed at \emph{two} points of time.\footnote{Importantly, our results do not assume that networks are drawn from the model's stationary distribution, an assumption that can be rejected from the data when observation of a network in two periods of time is available (see \cite{Auerbach2022} for a statistical test). } Given observation of several networks at two instants in time, we are able to provide researchers with a menu of identification strategies, ranging from nonparametric exclusion restrictions to parametric assumptions on preferences and agents' matching technology, to recover \emph{both} preference and meeting parameters. We then discuss how these parameters may be estimated through the expectation propagation approximate Bayesian computation (EP-ABC) algorithm \citep{Barthelme2014,Barthelme2018}, a likelihood-free Bayesian approach that can deliver estimates relatively quickly.\footnote{\cite{Battaglini2021} use a variation of the standard ABC algorithm (see \Cref{estimation.preference.meeting}) in order to estimate a game of network formation. The authors use the structure of the game to speed up their implementation. In contrast, we consider a variation of the ABC algorithm that uses the structure of the data to reduce the computational toll of estimation.} Finally, we provide an empirical illustration on how to use these estimates to assess the counterfactual effects of changes in the meeting technology between agents -- something that previous has been unable to do.\footnote{In fact, as we show in Supplemental Appendix \ref{Supplement-identification_stationary}, meeting parameters are unidentified in the setting of \cite{Mele2017}.} 

A second set of important differences with previous work is that, by identifying and estimating \emph{both} meeting- and preference-related parameters, our methodology enables us to analyze the effects of policies \emph{along the transition} to a new steady state, and not just at the model's stationary (long-run) distribution.\footnote{For example, the analyses in \cite{Mele2020} and \cite{gaonkar2023model}, which are conducted using the identification and estimation approaches in \cite{Mele2017}, are only able to assess the long-run effects of counterfactual policies that aim to, respectively, promote integration in schools, and restrict or incentivize firm entry on inter-firm collaboration.} Our results also cover a larger -- and arguably less restrictive -- class of preferences and meeting processes than those of \cite{Mele2017}, who crucially assumes that meeting probabilities do not depend on the existence of a link between agents in the current network. By studying identification and estimation of general classes of preference- and meeting-related parameters in possibly off-stationary-equilibrium settings, we contribute to the model's applicability and empirical usefulness, especially in conducting counterfactual analyses.\footnote{In a recent paper, \cite{Chetty2022social} use Facebook data to propose a decomposition of homophily in socioeconomic status between an ``exposure bias'', the share of individuals with same socioeconomic status in the group (school, church) an individual participates vis-à-vis the share of individuals with same status in the overall population, and a friending bias, the share of same-status friendships within the group vis-à-vis the share of same-status individuals in the group. While they suggest these measures could be used to provide  assessments of the effects of policies that aim to reduce segregation via changes in either exposure or friending bias, they do recognize that these measures need not be invariant to policy changes, e.g. a policy that is expected to reduce friending bias by $x$ p.p. may have a different overall effect than a calculation which treats exposure bias as fixed may suggest, inasmuch as it alters the incentives for group participation.  In contrast, by working with a structural model, we are able to analyse more complex counterfactuals, where changes in exposure may interact with group participation dynamically. Our model also allows for welfare analyses and, by coupling a peer effects model to it, enables the analysis of the effects of network policies on outcomes (see our empirical application in \Cref{application} and Supplemental Appendix \ref{Supplement-model_peer} for an example).}

\paragraph{Overview of identification results} In this paper, identification of meeting and preference parameters relies on different types of exclusion restrictions that limit some sources of variation to affect either the meeting process or agents’ preferences, but not both. We believe these conditions to be plausible in several empirical settings, as we seek to illustrate below.

In many environments, the opportunity to interact with other agents is shaped by institutional or logistical constraints that need not coincide with the determinants of agents’ preferences. In school settings, for example, meeting opportunities may be influenced by factors such as classroom seating arrangements, group assignments, schedule structures, or teacher-imposed activity groups. These features affect which students are more likely to interact during the school day, but need not directly enter the utility students derive from maintaining a friendship with a particular peer.

Conversely, some characteristics may affect agents’ preferences over relationships without substantially affecting the likelihood of interaction. For example, similarity in interests, academic ability, or personality traits may shape the utility students derive from friendships once interactions occur, even if these traits do not systematically influence the institutional structure determining who encounters whom.

In  our identification analysis, we provide researchers with a menu of identifying assumptions that precisely exploit these ideas. As we show, there will be often a tension between the stringiness of the exclusion restriction and the degree of parametrization the researcher is willing to impose on preferences and meetings. In our model, fully nonparametric identification relies on the existence of several excluded ``instruments'' with a large support. In contrast, the adoption of parametric forms for preferences and meetings will allow us to considerably limit both the number of needed instruments and the support requirements, while still allowing for sufficient flexibility so as to capture known empirical regularities in network data, as well as ``intuitive'' patterns in institutional and individual behavior. 

\paragraph{Overview of empirical application} As an empirical application, we study how ``homophilies'' structure network formation in primary schools in Brazil \citep{Pinto2017}. We consider 30 municipal elementary schools in Recife, Pernambuco, for which baseline (middle 2014) and follow-up (late 2014) data on 3rd- and 5th-grade intra-classroom friendship networks was collected. Using this information, we structurally estimate our model, using the distance between students in the alphabetically-ordered class-list as an instrument entering the meeting process but excluded from preferences. We then assess how changes in the meeting technology between classmates impact homophily in friendships. Our results suggest that removing biases in meeting opportunities (shutting down homophily in the meeting process) does not decrease observed homophily patterns in students' cognitive skills. By contrast, in a counterfactual scenario where the role of preferences is excluded from the network formation process, the probability that a student maintains a friendship with a classmate with a different level of cognitive skills increases. The results provide evidence that both types of homophily are important determinants of the edges in our networks; although homophily due to preferences does appear to be more important. In this context, a tracking policy that reallocates students between classrooms according to their cognitive skills leads to welfare improvements -- as students benefit from connecting with similar individuals -- though this benefit appears to diminish (vis-à-vis leaving the network process unchanged) in the long run. Given the opportunity to connect with similar people, students experience a positive jump in welfare in the short run. However, after they make their new friendships, they tend to keep these links, and the relative impact of the policy decreases in the long run -- inasmuch that, by the end of the school year, current welfare is roughly the same in both the base and counterfactual scenarios.

\paragraph{Structure of the paper} In the next sections, we introduce the network formation game under consideration (\Cref{setup}); explore identification when information on the network structure is available at two distinct points of time (\Cref{identification}); and discuss estimation (\Cref{estimation}). \Cref{application} presents the results of our application. \Cref{conclusion} concludes.

\section{Setup}
\label{setup}

The setup builds upon the work of \cite{Mele2017}. We consider a network game with a finite set of agents $\mathcal{I} \coloneqq \{1,2\ldots N\}$. Each agent $ i \in \mathcal{I}$ is endowed with a $k \times 1$ vector of exogenous characteristics $W_i$. These vectors are stacked on matrix $X \coloneqq \begin{bmatrix} W_1 & W_2 & \cdots & W_N \end{bmatrix}'$. Agents' characteristics are drawn according to law $\prob_X$ before the game starts and remain \emph{fixed} throughout. We denote the support of $X$ by $\mathcal{X}$ and a realization of $X$ by an element $x \in \mathcal{X}$.

Time is discrete. At each round $t \in \mathbb{N}$ of the network formation process, agents' relations are described by a \textit{directed network}. Information on the network is stored on an $N \times N$ \emph{adjacency matrix}, with entry $g_{ij} = 1$ if $i$ lists $j$ as a friend and $0$ otherwise.\footnote{Our model and main results are readily extended to the case of an \textit{undirected} network where friendships are forcibly symmetric.} By assumption, $g_{ii} = 0$ for all $i \in \mathcal{I}$. We denote the set of all $2^{N(N-1)}$ possible adjacency matrices by $\mathcal{G}$.

Agent $i$'s  utility from a network $g$ when covariates are $X = x$ is described by a utility function $u_i: \mathcal{G} \times \mathcal{X} \mapsto \mathbb{R}$, $u_i(g,x)$. The utility may depend on the entire network and the entire set of agents' covariates.

Agents are myopic, i.e., they form, maintain, or sever relationships based on the current utility these bring. In each round, a \emph{matching process} $m^t$ selects a pair of agents $(i,j)$.\footnote{We use the wording ``matching process'' or ``meeting process'' interchangeably throughout the paper.} The matching process $m^t$ is a stochastic process $\{m^t: t \in \mathbb{N}\}$ over $\mathcal{M} \coloneqq \{(i,j) \in \mathcal{I}\times \mathcal{I}: i \neq j\}$.  If the pair $(i,j)$ is selected, agent $i$ will choose whether to form/maintain or not form/sever a relationship with $j$. After the matching process selects a pair of agents, a pair of choice-specific idiosyncratic shocks $(\epsilon_{ij,t}(0), \epsilon_{ij,t}(1))$ are drawn, where $\epsilon_{ij,t}(1)$ corresponds to the taste shock in forming/maintaining a relationship with $j$ at time $t$. These shocks are unobserved by the econometrician and enter additively in the utility of each choice.\footnote{Additive separability of unobserved shocks is a common assumption in the econometric literature on discrete choice and games \citep{Aguirregabiria2010}, though it is not innocuous. In our setting, it precludes factors unobserved by the econometrician from affecting the marginal effect of covariates and network characteristics on utility (homophily in preferences), as taste shocks act as pure location shifts.} Given that choice is myopic, agent $i$ forms/maintain a relation with $j$ if, and only if

\begin{equation}
    u_i([1, g_{-ij}], X) + \epsilon_{ij,t}(1) \geq u_i([0, g_{-ij}], X) + \epsilon_{ij,t}(0)\,,
\end{equation}where $[a, g_{-ij}]$ denotes an adjacency matrix with all entries equal to matrix $g$ except for entry $ij$, which equals $a$.

The following two assumptions constrain the meeting process and the distribution of shocks.

\begin{assumption}
\label{ass.matching}
The meeting process $\{m^t: t \in \mathbb{N}\}$ is described by a \emph{time-invariant} matching function $\rho: \mathcal{M} \times \mathcal{G} \times \mathcal{X} \mapsto [0,1]$ , where $\rho((i,j), g, x)$ is the probability that $(i,j)$ is selected when covariates are $X = x$ and \emph{the previous-round network was $g$}. Moreover, for all $g \in \mathcal{G}$, $x \in \mathcal{X}$, $(i,j) \in \mathcal{M}$,  $\rho((i,j), g, x) > 0$.
\end{assumption}

Assumption \ref{ass.matching} constrains the matching function to assign positive probability to all possible meetings under all possible values of covariates and previous-round networks. %Note that this allows for dependence on the existence of previous-round links, which was not permitted by \cite{Mele2017}.\footnote{Do also note that, unlike \cite{Mele2017}, we do \emph{not} assume that utilities admit a potential function. See \Cref{identification_network_structure} for a discussion.}

\begin{assumption}
\label{ass.shocks}
    Shocks are iid draws across pairs and time, independent from $X$, from a \emph{known} distribution $(\epsilon(0), \epsilon(1))' \sim F_{\epsilon}$ which is absolutely continuous with respect to the Lebesgue measure on $\mathbb{R}^2$ and which has a positive density almost everywhere.
\end{assumption}

Conditional on $X=x$, we have that, under Assumptions \ref{ass.matching} and \ref{ass.shocks} -- and given an initial distribution $\mu_0(x)  \in \Delta(\mathcal{G})$\footnote{We denote by $\Delta(\mathcal{G})$ the set of all probability distributions on $\mathcal{G}$.} --,  the network game just described induces a homogenous Markov chain $\{g^t : t \in \mathbb{N} \cup \{0\}\}$ on the set of netwotks $\mathcal{G}$. The $2^{N(N-1)} \times 2^{N(N-1)}$ transition matrix $\Pi(x)$ has entries $\Pi(x)_{gw}$, $g, w \in \mathcal{G}$, which specify the probability of transitioning to $w$ given the current period network $g$.

For each $g \in \mathcal{G}$, define  $N(g) \coloneqq \{w \in \mathcal{G} \setminus \{g\}: \exists ! \ (i,j) \in \mathcal{M} , g_{ij} \neq w_{ij} \} $ as the set of networks that differ from $g$ in \emph{exactly} one edge. Entries of $\Pi(x)$ take the form
{\footnotesize
\begin{equation}
\label{eq.trans.matrix}
    \Pi(x)_{gw} = \begin{cases}
             \rho((i,j), g, x) F_{\epsilon(g_{ij}) - \epsilon(w_{ij})}( u_i(w,x) - u_i(g,x)) & \text{if}\ w \in N(g), \ g_{ij} \neq w_{ij} \\
             \sum_{(i,j) \in \mathcal{M}}  \rho((i,j),g, x) F_{\epsilon(1- g_{ij}) - \epsilon(g_{ij})}\left(u_i(g,x) - u_i([1-g_{ij}, g_{-ij}], x)\right) & \text{if} \ g = w \\
             0 & \text{elsewhere}
    \end{cases}\,,
\end{equation}} where $F_{\epsilon(1) - \epsilon(0)}$ and $F_{\epsilon(0) - \epsilon(1)}$ denote the distribution function of the difference in shocks.

\begin{remark}
	\label{rmk_irr}
    The transition matrix is irreducible and aperiodic. By Assumptions \ref{ass.matching} and \ref{ass.shocks}, the first and second cases in \eqref{eq.trans.matrix} are always positive for any $g \in \mathcal{G}$. We can thus always reach any other network $w$ starting from any $g$ with positive probability in finite time (irreducibility). Since the chain is irreducible and contains a self-loop ($\Pi(x)_{gg}>0$), it is also aperiodic.
\end{remark}

We next look for (conditional) stationary distributions. A stationary distribution is an element $\pi(x) \in \Delta (\mathcal{G}) $ satisfying $\pi(x) = \Pi(x)' \pi(x)$.

\begin{remark}
\label{rmk.existence}
    The transition matrix $\Pi(x)$ admits a unique stationary distribution, which is a direct consequence of the Perron-Froebenius theorem for nonnegative irreducible matrices \citep[Theorem 8.4.4]{Horn2012}.  Moreover, as the chain is irreducible and aperiodic, we have that, for any $\pi_0 \in \Delta(\mathcal{G})$, $\lim_{t \to \infty} (\Pi(x)^t)'\pi_0 = \pi(x)$ \citep[Theorem 1.8.3]{Norris1997}, so we may interpret the invariant distribution as a ``long-run'' distribution \citep{Mele2017}.
\end{remark}

\begin{remark}
\label{rmk.pos.distr}
   The stationary distribution puts positive mass over all network configurations. Indeed, since $\pi(x)$ is a distribution, there exists some $g_0 \in \mathcal{G}$  such that $\pi(g_0|x) > 0$.  Fix $w \in \mathcal{G}$. Since the chain is irreducible, there exists $k \in \mathbb{N}$ such that $(\Pi(x)^k)_{g_0,w}>0$, where $(\Pi(x)^k)_{g_0,w}>0$ denotes the $(g_0,w)$ entry of $\Pi(x)^k$. However $\pi(x) = (\Pi(x)^k)' \pi(x) \implies \pi(w|x) > 0$.
\end{remark}

\section{Identification}
\label{identification}
\subsection{Statement of the identification problem}
\label{id_state}
In this section, we study identification under many-network asymptotics. Specifically, we assume that we have access to a random sample (iid across $c$) of $C$ networks $\{G^{T_0}_c, G^{T_1}_c, X_c\}_{c=1}^C$ stemming from the network formation game described in \Cref{setup}.\footnote{Under large-network asymptotics, we have access to a single (a few) network(s) with a large number of players. Identification in this setting consists of providing conditions under which any two sequences of elements of the structural parameter space (which is now indexed by the number of players) with pairwise distinct values lead to asymptotically distinguishable (in a suitable metric) network distributions. See \cite{Mele2017} for further discussion.} In this context, $G^{T_0}_c$ and $G^{T_1}_c$ are observations of network $c$ over two (possibly nonconsecutive) periods\footnote{In Supplemental Appendix \ref{Supplement-identification_stationary}, we briefly discuss (non)identification when only one period of data stemming from the stationary distribution of the network formation game is available.} (labelled \emph{first} and \emph{second}); and $X_c$ is the set of covariates associated with network $c$. We recall the law of $X_c$ is $\prob_{X}$, as $X_c$ is a copy of $X$ (i.e., a random variable with the same law as $X$). As in the previous section, realizations of $X$ are denoted by lower-case letters, i.e., elements $x \in \mathcal{X}$.

Denote by $\Pi(X; \theta_0)$ the transition matrix under covariates $X$; where $\theta_0 \coloneqq ((u_i)_{i=1}^N, \rho)$ are the ``true'' parameters (functions). Further, write $\tau_0$ for the number of \emph{rounds} of the network formation game that has taken place between the first and second periods.  We can identify $\Pi(X;\theta_0)^{\tau_0}$, the transition matrix to the power of the number of rounds of the network formation game which took place between the first and second period ($\tau_0$), provided that the first-period conditional distribution, which we denote by $\pi_0(X)$, is such that $\pi_0(X) >> 0$ $\mathbb{P}_X$-a.s. To see this more clearly, suppose $X$ is empty. In this case, we could consistently estimate $\left(\Pi^{\tau_0}\right)_{gw}$, $g, w \in \mathcal{G}$, by $(\widehat{\Pi^{\tau_0}})_{gw} = \sum_{c=1}^C \mathbbm{1}\{G^{T_0}_c = g, G^{T_1}_c = w\}/\sum_{c=1}^C \mathbbm{1}\{G^{T_0}_c = g \}$,\footnote{In our setting, an agent is described by her set of exogenous characteristics $W_i$. When no such traits are included in the model, two adjacency matrices $g$ and $g'$ are deemed equal if they are equal up to relabeling of agents. When covariates are included, one compares adjacency matrices for a given labeling of agents -- and such labeling will simultaneously define both the distance between the covariate matrices X and X', and between $g$ and $g'$.} provided $\prob[G^{T_0}_c = g] > 0$.\footnote{The case where $X$ has discrete support is similar to the case where no covariates are included: for each $x \in \mathcal{X}$, we consider observations $c$ such that $X_c = x$ for some labelling of agents in $c$. We then use these observations, with the adopted labelling of the agents, to compute the transition probability given $X=x$. When $X$ contains continuous covariates, a consistent estimator is given by a kernel estimator \citep{LiRacine2006}. These estimators would behave poorly in most practical settings (even with few players). We do not suggest using them in practice, though; we rely on consistency only as an indirect argument for establishing identification of the conditional likelihood, i.e. for showing that a function of the distribution of observables recovers these objects.} The next assumption summarizes this requirement.

\begin{assumption}[Full support]
\label{ass.full.support.X}
$\pi_0(g|X) > 0$ for all $g \in \mathcal{G}$ $\prob_X$-a.s.
\end{assumption}

Since $\Pi(X;\theta_0)^{\tau_0}$ is identified from the data under \Cref{ass.full.support.X}, the identification problem subsumes to (denoting by $\Theta$ the parameter space):\footnote{Abstracting from measurability concerns, this is the set of utilities and matching functions that satisfy the assumptions in \Cref{setup}.}
\begin{equation}
	\label{eq_id_condition}
\forall (\theta, \tau) \in \Theta \times \mathbb{N}, \quad  (\theta, \tau) \neq (\theta_0,\tau_0) \implies \Pi(X;\theta)^{\tau} \neq \Pi(X;\theta_0)^{\tau_0}\, ,	
\end{equation}
 where the inequality must hold with positive probability over the distribution of $X$. Requirement \eqref{eq_id_condition} is equivalent to identification off from the model's conditional (on $G_0$ and $X$) likelihood, i.e. that the expected conditional log-likelihood is uniquely maximized at $(\theta_0,\tau_0)$ \citep{Newey1994}.

\begin{remark}[On the assumption of random sampling]
	In our identification analysis, we consider that the researcher observes a random sample of networks. This restricts the number of players to be constant in the population from which our sample is drawn. In a setting where the number of players may actually vary in the population of interest, our results can be alternatively interpreted as providing identification results for structural parameters off from the population distribution, \emph{conditional on the number of players}. Specifically, in a setting where group sizes vary in the population, considering identification of model parameters from the distribution of networks conditional on a group size means our identification results do not require structural parameters to be constant across different network sizes. Indeed, identification conditional on network size allows structural parameters to vary freely across group sizes (provided they respect the identifying restrictions we will later impose). Such flexibility is often desirable for identification purposes, as it dispenses with the identifying power of extrapolating features from one group size to the other that would be inevitably available when the assumption of constancy of (some) structural parameters is introduced in the analysis.\footnote{See \cite{Lee2007,Davezies2009}a dn \cite{Graham2009} for explorations of the identifying power of such cross-group-size restrictions in the context of peer effect models.} %However, some degree of extrapolation is inevitable for estimation purposes due to the limited sample size. Indeed, in our empirical application in \Cref{application}, we consider parametric forms for preferences and meeting opportunities and assume the underlying parameters to be constant across networks of different sizes.
\end{remark}

\subsection{Identification of the number of rounds}
In our statement of the identification problem, the number of rounds in the network formation game is assumed to be unknown. The researcher has no reason to expect $\tau_0$, the true number of rounds, to be known \textit{a priori} unless the network formation algorithm has a clear empirical interpretation. Nonetheless, it is still possible to identify $\tau_0$ under some assumptions. For all $\theta \in \Theta$ and $x \in \mathcal{X}$, $\Pi(x;\theta)$ is irreducible and has a strictly positive main diagonal. It is clear, then, that the number of strictly positive entries in  $\Pi(x;\theta)^\tau$ is nondecreasing in $\tau$. Moreover, this number is strictly increasing for $\tau \leq N(N-1)$\footnote{$N(N-1)$ is the minimum number of rounds required for the probability of transitioning from a ``fully empty'' network ($g_{ij}=0$ for all $ij$) to a ``fully connected'' network ($g_{ij}=1$ for all $ij$) to be strictly positive.} and does not depend on the choice of $(x,\theta)$. Thus, provided that $\tau_0 \leq N(N-1)$, we can identify $\tau_0$ by ``counting''\footnote{We provide a consistent estimator for $\tau_0$ in \Cref{estimation.tau0}.} the number of positive entries in $\Pi(X;\theta_0)^{\tau_0}$.

\begin{assumption}[Upper bound on $\tau_0$]
\label{ass.upperbound.tau0}
The number of rounds that took place in the network formation game between the first and second period ($\tau_0$) is smaller than or equal to $N(N-1)$.
\end{assumption}
\begin{claim}
	\label{claim_bound}
Under Assumptions \ref{ass.matching}, \ref{ass.shocks}, \ref{ass.full.support.X} and \ref{ass.upperbound.tau0}, $\tau_0$ is identified.
\begin{proof}
	See \Cref{proof_bound}.
\end{proof}
\end{claim}

A similar assumption is considered in \cite{Christakis2010}, where the authors assume that $\tau_0 = N(N-1)/2$ (they work with an undirected network, so the set of available matches is divided by two) and that all meeting opportunities are played (though in an unknown order). In their setting, however, the assumption's primary purpose is to reduce the computational toll of evaluating the model likelihood (see \Cref{estimation.preference.meeting} for a similar discussion). In our case, we require it for identification. We also emphasize that the bound in \Cref{ass.upperbound.tau0} is more or less restrictive, depending on the setting. Knowledge of the particular application in mind should help to assess its appropriability. {For example, one may know the number of time periods $\mu$ (e.g. schooldays) between $T_0$ and $T_1$, and expect that, in a particular setting, a maximum of $\ell$ meetings could have taken place at each period. In this case, one can check whether  $N \cdot (N-1)$ is smaller than $\mu \cdot \ell  $ to assess the plausibility of the bound. Finally, we note that our identification results on preferences and meeting parameters hold irrespective of the bound -- it suffices that $\tau_0$ is either known \textit{a priori} or identified.}

In the following subsection, we discuss the identification of the vector of preference and meeting parameters $\theta_0$, taking the number of rounds as identified.  

\subsection{Identification of preferences and meeting probabilities}
\label{sec_ident_pref_meeting}

This section considers the power of exclusion restrictions in identifying preferences and meeting parameters. Our approach is motivated by the results in Supplemental Appendix \ref{app_np}, which show that identification of preferences and meetings in the model defined by Assumptions \ref{ass.matching} and \ref{ass.shocks} is generally impossible absent further restrictions.\footnote{Specifically, Supplemental Appendix \ref{Supplement-identification_stationary} shows that, if networks are drawn from the model's stationary distribution and observed at only a single period of time, then the model is generally unidentified. This result motivates our statement of the identification problem in Section \ref{id_state}, which requires observation of the network at two periods of time. However, Supplemental Appendix \ref{Supplement-non_two} shows that, even in our setting where networks are observed at two points of time,  identification is generally impossible absent further restrictions on meetings and preferences.} We consider different types of exclusion restrictions that limit either the role of some covariate in shaping the meeting process, or that assume that some observed trait does not enter the decision of agents of, upon meeting, forming/mantaining a relationship status. We begin with the simplest cases, where $\tau_0 \in \{1,2\}$, or, alternatively, when the researcher imposes enough restrictions in the problem to ensure that  $\Pi(X;\theta_0)$ or $\Pi(X;\theta_0)^2$ is identified from $\Pi(X;\theta_0)^{\tau_0}$. We will then provide sufficient conditions for the latter later in this section.

\paragraph{Notation} To make the dependence on covariates explicit, we denote by $X^m(g)$ the set of relevant transformations of $X$ entering the meeting probabilities of agents under network $g$. We thus write $\rho((i,j),g,X)  = \rho_{ij}(g|X^m(g))$ for every $(i,j)\in \mathcal{M}$. Similarly, for every $g \in \mathcal{G}$ and $(i,j) \in \mathcal{M}$, we denote by $X_{ij}(g)$ the transformation of $X$ capturing the set of pay-off relevant covariates entering the decision of agent $i$ of altering her relationship status with $j$ under network $g$. We then write $u_i([1-g_{ij},g_{-ij}],X) - u_i(g,X) = \Delta u_{ij}(g|X_{i,j}(g))$.

\subsubsection{Exclusion restrictions in the meeting process}
Our first class of exclusion restrictions considers a situation where a covariate that is relevant for the pay-off of agent $i$ of moving from network $g$ to  $w \in N(g)$, $w_{ij}\neq g_{ij}$, is excluded from the covariates relevant for the meeting process under $g$, i.e. $X^m(g)$.

\begin{assumption}
	\label{ass_simple}
	Fix $g \in \mathcal{G}$, and $w \in N(g)$, $w_{ij} \neq g_{ij}$. For $\mathbb{P}_X$-almost every value $(x^m(g),x^m(w))$  of $(X^m(g),X^m(w))$, there exist known values $z$ amd $\tilde{z}$  in the support of $X_{i,j}(g)|(X^m(g),X^m(w))=(x^m(g),x^m(w))$ such that
	$\Delta u_{ij}(g|z) \neq \Delta u_{ij}(g|\tilde{z})$.
\end{assumption}

Under the above Assumption, we obtain the following result
\begin{proposition}
Under Assumptions \ref{ass.matching}, \ref{ass.shocks} and \ref{ass_simple}, if $\Pi(X;\theta_0)$ is identified, then $\Delta u_{ij}(g|X_{ij}(g))$, $\rho_{ij}(g|X^m(g))$ and $\rho_{ij}(w|X^m(w))$  are identified.
\begin{proof}
	Under Assumption \ref{ass_simple}, we are able to write, for any $(x^m(g),x^m(w))$ in the support of $(X^m(g),X^m(w))$:
	
	$$\frac{\Pi_{g,w}(\{x^m(g),\tilde z\}) - \Pi_{g,w}(\{x^m(g), z\})}{\Pi_{w,g}(\{x^m(w),\tilde z\}) - \Pi_{w,g}(\{x^m(w),z\})} = - \frac{\rho_{ij}(g|x^m(g))}{\rho_{ij}(w|x^m(w))}\, ,$$
	which identifies the ratio $\rho_{ij}(g|x^m(g))/\rho_{ij}(w|x^m(w))$. But then, it follows that, for any $s$ in the support of $X_{i,j}(g)|(X^m(g),X^m(w))=(x^m(g),x^m(w))$ that:
	$$\frac{\Pi_{g,w}(\{x^m(g),s\}) }{\Pi_{w,g}(\{x^m(w),s\})} =  \frac{\rho_{ij}(g|x^m(g))}{\rho_{ij}(w|x^m(w))}\frac{F_{\epsilon(g_{ij})-\epsilon(w_{ij})}(\Delta u_{ij}(g|s))}{1-F_{\epsilon(g_{ij})-\epsilon(w_{ij})}(\Delta u_{ij}(g|s))}\, .$$
	
	Since $u \mapsto \frac{F_{\epsilon(g_{ij})-\epsilon(w_{ij})}(u)}{1-F_{\epsilon(g_{ij})-\epsilon(w_{ij})}(u)}$ is strictly increasing by Assumption \ref{ass.shocks}, it follows that $\Delta u_{ij}(g|s)$ is identified. But then, $\rho_{ij}(g|x^m(g))$ and $\rho_{ij}(w|x^m(w))$ are also identified, since:
	
	\begin{equation}
		\begin{aligned}
				\rho_{ij}(g|x^m(g)) = \frac{\Pi_{g,w}(\{x^m(g),s\}) }{F_{\epsilon(g_{ij})-\epsilon(w_{ij})}(\Delta u_{ij}(g|s))} \\
				\rho_{ij}(w|x^m(w)) = \frac{\Pi_{w,g}(\{x^m(w),s\}) }{1- F_{\epsilon(g_{ij})-\epsilon(w_{ij})}(\Delta u_{ij}(g|s))}
		\end{aligned} \, .
	\end{equation}
	
\end{proof}
\end{proposition}

The above result establishes identification of preferences and meeting parameters pertaining to the decision of a pair $(i,j)$ moving to and from a network $g$. If Assumption \ref{ass_simple} holds for every pair $(i,j) \in \mathcal{M}$ and network $g \in \mathcal{G}$, then we are able to establish identification of all the relevant marginal utilities and meeting parameters. Under a location normalization of the \emph{level} of utilities, e.g. that there exists some $g_0 \in \mathcal{G}$ such that $u_{i}(g_0,X)=0$ for every $i \in \mathcal{N}$, we may then establish identification of utilities in levels.

The previous result considered the case where $\tau_0 = 1$ or $\Pi(X;\theta_0)$ is identified. For the case where $\tau_0=2$, or $\Pi(X;\theta_0)^{2}$ is identified, we require a stronger exclusion restriction, which we state in the following. In what follows, we define $N^s(g)$ as the set of networks that differ from $g$ in \emph{exactly} $s$ edges.

\begin{assumption}
	\label{ass_tau2}
	Fix $g \in \mathcal{G}$ and $h,s \in N(g)$, with $h_{ij} \neq g_{ij}$ and $s_{kl} \neq g_{kl}$, such that, with probability one:
	\begin{itemize}
		\item[a]	$\Delta u_{ij}(g|X_{ij}(g)) = \Delta u_{ij}(s|X_{ij}(s))$ and $\Delta u_{kl}(g|X_{kl}(g)) = \Delta u_{kl}(h|X_{kl}(h))$
	\end{itemize}

	 Let $w \in N^2(g)$ be such that $w_{ij} \neq g_{ij}$ and $w_{kl} \neq g_{kl}$. We assume that, for $\mathbb{P}_X$-almost every value $(x_{ij}(g), x_{kl}(g))$ of $(X_{ij}(g), X_{kl}(g))$:
	
	\begin{itemize}
		\item[b] There exists known points $(\tilde{x}^m(g),\tilde{x}^m(h),\tilde{x}^m(s),\tilde{x}^m(w), \tilde{x}_{kl}(g))$ in the support of $$(X^m(g),X^m(h),X^m(s), X^m(w),X_{kl}(g))|X_{ij}(g)=x_{ij}(g)$$ and a known sequence $(\tilde{z}_n)_{n\in \mathbb{N}}$ in the support of $$X_{i,j}(g)|(X^m(g),X^m(h),X^m(s), X^m(w),X_{kl}(g))=(\tilde{x}^m(g),\tilde{x}^m(h),\tilde{x}^m(s),\tilde{x}^m(w),\tilde{x}_{kl}(g))\,,$$ such that:
		
		$$\lim_{n\to \infty}\Delta u_{ij}(g|\tilde{z}_n) = \infty \, .$$
		\item[c] There exists known points $(\check{x}^m(g),\check{x}^m(h),\check{x}^m(s),\check{x}^m(w), \check{x}_{ij}(g))$ in the support of $$(X^m(g),X^m(h),X^m(s), X^m(w),X_{ij}(g))|X_{kl}(g)=x_{kl}(g)$$ and a known sequence $(\check{z}_n)_{n\in \mathbb{N}}$ in the support of $$X_{kl}(g)|(X^m(g),X^m(h),X^m(s), X^m(w),X_{ij}(g))=(\check{x}^m(g),\check{x}^m(h),\check{x}^m(s),\check{x}^m(w),\check{x}_{ij}(g))\,,$$ such that:
		
		$$\lim_{n\to \infty}\Delta u_{ij}(g|\check{z}_n) = \infty \, .$$
	\end{itemize}
	
\end{assumption}

	Assumption \ref{ass_tau2} considers network configurations $g \in \mathcal{G}$, and pairs of agents $(i,j)$ and $(k,l)$, such that, locally to $g$, the presence of a link between $k$ and $l$ ($i$ and $j$) does not affect the pay-off of $i$ and $j$ ($k$ and $l$) altering their relationship, i.e. we assume that this pay-off is the same under network $g$ and the network $[1-g_{kl},g_{-kl}] = s$ ($[1-g_{ij},g_{-ij}] = h$). This type of assumption is common in the network formation literature, where one usually assumes that the relationships ``far away'' from an agent's existing connections do not exert an effect over her preferences \citep[e.g.][]{dePaula2018a,Jochmans2023}. 
	
	For this subset of networks and pairs, parts (b) and (c) of Assumption \ref{ass_tau2} assume the existence of a ``large-support'' covariate entering the relative pay-off of $(i,j)$ (the pay-off of $(k,l)$) changing their relationship status under network $g$, but excluded from the relative pay-off of $(k,l)$ changing their status under $g$ (the relative pay-off of $(i,j)$ changing their status under $g$), as well as meeting probabilities locally to $g$. These covariates should admit sufficiently ``high'' realizations that would ensure that, upon meeting, agents would almost certainly change their relationship status. The idea is then to leverage information from pairs that would amost certainly change their relationship status to infer about the meeting process, separating it from the role of preferences. This type of ``identification-at-infinity'' argument is commonly employed in the literature on the estimation of games and Industrial Organization \citep{Tamer2003, Bajari2010, Colas2020}, as well as on recent research seeking to disentangle different sources of discrimination \citep{Hull2022,Baron2024}.

	Under Assumption \ref{ass_tau2}, we establish the following result.
	\begin{proposition}
		\label{prop_tau2_pref}
		Under Assumptions \ref{ass.matching}, \ref{ass.shocks} and \ref{ass_tau2}, if $\Pi(X;\theta_0)^2$ is identified, then $\{\Delta u_{ij}(v|X_{ij}(v)), \Delta u_{kl}(v|X_{kl}(v))\}$ are identified, for $v \in \{g,h,s,w\}$. Moreover, for  support points satisfying:
		\begin{equation}
		\label{eq_homoge}
		\rho_{ij}(g|x^m(g)) = \rho_{kl}(g|x^m(g)), \quad \text{ and } \quad \rho_{ij}(w|x^m(w)) = \rho_{kl}(w|x^m(w))	\, ,	\end{equation} 
		the identification region of  $\{\rho_{ij}(v|x^m(v)),\rho_{kl}(v|x^m(v))\}$, $v \in \{g,h,s,w\}$ is given by:
		$$(0,1)^{8} \cap \exp(\Psi)$$
		where $\Psi$ is a one-dimensional affine subspace of $\mathbb{R}^8$. 
		\begin{proof}
			See Appendix \ref{app_proof_tau2_pref}.
		\end{proof}
	\end{proposition} 
	
	Proposition \ref{prop_tau2_pref} establishes point identification of marginal utilities under the exclusion restriction in Assumption \ref{ass_tau2}. It also shows that, under the additional assumption that both pairs under consideration have the same probability of meeting under network configurations $g$ and $w \in N^2(g)$, meeting probabilities are partially identified, with one additional log-linear restriction being sufficient to establish point identification. For example, one can derive additional restrictions -- and thus identifying power -- by specifying parametric forms on preference and meeting parameters. These parametric forms can also be used to \emph{extrapolate} identification results from pairs satisfying Assumption \ref{ass_tau2} to other agents in the network. We discuss the role of parametric forms in extrapolating identified patterns in Section \ref{sec_extrap}.
	
	To conclude, we consider an alternative type of exclusion restriction in the meeting process. Following \cite{Mele2017}, we consider a setting where meeting probabilities do not depend on the existence of a link between agents, i.e. the meeting technology does not exhibit differential behavior on the basis of prior relationships. We show that, in our two-period setting, this assumption has identifying power, being able to recover preferences and meeting probabilities. This is in contrast to a setting with one period of data, wherein  Supplemental Appendix \ref{Supplement-identification_stationary} shows that the assumption does not bring identifying power.
	\begin{proposition}
		\label{prop_exclude_meeting}
		Suppose that Assumptions \ref{ass.matching} and \ref{ass.shocks} hold, and that, for every $g \in \mathcal{G}$ and $h \in N(g)$, $g_{ij}\neq h_{ij}$, we have, with probability one:
		
		$$\rho_{ij}(g|X^m(g))=\rho_{ij}(h|X^m(h)) \, .$$
		
		We then have that:
		
		\begin{enumerate}
			\item[a] Marginal utilities and meeting probabilities are identified from $\Pi(X;\theta_0)$, for every pair of agents and network configuration.
			\item[b] The marginal utilities are identified from $\Pi(X;\theta_0)^2$ for every pair  of agents and network configuration. Moreover, meeting probabilities are identified for every configuration of pairs  $(i,j)\neq (k,l)$, networks $g \in \mathcal{G}$, and support points satisfying:
			
			$$\rho_{ij}(g|x^m(g)) = \rho_{kl}(g|x^m(g)) = \rho_{ij}(w|x^m(w)) = \rho_{kl}(w|x^m(w))\, ,$$
			where $w \in N^2(g)$ with $g_{ij} \neq w_{ij}$ and $g_{kl} \neq w_{kl}$.
		\end{enumerate}
		\begin{proof}
			See Appendix \ref{proof_prop_exclude_meeting}.
		\end{proof}
	\end{proposition} 
\subsubsection{Exclusion restriction in preferences}

In this section, we consider a situation where large support covariates are included in the meeting process, but excluded from preferences. For the case where $\Pi(X;\theta_0)$ is identified, we consider the following assumption:

\begin{assumption}
\label{ass_meet_tau1}
	Fix $g \in \mathcal{G}$ and $ij \in \mathcal{M}$. For $\mathbb{P}_X$ every support point $x_{ij}(g)$ of $X_{ij}(g)$, there exists a known sequence $(z_n)_{n \in \mathbb{N}}$ in the support of $X^m(g)|X_{ij}(g)=x_{ij}(g)$ such that:
	
	$$\lim_{n \to \infty }\rho_{ij}(g|z_n) = 1\, .$$
\end{assumption}

Assumption \ref{ass_meet_tau1} considers a situation where it is possible to find a covariate that affects the meeting probability of agents under network $g$, but is excluded from the relative pay-off of $(i,j)$ changing their relationship status under network $g$. Assumption \ref{ass_meet_tau1} requires this covariate to have a large support, insofar that it is possible to find a sequence of support points of $X^m(g)$ where the probability of agents $ij$ meeting under $g$ converges to one, while the relative pay-off of the pair changing status upon meeting remains unchanged. 

For the case where $\Pi(X;\theta_0)^2$ is identified, we require a strengthening of Assumption \ref{ass_meet_tau2}. First, we require the existence of a large support covariate entering $X^m(g)$ that is also be exluded from the covariates entering meeting probabilities in the ``adjacent network'' $w = [1-g_{ij},g_{-ij}]$. Second, we also require the existence of a limit where the relative probability of pair $ij$ meeting under networks $w$ \emph{and} $g$ converges to one.

\begin{assumption}
	\label{ass_meet_tau2}
	Fix $g \in \mathcal{G}$ and $ij \in \mathcal{M}$, and let $w = [1-g_{ij},g_{-ij}]$. For $\mathbb{P}_X$-almost every support points $(x_{ij}(g),x^m(w))$ in the suppport of $(X_{ij}(g),X^m(w))$, there exists a known sequence $(z_n)_{n \in \mathbb{N}}$ in the support of the conditional distribution $X^m(g)|X_{ij}(g)=x_{ij}(g),X^m(w)=x^m(w)$ such that:
	
	$$\lim_{n \to \infty }\rho_{ij}(g|z_n) = 1\, .$$
	
	Moreover, there exists a sequence $(a_n,b_n)$ in the support of $(X^m(g), X^m(w))|X_{ij}(g)=x_{ij}(g)$ such that:
	$\lim_{n \to \infty }\rho_{ij}(g|a_n)/\rho_{ij}(g|b_n) = 1$.
\end{assumption}

Using the above restrictions, we establish the following identification result

\begin{proposition}
	\label{prop_meet_excl}
	Suppose that assumptions \ref{ass.matching} and \ref{ass.shocks} hold. We then have that:
	
	\begin{enumerate}
		\item[a] If $\Pi(X;\theta_0)$ is identified and Assumption \ref{ass_meet_tau1} holds, then $\rho_{ij}(g|X^m(g))$, $\rho_{ij}(w|X^m(w))$ and $\Delta u_{ij}(g|X_{ij}(g))$ are identified.
		\item[b] If $\Pi(X;\theta_0)^2$ is identified and Assumption \ref{ass_meet_tau2} holds, then $\rho_{ij}(g|X^m(g))$, $\rho_{ij}(w|X^m(w))$ and $\Delta u_{ij}(g|X_{ij}(g))$ are identified.
	\end{enumerate}
	\begin{proof}
		See Appendix \ref{app_meet_excl}.
	\end{proof}
\end{proposition}
\subsubsection{Extending to the case where $\tau_0 > 2$}
The previous identification results considered the case where $\Pi(X;\theta_0)$, or $\Pi(X;\theta_0)^2$, is identified. We now provide sufficient conditions for these objects to be identified from $\Pi(X;\theta_0)^{\tau_0}$. For that, we impose the following structure.

\begin{assumption}[Distribution of shocks]
	\label{ass_distribution}
	Taste shocks $(\epsilon(0), \epsilon(1))$ are independent draws from an extreme-value type 1 distribution.
\end{assumption}

\begin{assumption}[Potential function in preferences]
		\label{ass_pot_pref}
	There exists a function $Q: \mathcal{G} \times \mathcal{X} \mapsto \mathbb{R}$ such that, for every $g \in \mathcal{G}$, $w \in N(g)$ with $g_{ij} \neq w_{ij}$ and $x \in \mathcal{X}$.
	
	$$Q(w,x)-Q(g,x) = \Delta u_{ij}(g|x_{ij}(g))\, .$$
\end{assumption}

\begin{assumption}[Potential function in meetings]
			\label{ass_pot_meet}
		There exists a function $Q: \mathcal{G} \times \mathcal{X} \mapsto \mathbb{R}$ such that, for every $g \in \mathcal{G}$, $w \in N(g)$ with $g_{ij} \neq w_{ij}$ and $x \in \mathcal{X}$:
		
		$$\log\rho_{ij}(w|x^m(w)) - \log\rho_{ij}(g|x^m(g)) = M(w,x) - M(g,x)$$
\end{assumption}

The assumption that taste shocks follow an extreme value type 1  distribution (Assumption \ref{ass_distribution}) is a staple in structural models featuring discrete choices since the seminal work of \cite{McFadden1974}. Assumption \ref{ass_pot_pref} requires that preferences admit a potential function. It is satisfied by the class of preferences considered by \cite{Mele2017}, which we also adopt in our empirical application in Section \ref{application}.\footnote{More generally, if agents internalize the externality of their individual choices on other agents' pay-offs, then preferences will admit a potential function. Such representation is not restricted to these settings, though.} Finally, Assumption \ref{ass_pot_meet} requires that the effect of a relationship on the meeting probability of a pair be representable by a ``potential'' meeting function. This assumption is a strict relaxation over \citeauthor{Mele2017}'s setting, where meeting probabilities do not depend on the existence of a link between agents, i.e. $M$ is a constant function. We provide an example where the assumption holds (approximately) with a nontrivial $M$ in the next section, which we also adopt in our empirical application.

Under the above assumptions, we are able to state the following result.

\begin{proposition}
	\label{prop_spectral}
	Suppose that Assumptions  \ref{ass.matching}, \ref{ass.shocks}, \ref{ass_distribution}, \ref{ass_pot_pref} and \ref{ass_pot_meet} hold. We then have that:
	
	\begin{itemize}
		\item[(a)] If $\tau_0$ is odd, then $\Pi(x;\theta_0)$ is identified from $\Pi(x;\theta_0)^{\tau_0}$, for any $x \in \mathcal{X}$.
		\item[(b)]If $\tau_0$ is even, then $\Pi(x;\theta_0)^2$ is identified from $\Pi(x;\theta_0)^{\tau_0}$, for any $x \in \mathcal{X}$. Moreover, $\Pi(x;\theta_0)$ is identified from $\Pi(x;\theta_0)^{\tau_0}$, for every support $x$ of $\mathcal{X}$ where $\operatorname{det}(\Pi(x;\theta_0)^2) > \exp\left(-2\pi\left[\frac{1}{\sqrt{3}}\mathbf{1}\{N>2\} + \mathbf{1}\{N=2\}\right]\right)$.
	\end{itemize}
	\begin{proof}
		See Appendix \ref{proof_prop_spectral}.
	\end{proof}
\end{proposition}

Proposition \ref{prop_spectral} leverages results on the spectral theory of Markov matrices to provide conditions under which it is possible to recover $\Pi(X;\theta_0)$ or $\Pi(X;\theta_0)^2$ from the identified transition matrix $\Pi(X;\theta_0)^{\tau_0}$. Part (a) shows that, if the number of rounds is odd, then $\Pi(x;\theta_0)$ is recoverable from $\Pi(x;\theta_0)^{\tau_0}$, for any $x$ in the support of $X$. Therefore, if $\tau_0$ is identified to be an odd number, one can leverage the identification results in previous sections that assumed $\Pi(X;\theta_0)$ to be identified in order to recover preferences and meeting probabilities. Similarly, part (b) shows that, if $\tau_0$ is even, then $\Pi(x;\theta_0)^2$ is recoverable from $\Pi(x;\theta_0)^{\tau_0}$. In this case, identification results that assumed  $\Pi(X;\theta_0)^2$ to be known can be used to recover $\theta_0$. As an alternative to the latter, the second statement in item (b) shows that, with an even number of rounds, it is still possible to recover $\Pi(x;\theta_0)$ for the support points $x$ of $X$ where the transition matrix exhibits sufficient variability. If the identification conditions in previous subsections that assumed $\Pi(x;\theta_0)$ to be known can be shown to hold in this restricted region of the support, then it would be possible to identify preferences and meetings for this region of the support without relying on $\Pi(x;\theta_0)^2$; and then to extrapolate these patterns to the remainder of the support under additional assumptions. While potentially promising, we do not further exploit this alternative strategy in this paper. In the next subsection, we explore parametric forms and the role of extrapolation in aleviating the identification conditions when $\Pi(x;\theta_0)$ or $\Pi(x;\theta_0)^2$ are identified for every support point $x$ of $X$.

\begin{remark}[Relaxing assumptions \ref{ass_distribution}, \ref{ass_pot_pref} and \ref{ass_pot_meet}]
	Supplemental Appendices \ref{identification_covariates} and \ref{Supplement-ident.exclusion.matching} provide identification results for preference and meeting parameters directly from $\Pi(x;\theta_0)^{\tau_0}$. In this case, we may dispense with Assumptions \ref{ass_distribution}, \ref{ass_pot_pref} and \ref{ass_pot_meet}, as we need not infer $\Pi(x;\theta_0)$ or $\Pi(x;\theta_0)^2$ from $\Pi(x;\theta_0)^{\tau_0}$ beforehand to proceed with the identification analysis. Specifically, Supplemental Appendix \ref{identification_covariates} provides identification results that leverage large-support covariates entering preferences, but that are excluded from the meeting process. Symmetrically, Supplemental Appendix \ref{Supplement-ident.exclusion.matching} establishes identification under large support covariates entering the meeting process, but excluded from preferences. Given direct reliance on $\Pi(x;\theta_0)^{\tau_0}$ for identification, these results require more stringent exclusion restrictions, as well as the existence of a larger set of large-support limits, than the identification-at-infinity arguments introduced in this and earlier sections. We thus view these more stringent assumptions as the price of being robust to Assumptions \ref{ass_distribution}-\ref{ass_pot_meet}.
	
\end{remark}
\subsubsection{Parametric forms and the role of extrapolation}
\label{sec_extrap}

Finally, we consider the role of parametric restrictions in relaxing the exclusion restrictions previously considered. Following \cite{Mele2017}, we consider the following parametric form for preferences.

\begin{equation} 
	\label{eq_formulation}
	\begin{aligned}
		u_i(g,X) = \underbrace{\sum_{j \neq i} \beta_{ud}' \begin{pmatrix} 1 \\ W_{ij} \end{pmatrix} g_{ij}}_{\text{direct links}} + \underbrace{\sum_{j \neq i}  \beta_{ur} ' \begin{pmatrix} 1 \\ W_{ij} \end{pmatrix} g_{ij} g_{ji}}_{\text{mutual links}} + \\ +  \underbrace{\sum_{j \neq i} g_{ij} \sum_{\substack{l \neq i\\ l \neq j}} \beta_{un} ' \begin{pmatrix} 1 \\ W_{il} \end{pmatrix} g_{jl}}_{\text{indirect links}} +  \underbrace{\sum_{j \neq i} g_{ij} \sum_{\substack{l \neq i\\ l \neq j}} \beta_{up} '\begin{pmatrix} 1 \\ W_{jl}  \end{pmatrix}g_{li}}_{\text{popularity}} \,,
	\end{aligned} 
\end{equation}
where $W_{ij}$ is a vector of observed pair-level traits between agents $i$ and $j$, e.g. measures of pairwise distance in their observed characteristics. The first term of the specification allows agents $i \in \mathcal{I}$ to accrue a direct benefit from their connections $g_{ij}$, $j \neq i$. Moreover, this gain may vary according to the observed differences in agents' traits, i.e. one allows for homophily in direct connections. The second term of the specification further allows agents to accrue a differential gain from reciprocal relationships, i.e. to have an additional gain from establishing a connection $g_{ij}$ with a pair such that $g_{ji} = 1$. The third term allows agents to derive utility from their friends' connections, whereas the fourth term captures a ``popularity'' effect (individuals derive utility of serving as a ``bridge'' between agents).\footnote{The inclusion of preferences over indirect links and popularity may be seen as providing a rationale for observed clustering patterns that are common in friendship networks (see \cite{Badev2017}; also \cite{Jackson2007}). } Under the Assumption that $\beta_{\text{un}} = \beta_{\text{up}}$, \cite{Mele2017} shows that the utility function admits a potential function $Q$ as in Assumption \ref{ass_pot_pref}.

Next, we consider the following parametrization for the meeting process:
\begin{equation}
	\label{eq_pair}
	\rho_{ij}(g|X) = \frac{\exp(\gamma'W_{ij} + \delta g_{ij} + g_{ij}\psi'Z_{ij}) }{\sum_{(k,l) \in \mathcal{M}} \exp(\gamma'W_{ij} + \delta g_{kl} + g_{kl}\psi'Z_{kl}) }\, , 
\end{equation}
where $Z_{ij}$ is a vector of pair-level traits that satisfties a certain exclusion restriction with respect to $W_{ij}$. Specification \eqref{eq_pair} allows homophiliy in meeting probabilities with respect to observed traits $W_{ij}$, as well as dependence on the meeting process with respect to prior relationships. For example, if $\delta > 0$, the meeting process is biased towards agents who are already friends, as these are more likely to meet than those pairs of individuals who do not have a relationship. Moreover, our specification allows this effect to be heterogeneous with respect to the traits $Z_{ij}$.

Under the parametric form \eqref{eq_pair}, if we assume that $\mathcal{M}$ is sufficiently large relatively to the value of the coefficients $\delta$ and $\psi$, insamuch that the percentage impact of changing the value of one of the edge indicators over the \emph{denominator} of \eqref{eq_pair} is negligible, then we have that, for every $g \in \mathcal{G}$ and $w \in N(g)$ with $g_{ij} \neq w_{ij}$:

$$\log(\rho_{ij}(g|X)) - \log(\rho_{ij}(w|X)) \approx \delta (w_{ij} - g_{ij}) + (w_{ij} -g_{ij}) \psi'Z_{ij} \,   $$
which shows that, up to approximation error,  meeting probabilities admit the following potential meeting function $M$:

\begin{equation}
	\label{approx_meeting}
	M(g,x) = \sum_{(i,j) \in \mathcal{M}} g_{ij}(\delta+\psi'Z_{i,j})\, .
\end{equation}

The following proposition provides identification results for the parametric model defined by equations \eqref{eq_formulation} and \eqref{eq_pair}, under the assumption of extreme-value type 1 taste shocks and the approximation \eqref{approx_meeting}.

\begin{proposition}
	\label{proposition_parametric}
	Consider the parametrization of preferences given by \eqref{eq_formulation}, with $\beta_{\text{un}} = \beta_{\text{up}}$, and the parametrization of meetings given by \eqref{eq_pair}. Suppose that taste shocks satisfy Assumption \ref{ass_distribution}, and assume the potential meeting function approximation \eqref{approx_meeting} holds without error. If the following rank condition is satisfied:
	\begin{itemize}
		\item[i] There exists a pair $(i,j) \in \mathcal{M}$ such that:
		$$\mathbb{E}[L_{ij}L_{ij}']$$ has full rank, where $L_{ij} = \begin{bmatrix}
			1 & Z_{ij}' & W_{ij}'
		\end{bmatrix}$ ;
	\end{itemize}
we then have that $\psi, \beta_{\text{ur}}, \beta_{\text{un}}, \beta_{\text{up}}$, and every entry of $\beta_{\text{ud}}$ except the first one,  are identified from $\Pi(X;\theta_0)^{\tau_0}$. Moreover, the remaining parameters, namely $\gamma$, $\delta$, and the first entry of $\beta_{\text{ud}}$, are identified from $\Pi(X;\theta_0)^{\tau_0}$ if, in addition:
	\begin{itemize}
		\item[ii] If $\tau_0$ is odd, there exists pairs $(a,b)$ and $(c,d)$ such that the following rank conditions hold for $\Delta \tilde{W} \coloneqq (W_{a,b} - W_{c,d})$:
		
		\begin{itemize}
			\item (Condition I)  $\mathbb{E}[\Delta \tilde{W}\Delta \tilde{W}']$ has full rank.
			\item (Condition II) For every $d \in \mathbb{R}$, we have that:
			
			\begin{align*}
				\mathbb{E}\left[ \left(\operatorname{Logit}(c + \beta_{ud}' [1 \ W_{a,b}']')- \operatorname{Logit}(c + \beta_{ud}' [1 \ W_{c,d}']'))\right)\right]\neq\\ 
				 \mathbb{E}[\Delta \tilde{W}]'	 \mathbb{E}[\Delta \tilde{W}\Delta \tilde{W}']^{-1}\mathbb{E}\left[\Delta \tilde{W} \left(\operatorname{Logit}(c + \beta_{ud}' [1 \ W_{a,b}']')- \operatorname{Logit}(c + \beta_{ud}' [1 \ W_{c,d}']'))\right)\right]
			\end{align*}
			
		\end{itemize}
		\item[iii] If $\tau_0$ is even, there exists three pairs $(a,b)$, $(c,d)$ and $(e,f)$, $e\notin \{b,d\}$ and $f \notin \{a,c\}$,  such that there exists a suport point $z_{e,f}$ of $Z_{e,f}$, a known direction $\overrightarrow{t} \in \mathbb{R}^{\dim(Z_{e,f})}$, and $\bar{\epsilon} > 0$: such that:
		
		\begin{itemize}
			\item (Continuous variation) The event $A_{\epsilon} = \{ Z_{e,f}  \in [z_{e,f} - \epsilon \overrightarrow{t} , z_{e,l} + \epsilon \overrightarrow{t}] \}$ has positive probability, for every $\epsilon < \bar{\epsilon}$, conditionally on $\{W_m: m \in \mathcal{M}\}$.
			\item (Relevance) $\psi' \overrightarrow{t} \neq 0$;
			\item (Rank condition) Denoting by $\Delta \tilde{W} \coloneqq (W_{a,b} - W_{c,d})$, the following conditions hold:
			\begin{itemize}
				\item The smallest eigenvalue of  $\mathbb{E}[(W_{a,b} - W_{c,d})(W_{a,b} - W_{c,d})'|A_\epsilon]$ is bounded away from zero, uniformly over $0<\epsilon < \bar{\epsilon}$.
				
				 \item For every $d \in \mathbb{R}$, there exists a $v >0$ such that, for every $0<\epsilon < \bar{\epsilon}$

			\end{itemize}

		\end{itemize}
	\end{itemize}
			\begin{align*} \footnotesize 
		\Big|\mathbb{E}\left[ \left(\operatorname{Logit}(c + \beta_{ud}' [1 \ W_{a,b}']')- \operatorname{Logit}(c + \beta_{ud}' [1 \ W_{c,d}']'))\right)|A_\epsilon\right] -\\ 
		\mathbb{E}[\Delta \tilde{W}|A_\epsilon]'	 \mathbb{E}[\Delta \tilde{W}\Delta \tilde{W}'|A_\epsilon]^{-1}\mathbb{E}\left[\Delta \tilde{W} \left(\operatorname{Logit}(c + \beta_{ud}' [1 \ W_{a,b}']')- \operatorname{Logit}(c + \beta_{ud}' [1 \ W_{c,d}']'))\right)\Big|A_\epsilon\right]\Big|  \\
		> v 
	\end{align*}
	\begin{proof}
		See Appendix \ref{app_parametric}.
	\end{proof}
\end{proposition}

Proposition \ref{proposition_parametric} provides identification results for the parametric model defined by equations \eqref{eq_formulation} and \eqref{eq_pair}. The first part of the proposition shows that every preference parameter except the ``intercept'' in the direct-links part of pay-offs, as well as the role of the covariates $Z_{ij}$ in the meeting process, are identified under a standard rank condition \emph{for a single pair}. The parametric restriction then effectively allow us to ``extrapolate'' these quantities to other pairs in the network. Notice that the rank condition effectively acts as an exclusion condition, since we do not allow $Z_{ij}$ to be collinear with $W_{ij}$.\footnote{While it is possible to allow some of the entries of $Z_{ij}$ to vary colinearly with $W_{i,j}$, this would require additional rank conditions in the second and third parts of the proposition, in order to be able to separately identify $\beta_{\text{ud}}$ from $\psi$. We opt not pursue this extension, as in our empirical specification $Z_{i,j}$ only includes ``excluded'' covariates.} The second and third parts of the proposition provide sufficient conditions for identification of the remainder parameters. Part (ii) shows that, when the number of rounds is odd, a two-part rank condition identifies the remaining parameters. The first part of the rank condition is implied by assumptions typically used in conditional logit analyses \citep[e.g.][]{McFadden1974}. We supplement this condition with an additional assumption that essentially requires the difference in the direct pay-offs of pairs $(a,b)$ and $(c,d)$ in forming a relationship to exhibit sufficient nonlinear variation so as \textbf{not} to be approximated, in expectation, by a linear function of $\Delta \tilde{W}$.\footnote{For example, this condition excludes the possibility of $\beta_{ud} =0$ and one of the differences between $W_{ab}-W_{cd}$ being constant.} Under the two-part rank condition, we are able to identify $\gamma$, and,  having done so, it is then possible to separate $\delta$, which captures the role of existing links in the meeting process, from the intercept in $\beta_{\text{ud}}$, which captures the ``base'' pay-off of a direct connection. When the number of rounds is \text{even}, additional variation is needed, though. Part (iii) states that, in this case, the two-part rank condition must be satisfied while allowing for marginal perturbations in the covariates $Z$ of a third pair. These perturbations should affect the odds of the pair meeting when a link is present (the relevance condition). Finally, we note that the first item in part (iii) implies further exclusion conditions on $Z_{e,f}$ than those implied by part (i), since we assume variation in some direction conditionally on $\{W_m: m \in \mathcal{M}\}$.

\section{Estimation}
\label{estimation}
In this section, we will analyze estimation. We have access to a sample of $C$ networks, $\{G^{T_0}_c, G^{T_1}_c, X_c\}_{c=1}^C$, stemming from the network formation game previously described.

\subsection{Estimating $\tau_0$}
\label{estimation.tau0}
We first propose to estimate $\tau_0$ as follows:

\begin{equation}
    \label{estimate.tau0}
    \hat{\tau} = \max_{c}\{\lVert G^{T_1}_c - G^{T_0}_c\rVert_{F}\} \,,
\end{equation}
where $\lVert \cdot \rVert_{F}$ is the Froebenius norm. This estimator is intuitive: it amounts to ``counting'' the number of different edges between periods in each network and then taking the maximum. It turns out that, under iid sampling and the bound in \ref{ass.upperbound.tau0}, $\hat{\tau} \overset{a.s.}{\to} \tau_0$.

\begin{lemma}
\label{lemma.est.tau0}
Suppose $\{G^{T_0}_c, G^{T_1}_c, X_c\}_{c=1}^C$ is a random sample (iid across $c$). Under Assumptions \ref{ass.matching}, \ref{ass.shocks}, and \ref{ass.upperbound.tau0}, $\hat{\tau} \overset{a.s.}{\to} \tau_0$.
\begin{proof}
See \Cref{tau_proof}.
\end{proof}
\end{lemma}

As we argue in  \Cref{tau_proof}, the previous lemma can be extended to accommodate a setting where the sequence of observations is independently drawn from games with common $\tau_0$, but where preferences and the meeting process, the distribution of covariates and the number of players may vary with $c$, provided that the number of rounds, $N_c$, is such that $\limsup_{c \to \infty} \mathbb{P}[N_c (N_c - 1) \geq \tau_0] > 0$ holds and that preferences, the meeting process and the distribution of covariates do not asymptotically concentrate on regions where the probability of a network changing by strictly less than $\tau_0$ edges is arbitrarily close to one. This extension is important, as in most practical settings, one expects variation in group sizes: indeed, this is the case in our empirical application in \Cref{application}. Notice that, in these settings, if the sample of networks is assumed to be randomly drawn from a population heterogeneous in $N$, then  $\limsup_{c \to \infty} \mathbb{P}[N_c (N_c - 1) \geq \tau_0] > 0$ is equivalent to  $\mathbb{P}[N (N - 1) \geq \tau_0] > 0$, i.e. the condition requires there exist a positive mass of networks whose number of players $N$ is such that $N (N - 1) \geq \tau_0$. In other words, we do not need all networks in the population to satisfy \Cref{ass.upperbound.tau0}, just that a positive mass do.

{Notice that, even though $\hat{\tau}$ is consistent for $\tau_0$ under the Assumptions required by Lemma \ref{lemma.est.tau0}, our proposed estimator approaches $\tau_0$ from below, thus being generally downward biased. We thus recommend researchers to assess the sensitivity of their results to the estimated value of $\tau_0$. This can be achieved by directly veryfing how estimates of preference and meeting-related parameters vary when one changes the estimated value of $\tau_0$ -- an approach we undertake in our empirical application in Section \ref{application}. Alternatively, if one adopts a Bayesian approach to estimation -- this is our recommended strategy to estimating preferences and meeting parameters in Section \ref{estimation.preference.meeting} --, one could choose a data-driven prior for $\tau_0$ that takes our estimator $\hat{\tau}$ as a lower endpoint to its support. See Remark \ref{rmk_bayes_tau0} in the next section for further discussion on the specification of priors for $\tau_0$ in a Bayesian estimation framework. }

\begin{remark}[On the rate of convergence of $\hat \tau$]
	The strong consistency of the estimator $\hat \tau$ in Lemma \ref{lemma.est.tau0}, combined with the fact that $\hat \tau$ takes values on a discrete set, implies that, for any sequence $(r_c)_{c \in \mathbb{N}}$ such that $\lim_{c \to \infty} r_c = \infty$, $r_C (\hat \tau - \tau_0) \overset{a.s.}{\to} 0 $. In other words, $\hat \tau$ has a faster rate of convergence than any estimator that converges at a rate $r_C$. This entails that, from a frequentist point of view, first-step estimation of the number of meeting rounds does not affect the asymptotic distribution of parametric estimators of preferences and meetings. To see this, denote by $\hat{\beta}_C(\tau)$ an estimator of a parametrization $\beta_0$ of meetings and preferences, when the number of rounds in the first step of estimation is fixed at $\tau \in \mathbb{N}$. Suppose that, for some sequence $(r_c)_{c \in \mathbb{N}}$ diverging to infinity and probability law $L$, $r_C(\hat \beta(\tau_0) - \beta_0) \overset{d}{\to} L$. We then have that $r_C(\hat \beta(\hat \tau) - \beta_0) = r_C(\hat \beta(\tau_0) - \beta_0) + e_C$, where $e_C \overset{a.s.}{\to}0$. This yields that $r_C(\hat \beta(\hat \tau) - \beta_0) \overset{d}{\to} L$, meaning that first-step estimation of $\tau_0$ does not affect the asymptotic distribution of the second step estimator.    \end{remark}

\subsection{Estimation of preference and meeting parameters}
\label{estimation.preference.meeting}

Let vector $\beta_0 \in \mathbb{B}  \subseteq \mathbb{R}^l$ encompass a parametrization of preferences and meetings, i.e. $u_{i}(g,X) = u_{i}(g,X;\beta_0)$ and $\rho_{ij}(g, X) = \rho_{ij}(g,X;\beta_0)$ for all $(i,j) \in \mathcal{M}$, $g \in \mathcal{G}$. If the adopted parametrization satisfies one of the identification conditions in Section \ref{identification}, we know that $\beta_0$ is the unique minimizer of the expectation of the conditional log-likelihood, i.e. $$\beta_0 = \operatorname{argmin}_{\beta \in \mathbb{B}} \mathbb{E}[\log (\Pi(X_c;\beta)^{\tau_0}_{G^{T_0}_c, G^{T_1}_c})|X_c, G^{T_0}_c]\, .$$

As a consequence, a natural estimator to be considered in this case is a conditional MLE that replaces the expectation operator  with its sample analog and $\tau_0$ with a consistent estimator of the number of rounds. To compute this estimator, we note that the log-likelihood of a second-period network $G^{T_1}_c$, conditional on $X_c$ and $G^{T_0}_c$, may be written as:

$$ l_c(G^{T_1}_c|G^{T_0}_c, X_c; \tau_0, \beta) = \sum_{g \in \mathcal{G}}  \mathbbm{1}\{G^{T_1}_c = g\} \ln \left((\Pi(X_c;\beta)^{\tau_0})_{G^{T_0}_c g}\right) \,, $$
and the sample log-likelihood, under an independent sequence of observations, is

\begin{equation}
	\label{eq_log_lkl}
	\mathcal{L}(\{G^{T_1}_c\}_{c=1}^C|\{G^{T_0}_c\}_{c=1}^C; \tau_0, \beta) = \sum_{c = 1}^C \sum_{g \in \mathcal{G}}  \mathbbm{1}\{G^{T_1}_c = g\} \ln \left((\Pi(X_c;\beta)^{\tau_0})_{G^{T_0}_c g}\right) \,. 
\end{equation}

The  second-step MLE estimator will thus be

$$\hat{\beta}_{\text{MLE}} \in \text{argmax}_{\beta \in \mathbb{B}} \mathcal{L}(\{G^{T_1}_c\}_{c=1}^C|\{G^{T_0}_c\}_{c=1}^C; \hat{\tau}, \beta) \,,$$ where $\hat{\tau}$ is the estimator discussed in the previous section. This formulation can also be easily modified to accommodate observations of networks with different numbers of players.

Numerically, computation of the likelihood is complicated by the fact that we need to sum over all walks between $G^{T_0}_c$ and $G^{T_1}_c$.\footnote{A walk between $g$ and $w$ in $\tau$ rounds is a sequence of networks $g_1, \ldots, g_{\tau}$ such that $g_1=g$, $g_\tau = w$ and $g_t \in N(g_{t-1})\cup \{g_{t-1}\}$ for all $t=2,\ldots,\tau$.} For a small estimate of $\tau_0$, this is feasible, but for higher estimated values of $\tau_0$, it becomes impractical. Indeed, for a given number of rounds $\tau \in \mathbb{N}$, there are $[N(N-1) + 1]^{\tau}$ walks starting from $G^{T_0}_c$ and ending in some network $g \in \mathcal{G}$. Evaluating the model likelihood requires summing over all walks ending in $G^{T_1}_c$. A ``recursive'' approach for evaluating the model likelihood would consist of, for each $c \in \{1,2\ldots C\}$, ``writing down'' the formula for each walk iteratively, i.e., starting from $G^{T_0}_c$, compute all possible $N(N-1)+1$ transitions in the first round; then, for each of these $N(N-1)+1$ possible transitions, compute the $N(N-1)+1$ transitions in the second round and multiply each of these probabilities by the probability of the associated transition in the first round, and so on; and then summing over all walks ending in $G^{T_1}_c$. Walks that ``strand off'' from $G^{T_1}_c$ in some round $r < \tau$ can be excluded from the next steps in the recursion,\footnote{By ``strand off'' we mean a path of realizations of the stochastic process $g^t_c$ up to round $r$ such that the probability of reaching $G^{T_1}_c$ in $ \tau-r$ rounds is 0.} which ameliorates the computational toll, but does not solve it.

Given the above difficulty, an interesting alternative is to work with simulation-based methods, which allows us to bypass direct evaluation of the model likelihood. We take a Bayesian perspective\footnote{From the frequentist point of view, a simulated method of moments estimator would be a possibility in our case, though the nonsmoothness of the objective function (which involves indicators of simulated network observations) as well as the poor properties of GMM estimators with many moment conditions (transition probabilities) in finite samples \citep{NeweySmith2004} are unappealing. An indirect inference approach is also unappealing, as low-dimensional sufficient statistics are unknown in our context.} and follow an approach known as likelihood-free estimation or approximate Bayesian computation (ABC) \citep{Sisson2011}.\footnote{From the Bayesian point of view, an alternative and well-known approach to simplify, but not bypass, a complicated model likelihood is data augmentation \citep{Hobert2011}. However, this alternative is not useful in our setting due to the dimensionality of the support of the meeting process. Thus, an approach that conditions the likelihood on the (unobserved) matching process (reducing the number of walks starting at $G^{T_0}_c$ from $[N(N-1) + 1]^{\tau}$ to $2^\tau$) is unworkable here in our case, since we still have to draw from the matching process distribution conditional on the data and the model parameters.} This method bears a close correspondence to nonparametric (frequentist) estimation \citep{Blum2010} and indirect inference \citep{Frazier2018}. The methodology requires the researcher to be able to draw a sample (or statistics thereof) from the model given the parameters. \Cref{alg:abc} outlines the simplest accept-reject ABC algorithm in our setting, where $S$ is the maximum number of iterations and $p_0(\beta,\tau)$ is a prior distribution over $\mathbb{B} \times \mathbb{N}$:\footnote{In practice, a few improvements can be made upon \Cref{alg:abc} \citep{Wentao2018a}. First, we can use importance sampling: instead of drawing from the prior, we may draw from a proposal distribution $q_0(\beta,\tau)$ such that $\text{supp} \ p_0 \subseteq \text{supp} \ q_0$. Accepted draws should then  be associated with weights  $w_s \coloneqq p_0(\beta_s, \tau_s)/q_0(\beta_s, \tau_s)$. \cite{Wentao2018a} provides a data-driven method to select the proposal. Second, we can use a ``smooth'' rejection rule, i.e., we accept a draw with probability $K(\lVert  T_s - T_{\text{obs}}\rVert/\epsilon)$, where $K(\cdot)$ is a rescaled univariate kernel such that $K(0) = 1$. See Supplemental Appendix \ref{Supplement-abc.proof} for details of this augmented algorithm.}

\begin{algorithm}
\caption{Basic accept-reject ABC algorithm}
\label{alg:abc}
\begin{algorithmic}
\State define some tolerance $\epsilon > 0$
\State define a vector of $m$ statistics $T: \mathcal{G}^C \mapsto \mathbb{R}^m$
\State compute the observed sample statistics $T_{\text{obs}} \coloneqq T(\{G^{T_1}_c\}_{c=1}^C)$
\For{$s \in \{1,2\ldots S\}$}
\State draw $(\beta_s, \tau_s) \sim p_0$
\State generate an artificial sample $\{\tilde{G}_c^{T_1}\}_{c=1}^C$ given $\{G^{T_0}_c, X_c\}_{c=1}^C$ and $(\beta_s, \tau_s)$
\State compute the simulated statistic $T_s \coloneqq T(\{\tilde{G}_c^{T_1}\})$
\State if $\lVert T_s - T_{\text{obs}} \rVert \leq \epsilon$, accept $(\beta_s, \tau_s)$
\EndFor
\end{algorithmic}
\end{algorithm}

In this methodology, the researcher must make two crucial choices. One is the tolerance parameter. Here, we can use the recommendations of \cite{Wentao2018a}: we may choose $\epsilon$ so the algorithm produces a ``reasonable'' acceptance rate. The second important choice is the vector of statistics. This is closely related to identification: for the proper working of the ABC algorithm, the chosen vector of statistics should be informative of the model's parameters \citep{Wentao2018a, Frazier2018}.\footnote{Both authors study the frequentist properties of ABC algorithms. In their setting, where a finite-dimensional vector of summary statistics is considered, it is crucial that the binding function $b(\beta_s,\tau_s) = \operatorname{plim}_{C \to \infty} T_s$ identifies model parameters.} As an example, \cite{Battaglini2021} use a version of the ABC algorithm to estimate a network formation game. In their setting, the authors use the characterization of the model equilibrium to construct the summary statistics used to assess the quality of a parameter draw.

In Supplemental Appendix \ref{Supplement-abc.proof}, we show that, if we take the vector of summary statistics as \emph{the whole second-period network data}, then, as  $\epsilon \to 0$ and $S \to \infty$, the mean of the accepted draws $h(\theta_s)$ converges in probability to the expectation of $h(\cdot)$ with respect to the posterior computed using the conditional likelihood yielded by \eqref{eq_log_lkl} in the Bayes rule, where $h(\cdot)$ is a function with finite moments (with respect to the prior distribution). Using the entire dataset as the vector statistics is important, as it ensures we do not discard any information in the likelihood that, under our assumptions, enables point identification of the model parameters.\footnote{As remarked in an earlier footnote, our model does not yield any obvious low-dimensional vector of sufficient statistics to be used in the analysis without compromising identifiability.} The result in Supplemental Appendix \ref{Supplement-abc.proof} motivates the computation of approximations to the posterior mean and credible intervals using the accepted simulated draws. However, it should be noted that, if there are many networks in the sample, it will generally be very difficult to generate a good approximation to posterior quantities without running a very large (and often computationally unfeasible) number of simulations.  Intuitively, with many networks, it will take many tries to get a draw that approximates the entire vector of observed data well. With a fixed number of simulations, this means that, to have a reasonably small Monte Carlo variance, we tolerate high levels of bias in the ABC approximation to the true posterior.\footnote{This ``curse of dimensionality'' in the dimension of the vector of statistics used in ABC methods bears resemblance to the notion of a ``curse of dimensionality'' in the rate of convergence in nonparametric kernel estimation methods, as the tolerance may be seen as a type of bandwidth. See \citet{Blum2010} for further exploration of this connection.}

To ameliorate the curse of dimensionality associated with taking the whole second-period data as the vector of summary statistics, we propose to use an alternative version of the ABC algorithm: the Expectation Propagation (EP) ABC \citep{Barthelme2014, Barthelme2018}. In this method, we leverage the assumption of an independent sample of networks to factor the posterior distribution as:
\begin{equation}
	\begin{aligned}
d F(\beta )\times \mathbb{P}[	\mathbf{G}_1  | \beta, \mathbf{G}_0, \mathbf{X}]   \overset{\text{independence}}{=}
	 	dF(\beta)  \times \prod_{c=1}^C  \mathbb{P}[G^{T_1}_c|\beta,G^{T_0}_c, X_c] \,,
	\end{aligned}
\end{equation}
where $\mathbf{G}_{0} =(G_c^{T_0})_{c=1}^C$; $\mathbf{G}_{1} =(G_c^{T_1})_{c=1}^C$; $\mathbf{X} = (X_c)_{c=1}^C$; and the prior $F$ only ranges over utility and meeting parameters because we assume that $\tau_0$ is either fixed or estimated in a first step via \eqref{estimate.tau0} (see \Cref{application} for further discussion). EP-ABC assumes a Gaussian prior for $\beta_0$, and further considers a Gaussian approximation with mean $\mu_c$ and covariance matrix $\Omega_c$ to each network likelihood function $\beta\mapsto \mathbb{P}[G^{T_1}_c|\beta,G^{T_0}_c, X_c]$. It then proceeds by sequentially updating the approximations as follows. Suppose we are currently at network $c$. The goal is to find  $(\mu_{c}, \Omega_c)$ to minimize the Kullback-Leibler (KL) divergence between (i) a hybrid approximation that uses the true density at the $c$-th network and the Gaussian approximation at the remaining networks; and (ii) the ``full'' Gaussian approximation. Let us denote the prior distribution by $l_0(\beta)$ ($dF(\beta)\eqqcolon l_0(\beta)$), the "true" likelihood function at each network by $l_c(\beta)$ ($\mathbb{P}[G^{T_1}_c|\beta, G^{T_0}_c, X_c] \eqqcolon l_c(\beta)$), and let $f_c(\beta)$ be a Gaussian density with parameters $(\mu_c,\Omega_c)$.  \cite{Barthelme2014} show that the choice of $(\mu_{c}, \Omega_c)$ that minimizes this KL divergence is
\begin{equation}
	\begin{aligned}
	Z_c = \int  l_0(\beta)  \times \prod_{i \neq c}^C f_i(\beta)  \times l_c(\beta)   \ d \beta \, , \\
	\mu_c =  \frac{1}{Z_c}\int  l_0(\beta)  \times \prod_{i \neq c}^C f_i(\beta)  \times l_c(\beta) \times \beta  \ d \beta \, , \\
	\Omega_c =  \frac{1}{Z_c}\int  l_0(\beta)  \times \prod_{i \neq c}^C f_i(\beta)  \times l_c(\beta) \times \beta \beta' \ d \beta -  \mu_c \mu_c'\, .
\end{aligned}
\end{equation}

These quantities are approximated by running a variant of the ABC algorithm \eqref{alg:abc} with prior $ l_0(\beta)  \times \prod_{i \neq c}^C f_i(\beta)$, which is Gaussian and can be evaluated; and likelihood $l_c(\beta)= \mathbb{P}[G^{T_1}_c|\beta, G^{T_0}_c, X_c]$, which involves simulating the second-period adjacency matrices \textbf{only} in network $c$. This ameliorates the curse of dimensionality since, at any given step, a draw must be good at reproducing features of \textbf{the current network}. Starting from initial guesses for the $(\mu_c,\Omega_c)$ equal to the prior parameters, the algorithm proceeds by sequentially going through all networks $c=1,\ldots C$ and updating $(\mu_c,\Omega_c)$ until meeting a convergence criterion or a full number of passes through all networks.\footnote{In Supplemental Appendix \ref{Supplement-EPlocal}, we describe another method: ``expectation propagation with `local' summary statistics'', as an alternative to the EP-ABC described above. To further reduce dimensionality, this method replaces the vector of network edge indicators in the EP-ABC approach with data-driven ``local'' summary statistics.}

\begin{remark}[On estimation of $\tau_0$]
\label{rmk_bayes_tau0}	Note that Algorithm \ref{alg:abc} requires the researcher to specify a prior for $\tau_0$. A fully Bayesian analysis may \emph{hyperparametrize} this distribution, by allowing it to depend on a set of parameters $\omega$ over which one then assigns a new prior. If the researcher then changes Algorithm \ref{alg:abc} to allow a different draw of $\tau_0$ for each network $c$, one may interpret the hyperparametrized approach as estimating the parameters $\omega$ of the population distribution of $\tau_0$, in a setting where $\tau_0$ varies across networks. While this relaxes the assumption of a common number of rounds across networks adopted in our discussion, it severely complicates the identification analysis. An intermediate approach would be to partition the networks into a known set of clusters, with the property that networks in the same cluster have a common number of rounds, and observations in different clusters have a different number of rounds. In this case, if the number of networks in each cluster is large, one may use the estimator in \Cref{estimation.tau0} to estimate $\tau_0$ in each cluster, and use these as degenerate priors in the analysis. Alternatively, the researcher may impose a prior for each cluster. In both cases, a sufficient condition for identification of meeting and preference-related parameters is that the assumptions underlying one of the identification assumptions discussed in \Cref{sec_ident_pref_meeting} hold in the population defined by each cluster. Moreover, if we restrict preference- and meeting-related parameters to be constant across clusters, then this condition may be relaxed, since one then is able to \emph{extrapolate} information across clusters. 
\end{remark}

\section{Application}
\label{application}

Our application considers data on friendship networks from \cite{Pinto2017}. The dataset comprises information on 3rd- and 5th-graders from 30 elementary schools in Recife, Brazil. Data on students' traits and intraclassroom friendship networks was collected at the middle (baseline) and end (follow-up) of the 2014 school year.\footnote{Specifically, students could nominate up to 8 classmates in each of the three categories: classmates with whom they would (i) study, (ii) talk, or (iii) play. We consider an individual a friend if she appears on at least one of the three lists. No student exceeds eight friends since, in most cases, the three criteria coincide. In our raw dataset, only 1.01\% of students reported eight friends at the baseline, which falls to 0.3\% at the follow-up. For a complete dataset description, see \cite{Pinto2017}.} Once missing observations are removed, our working sample comprises 161 classrooms (networks), totaling 1,589 students.\footnote{ Figure \ref{Supplement-fig:network} in Supplemental Appendix \ref{Supplement-application_tables} plots one such network.} The median number of students in each classroom is $10$.

As a first step in our analysis, we attest that homophily is a salient feature of our data. For that, we run the following regression model:

\begin{equation}
\label{eq.xpec.reduced.form}
    g_{ij,c,1} = \beta'W_{ij,c,0} + \alpha_i + \gamma_j + \epsilon_{ij,c} \,, 
\end{equation}
where $g_{ij,c,1}$ equals $1$ if, in classroom $c$, individual $i$ nominates $j$ as a friend \emph{at the followup period}. Vector $W_{ij,c,0}$ consists of pairwise distances in gender, age (in years), and the logarithm of measures of cognitive and non-cognitive skills between $i$ and $j$ \emph{at the baseline period}.\footnote{Table \ref{Supplement-tab_summary_dyad} in Supplemental Appendix \ref{Supplement-application_tables} presents summary statistics of our dyad-level covariates. See \cite{Pinto2017} for details on the construction of the measures of cognitive and noncognitive skills.} The specification controls for sender ($\alpha_i$) and receiver ($\gamma_j$) fixed effects. We cluster standard errors at the classroom level.

Column (1) in \Cref{tab_reducedform_all} reports estimates obtained from running the above specification. Results indicate that homophily is pervasive, e.g., same-sex classmates are, on average, 19.8 pp more likely to be friends than boy-and-girl pairs.

% Table created by stargazer v.5.2.2 by Marek Hlavac, Harvard University. E-mail: hlavac at fas.harvard.edu
% Date and time: Sun, Jul 14, 2019 - 22:38:53
\begin{table}[H] \centering
  \caption{Dyadic regressions}
  \label{tab_reducedform_all}
\begin{adjustbox}{width = 0.6\textwidth}
\begin{tabular}{@{\extracolsep{0pt}}lcc}
\\[-1.8ex]\hline
\hline \\[-1.8ex]
 & \multicolumn{2}{c}{\textit{Dependent variable}} \\
\cline{2-3}
\\[-1.8ex] & \multicolumn{2}{c}{edge} \\
\\[-1.8ex] & (1) & (2)\\
\hline \\[-1.8ex]
distance in class list  &  & $-$0.002$^{***}$ \\
  &  & (0.0005) \\
distance in age & $-$0.018$^{***}$ & 0.002 \\
  & (0.006) & (0.003) \\
distance in gender & $-$0.198$^{***}$ & $-$0.186$^{***}$ \\
  & (0.006) & (0.006) \\
distance in cognitive skills & $-$0.254$^{***}$ & $-$0.127$^{***}$ \\
  & (0.049) & (0.034) \\
  distance in conscientiousness & $-$0.031$^{***}$ & $-$0.019$^{***}$ \\
  & (0.009) & (0.006) \\
  distance in neuroticism & $-$0.017$^{**}$ & $-$0.023$^{***}$ \\
  & (0.008) & (0.005) \\
 \hline \\[-1.8ex]
Sender fixed effects? & Yes & No \\
Receiver fixed effects? & Yes & No \\
Time effect? & Yes & No \\
\hline \\[-1.8ex]
Observations & 17,736 & 17,736 \\
R$^{2}$ & 0.323 & 0.064 \\
Adjusted R$^{2}$ & 0.185 & 0.063 \\
\hline
\hline \\[-1.8ex]
\multicolumn{1}{l}{\textit{Note:}} & \multicolumn{2}{r}{$^{*}$p$<$0.1; $^{**}$p$<$0.05; $^{***}$p$<$0.01} \\
 \multicolumn{3}{l}{Standard errors clustered at the classroom level in parentheses.} \\
\end{tabular}
\end{adjustbox}
\end{table}

In implementing our model of network formation, we consider the parametrisation of preferences given by \eqref{eq_formulation}, with $\beta_{\text{un}} = \beta_{\text{up}}$, and $W_{i,j}$  taken to be the vector  $W_{ij,c,0}$ of pair-level covariates  \emph{at the baseline}. The difference in preference shocks is drawn from a logistic distribution. For the meeting process, we consider a similar specification to \eqref{eq_pair}, where, for individuals $i$ and $j$ in classroom $c$, we assume that:
\begin{equation*}
    \begin{aligned}
	\rho_{ij}(g|X_c) \propto  \exp \left( \beta_{m}' W_{ij,c,0} + \delta g_{ij} +\psi (1-g_{ij}) Z_{ij,c,0} \right)
\end{aligned} \, ,
\end{equation*}
with $Z_{ij,c,0}$  a pair-level trait that, according to the discussion in Proposition \ref{proposition_parametric}, should be crucially excluded from pair-level covariates $\{W_{m,c,0}\}$, and affect network formation ($\psi\neq0$).\footnote{Even though, in the specification discussed in Section \ref{sec_extrap}, $Z_{ij}$ appears interacted with $g_{ij}$, it is immediate to adapt Proposition \ref{proposition_parametric} to hold in the case where this interaction occurs with $(1-g_{ij})$, as is the case our empirical specification.} In our implementation, we take this covariate to be the distance between two students in the alphabetically-ordered class list. We do so because we expect the distance in the classlist to affect the odds of a pair meeting, e.g. through in-class activities conducted in alphabetically-arranged groups. Still, we do not expect this variable to impact the utility of individuals over social relations, especially since, in specifying preferences, we control for distances in age, gender, and cognitive and noncognitive skill measures. Finally, our proposed specification also embodies the assumption that the class list meeting mechanism is only important for pairs that are not currently friends.

Column (2) in \Cref{tab_reducedform_all} provides reduced-form evidence of the relevance of our class list distance variable. Again, we run the specification in \eqref{eq.xpec.reduced.form}, but include our class list distance variable and exclude sender and receiver fixed effects. The covariate is statistically significant at the 1\% level, with the expected sign: all else equal, classmates ``one more student away'' in the class list is 0.2 pp less likely to be friends.

{We estimate our network formation model using a two-step approach. In the first step, we assume $\tau_0$ to be constant across classrooms and estimate it through \eqref{estimate.tau0}. Our first-step estimate of $\tau_0$ leads to 76 rounds. This estimate satisfies the inequality $\hat{\tau} < N_c (N_c - 1)$ for 81 out of 161 networks in our sample, indicating that, for about half of the classrooms, the estimated number of rounds is such that it would be impossible for the existing network in mid 2014 to be completely overturned by the end of the school year. We also find our estimate to be consistent with contextual information on the number of school days between the baseline and followup. Indeed, collection of baseline data started on July 21st, whereas the followup ended on December 12th, totalling around 100 school days. Our estimate would thus imply that three opportunities to revise friedship status -- our model notion of meetings -- occur every four days. As a sensitivity test, Supplemental Appendix \ref{Supplement-double_tau} reports estimates of preference and meeting parameters when we double our estimate of the number of rounds -- i.e. when we set $\hat \tau = 152$, or equivalently that three meeting opportunities occur every two days. Qualitatively, our main results are mostly unchanged.}

In the second step of our estimation approach, we use the EP-ABC method described in \Cref{estimation.preference.meeting} to approximate the posterior of preference- and matching-related parameters, where we take the first-step estimate of $\tau_0$ as given.\footnote{From a frequentist perspective, we can motivate our estimator by appealing to some Bernstein-von-Mises theorem \citep[chapter  10]{Vaart1998} which ensures posterior asymptotic normality as $C \to \infty$ (see also the discussion of \cite{Frazier2018}). From a Bayesian point of view, our estimator imposes a degenerate prior on $\tau_0$ at the frequentist estimator \eqref{estimate.tau0}.} We adopt independent zero-mean Gaussian distributions with a standard deviation equal to two as priors for both meeting and preference parameters.\footnote{The value of $2$ is chosen according to the reduced-form patterns in \Cref{tab_reducedform_all}, to cover with reasonable prior probability parameter values that enable either homophily in preferences by itself or homophily in meetings by itself to account for observed patterns.} We follow \citeauthor{Barthelme2014}'s (\citeyear{Barthelme2014}) method by making a single pass through the dataset, which appears to be sufficient for convergence. In the ABC step, we follow the suggestion of these authors and use a simple nonstochastic accept-reject rule for draws. We use the follow-up network data in the current classroom as the vector of statistics. Dimensionality, in this case, is smaller than using a standard ABC algorithm since we consider just a single classroom at each step. For each classroom, we simulate 100,000 draws and aim for an acceptance rate of 1\% (1,000 accepted draws). We use Halton random numbers to generate draws from the prior, as suggested by \cite{Barthelme2014}, to improve numerical stability.

\Cref{tab_coef_method3} presents the posterior mean of the coefficients of the utility function and matching obtained by the Expectation Propagation ABC method described above. We also report posterior quantiles and the posterior probability of a negative parameter.

% latex table generated in R 4.0.3 by xtable 1.8-4 package
% Fri Apr  9 16:55:12 2021
\begin{table}[H]
\centering
\caption{Posterior estimates - Expectation propagation}
\label{tab_coef_method3}
\begin{tabular}{lllll}
  \hline
  & Mean & Q 0.025 & Q 0.975 & Prob $<$ 0 \\
  \hline \multicolumn{5}{c}{ Meeting process } \\ \hline  \hline
distance in age &  0.2920 &  0.1596 &  0.4245 &  0.0000 \\
  distance in gender &  1.1646 &  0.8963 &  1.4328 &  0.0000 \\
  distance in cognitive skills &  0.1770 & -1.4080 &  1.7620 &  0.4134 \\
  distance in conscientiousness &  0.1809 & -0.1255 &  0.4873 &  0.1236 \\
  distance in neuroticism & -0.0473 & -0.4188 &  0.3241 &  0.5986 \\
  g\_ij &  1.9856 &  1.5925 &  2.3787 &  0.0000 \\
  (1-g\_ij)*distance in class list & -0.0966 & -0.1392 & -0.0539 &  1.0000 \\
   \hline \multicolumn{5}{c}{ Utility -- Direct Links } \\ \hline intercept &  1.0219 &  0.3494 &  1.6944 &  0.0014 \\
  distance in age & -0.1080 & -0.3992 &  0.1832 &  0.7663 \\
  distance in gender & -2.6392 & -3.3934 & -1.8851 &  1.0000 \\
  distance in cognitive skills & -2.3364 & -5.2264 &  0.5536 &  0.9435 \\
  distance in conscientiousness & -1.7369 & -2.4102 & -1.0637 &  1.0000 \\
  distance in neuroticism & -0.4752 & -1.0809 &  0.1305 &  0.9379 \\
   \hline \multicolumn{5}{c}{ Utility -- Reciprocity Links } \\ \hline intercept &  1.1757 & -0.0292 &  2.3806 &  0.0279 \\
  distance in age &  0.1337 & -0.6070 &  0.8744 &  0.3617 \\
  distance in gender & -0.7291 & -2.4036 &  0.9455 &  0.8033 \\
  distance in cognitive skills &  0.9659 & -2.1432 &  4.0750 &  0.2713 \\
  distance in conscientiousness &  4.2006 &  2.5797 &  5.8216 &  0.0000 \\
  distance in neuroticism &  1.7257 &  0.5708 &  2.8805 &  0.0017 \\
   \hline \multicolumn{5}{c}{ Utility -- Indirect/Popularity } \\ \hline intercept &  0.4427 &  0.1463 &  0.7391 &  0.0017 \\
  distance in age &  0.1295 & -0.0867 &  0.3457 &  0.1202 \\
  distance in gender & -3.2515 & -4.3049 & -2.1981 &  1.0000 \\
  distance in cognitive skills & -0.9512 & -2.7256 &  0.8232 &  0.8533 \\
  distance in conscientiousness & -0.5412 & -0.9177 & -0.1646 &  0.9976 \\
  distance in neuroticism & -0.8722 & -1.3541 & -0.3902 &  0.9998 \\
   \hline
\end{tabular}
\end{table}

The results indicate several mean estimates of utility parameters associated with covariates are negative. This suggests that homophily in preferences is pervasive and not restricted to direct links. We find high posterior probabilities of a negative effect for the role of gender in preferences, not only for the direct links but also for reciprocal links and popularity. Moreover, we find high posterior probabilities of a negative effect for the role of cognitive and noncognitive skills in different utility components. As for the matching parameters, the posterior probability of the coefficient associated with our instrument being negative is near 100\%, which is reassuring. We also find evidence of \textit{heterophilia} in the matching function with respect to age and gender and dependence of meeting opportunities on a previous link between agents. However, these results are not robust when we estimate our model using the alternative method discussed in Supplemental Appendix \ref{Supplement-EPlocal}.\footnote{See Supplemental Appendix \ref{Supplement-results_EPlocal} for results under this alternative method. The main conclusions of our counterfactual exercise remain essentially unchanged.}

Next, we proceed to counterfactual exercises. We consider the evolution of networks, starting from their baseline value, under four different sequences of matching parameters: (i)
when these are kept at their estimated value (\emph{base} case); (ii) when random unbiased matching is imposed across networks (\emph{random} meetings); (iii) when keeping grade and classroom size in schools fixed, we track students according to their cognitive skills (\emph{tracking} case); and (iv) when, upon meeting, friendships are formed at random with probability 1/2 (\emph{random} friendships). It is important to emphasize that we view counterfactuals (ii) and (iv) less as implementable policies, and more as a means of quantifying the role of preferences and meetings in network formation and welfare. Counterfactual (iii) illustrates how our model can be used in policy analysis.

\Cref{tab_ep_ols} reports posterior means and 95\% credible intervals of the projection coefficients of edge indicators at the followup period ($g_{ij,c,1}$) on an intercept and our main controls \emph{at the baseline}. Compared with the observed data, magnitudes in the base case are broadly in line with the frequentist reduced form (Column (2) in \Cref{tab_reducedform_all}, reproduced as Column ``Data'' in \Cref{tab_ep_ols}). Nonetheless, our model appears to overstate the role of homophily in age; and we also understate the average value of $g_{ij,c,1}$ in the data. Comparing the second and third columns, we see that imposing random unbiased matching increases observed homophily patterns in gender, cognitive skills, and conscientiousness. This indicates that shutting down biases in meeting opportunities does not lead to a decrease in observed homophily patterns. Moving on to the last column, when we close the channel of homophily due to preferences, the coefficients associated with homophily due to gender, cognitive skills, and conscientiousness decrease in magnitude compared to the base case. This is evidence that, in this example, homophily in preferences is more important than homophily in meetings. In the fourth column, tracking leads to a weak reduced-form estimate of homophily in cognitive skills. This is expected as students now interact in homogeneous groups. It also leads to a weaker pattern in the gender coefficient, which may be due to the correlation of this attribute with cognitive skills at the baseline.\footnote{Girls have, on average, $1.84 \%$ more cognitive skills at the baseline than boys, and this difference is statistically significant at the 1\% level.}

% latex table generated in R 4.0.3 by xtable 1.8-4 package
% Fri Apr  9 16:55:49 2021
\begin{table}[H]
\centering
\caption{Projection coefficients and edge statistics - Expectation propagation}
\label{tab_ep_ols}
\begin{adjustbox}{width=\textwidth}
\begin{tabular}{llllll}
  \hline
  & Data & Base case & Random matching & Tracking & Random friendship \\
  \hline \multicolumn{4}{c}{ Regression coefficients } \\ \hline  \hline
distance in class list & -0.0016 & -0.0016 & -0.0009 & -0.0030 & -0.0030 \\
    & [-0.0028;-0.0004] & [-0.0023;-0.0009] & [-0.0017;-0.0003] & [-0.0041;-0.0020] & [-0.0045;-0.0016] \\
  distance in age &  0.0018 & -0.0067 & -0.0004 & -0.0039 & -0.0073 \\
    & [-0.0048; 0.0083] & [-0.0143; 0.0010] & [-0.0074; 0.0071] & [-0.0104; 0.0029] & [-0.0126;-0.0017] \\
  distance in gender & -0.1864 & -0.1576 & -0.2221 & -0.1029 & -0.0884 \\
    & [-0.2072;-0.1657] & [-0.1701;-0.1456] & [-0.2383;-0.2061] & [-0.1144;-0.0918] & [-0.1001;-0.0780] \\
  distance in cognitive skills & -0.1275 & -0.1168 & -0.1883 & -0.0321 & -0.0404 \\
    & [-0.2172;-0.0378] & [-0.1805;-0.0486] & [-0.2502;-0.1260] & [-0.1076; 0.0426] & [-0.0968; 0.0171] \\
  distance in conscientiousness & -0.0193 & -0.0287 & -0.0371 & -0.0232 & -0.0195 \\
    & [-0.0332;-0.0053] & [-0.0397;-0.0181] & [-0.0493;-0.0238] & [-0.0325;-0.0140] & [-0.0295;-0.0097] \\
  distance in neuroticism & -0.0231 & -0.0102 & -0.0232 & -0.0131 & -0.0049 \\
    & [-0.0356;-0.0106] & [-0.0192;-0.0009] & [-0.0371;-0.0093] & [-0.0223;-0.0022] & [-0.0143; 0.0052] \\
   \hline \multicolumn{4}{c}{ Edge summary statistics (follow up)} \\ \hline edge indicator mean & 0.1788 &  0.0837 &  0.1637 &  0.0573 &  0.1201 \\
    & [0.168;0.1896] & [0.0775;0.0897] & [0.1562;0.1717] & [0.0521;0.0625] & [0.1034;0.1395] \\
  edge indicator stdev & 0.3832 &  0.2769 &  0.3700 &  0.2323 &  0.3248 \\
    & [0.374;0.3921] & [0.2674;0.2858] & [0.3630;0.3772] & [0.2222;0.2421] & [0.3045;0.3465] \\
   \hline
\end{tabular}
\end{adjustbox}
\begin{minipage}{0.8\textwidth}
	\vspace{1em}
	\footnotesize
	\textit{Notes:} In the first column, the output from a frequentist regression of the edge indicator $g_{ij}$ on pair characteristics, along with 95\% confidence intervals, are reported. Summary statistics on the edge indicator are also reported. Confidence intervals assume a Gaussian approximation and are constructed using standard  errors clustered at the classroom level. Confidence intervals on the standard deviation of the edge indicator additionally use the delta method. Columns 2-4 report mean estimates and 95\% credible intervals on the projection coefficients and summary statistics. These are obtained from 1,000 simulations of network data from draws of the posterior distribution under base and counterfactual modifications in parameter values.
\end{minipage}
\end{table}

\begin{figure}[H]
	\centering
	\caption{Random matching vs. Base case}
	\label{utilities_counterfactual_random_matching}
	\begin{subfigure}[b]{0.35\textwidth}
		\centering
		\includegraphics[width=\textwidth]{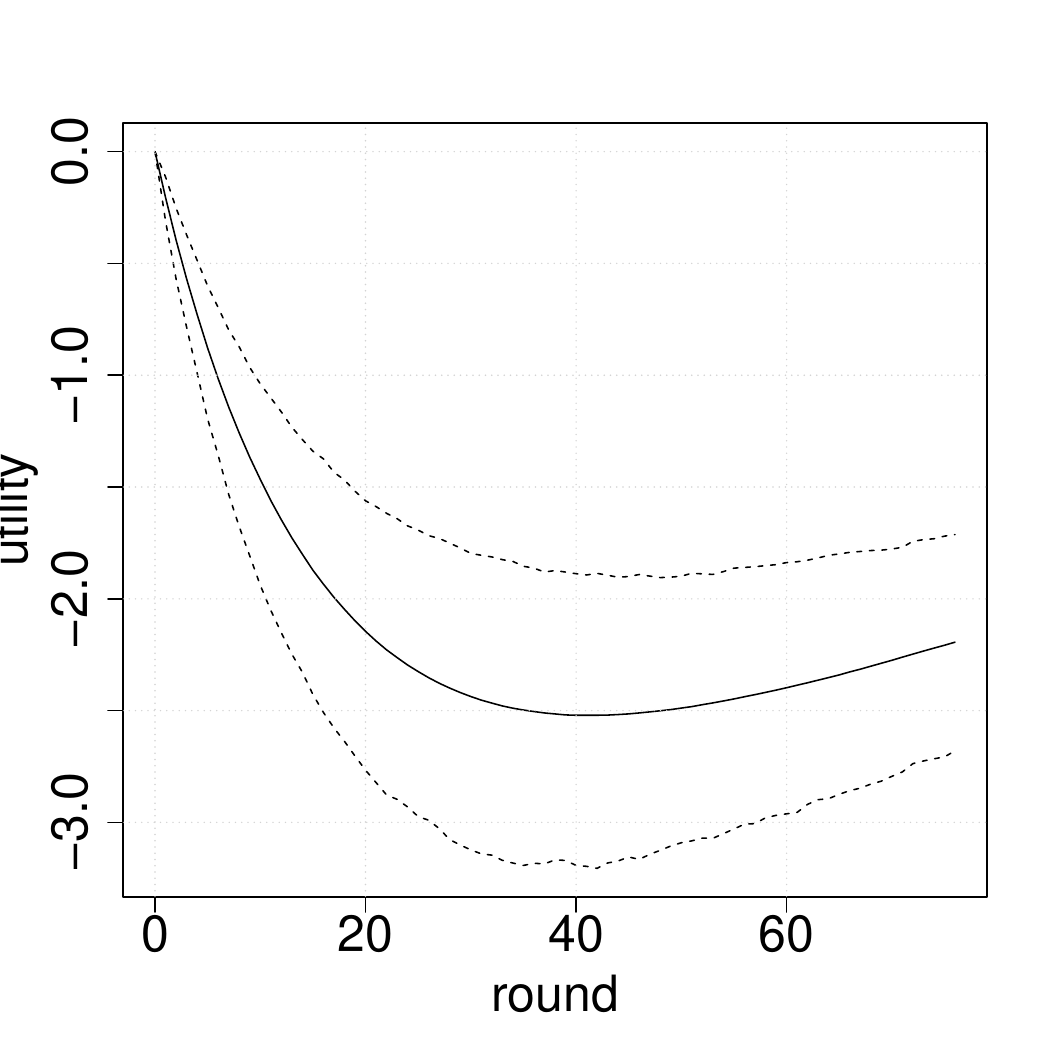}
		\caption{Total}
	\end{subfigure}
	\begin{subfigure}[b]{0.35\textwidth}
		\centering
		\includegraphics[width=\textwidth]{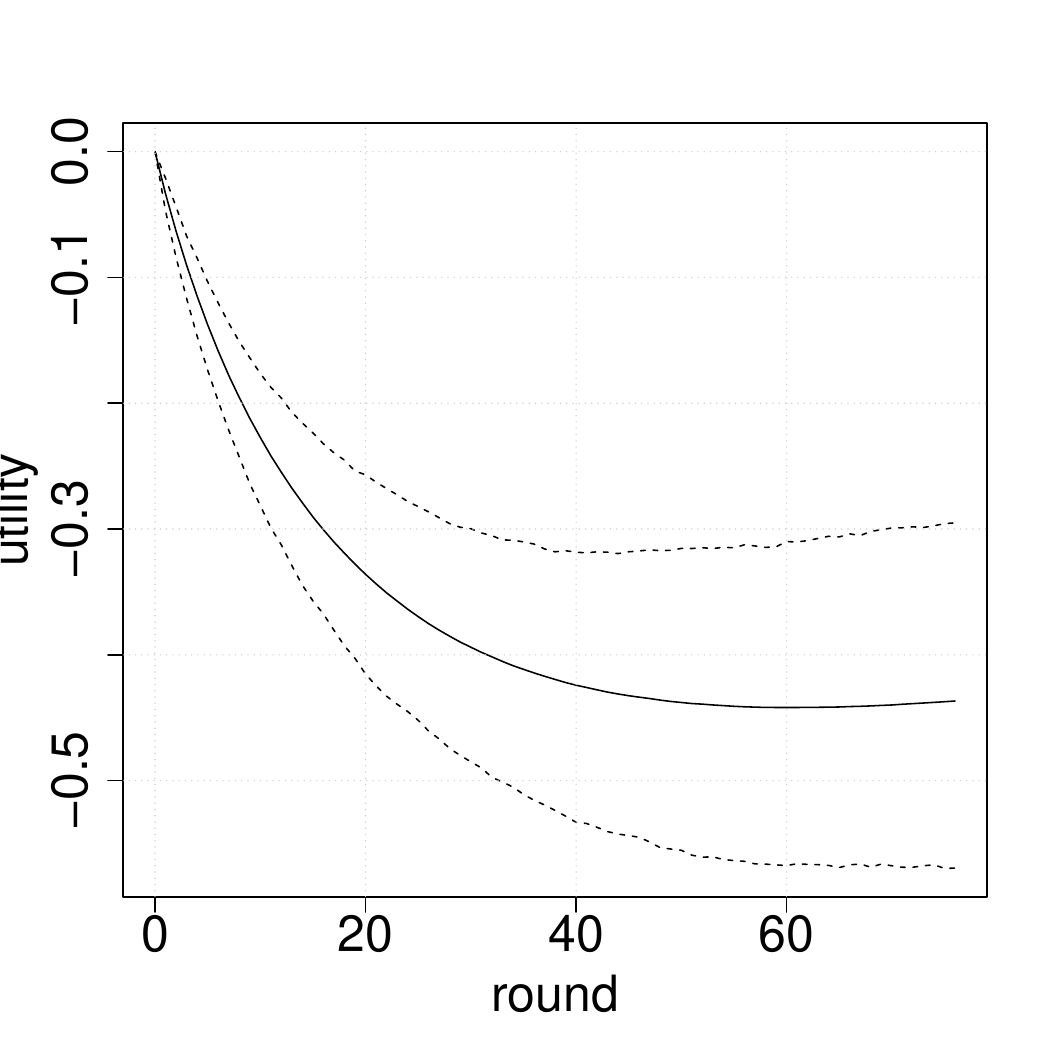}
		\caption{Direct}
	\end{subfigure}\\
	\vspace{-0.6em}
	\begin{subfigure}[b]{0.35\textwidth}
		\centering
		\includegraphics[width=\textwidth]{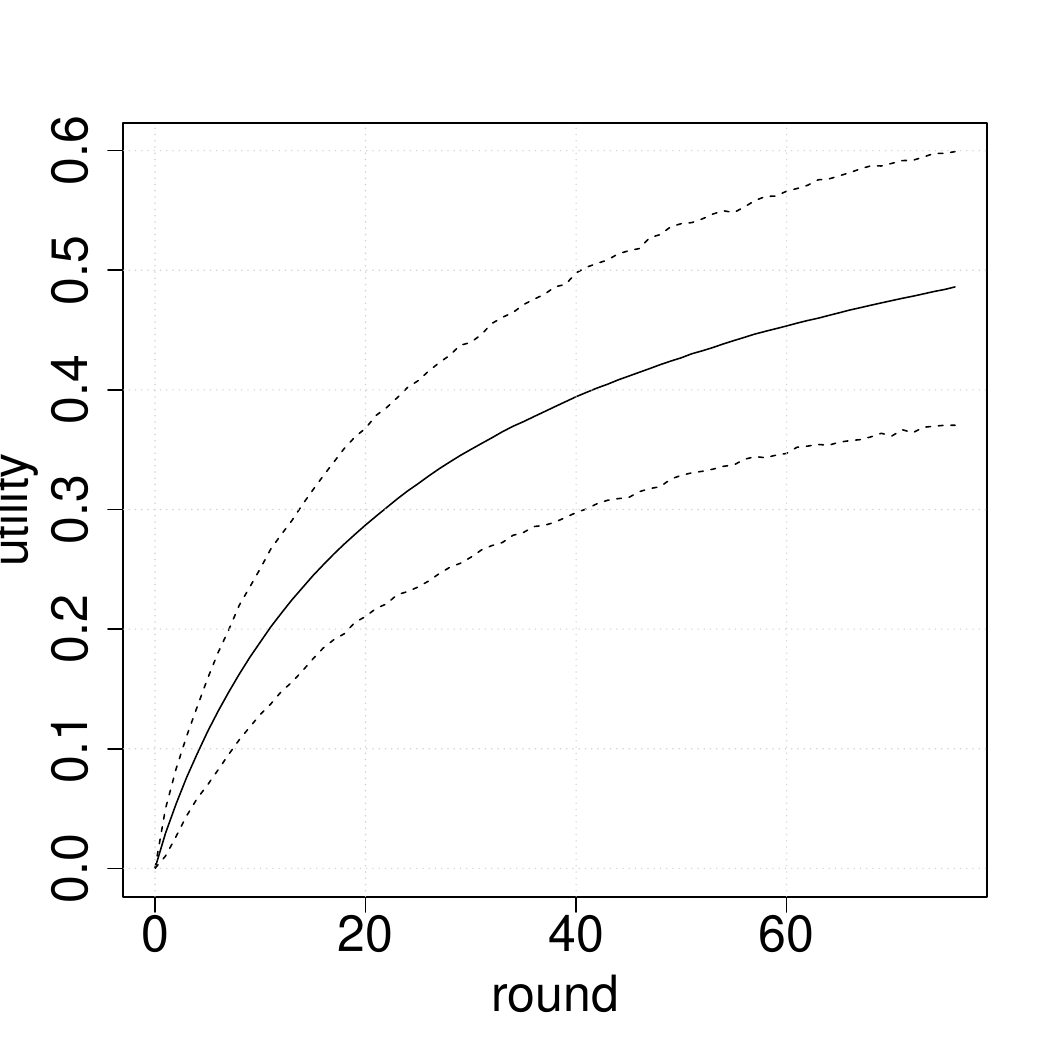}
		\caption{Mutual}
	\end{subfigure}
	\begin{subfigure}[b]{0.35\textwidth}
		\centering
		\includegraphics[width=\textwidth]{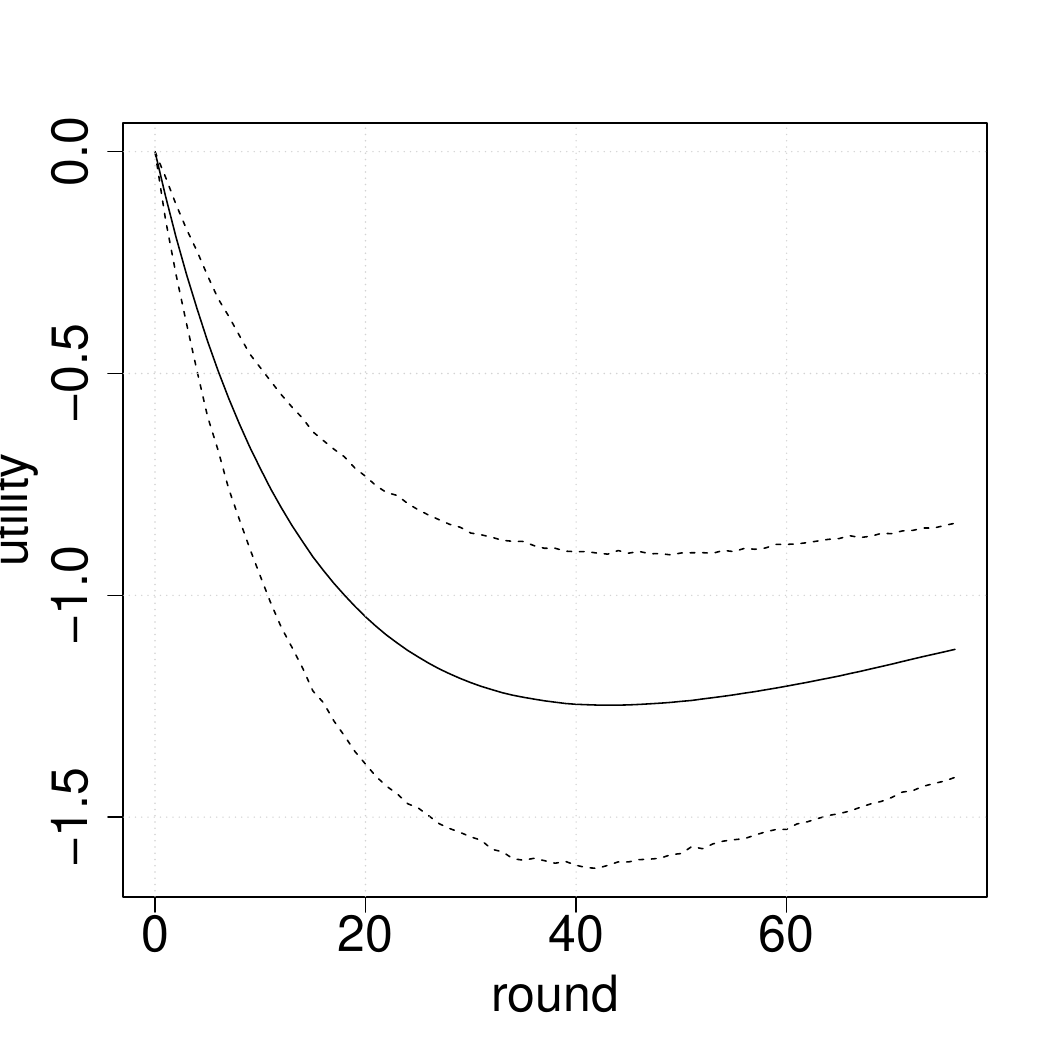}
		\caption{Indirect/Popularity}
	\end{subfigure}

\end{figure}

\begin{figure}[H]
	\centering
	\caption{Tracking vs. Base case}
		\label{utilities_counterfactual_tracking}
	\begin{subfigure}[b]{0.35\textwidth}
		\centering
		\includegraphics[width=\textwidth]{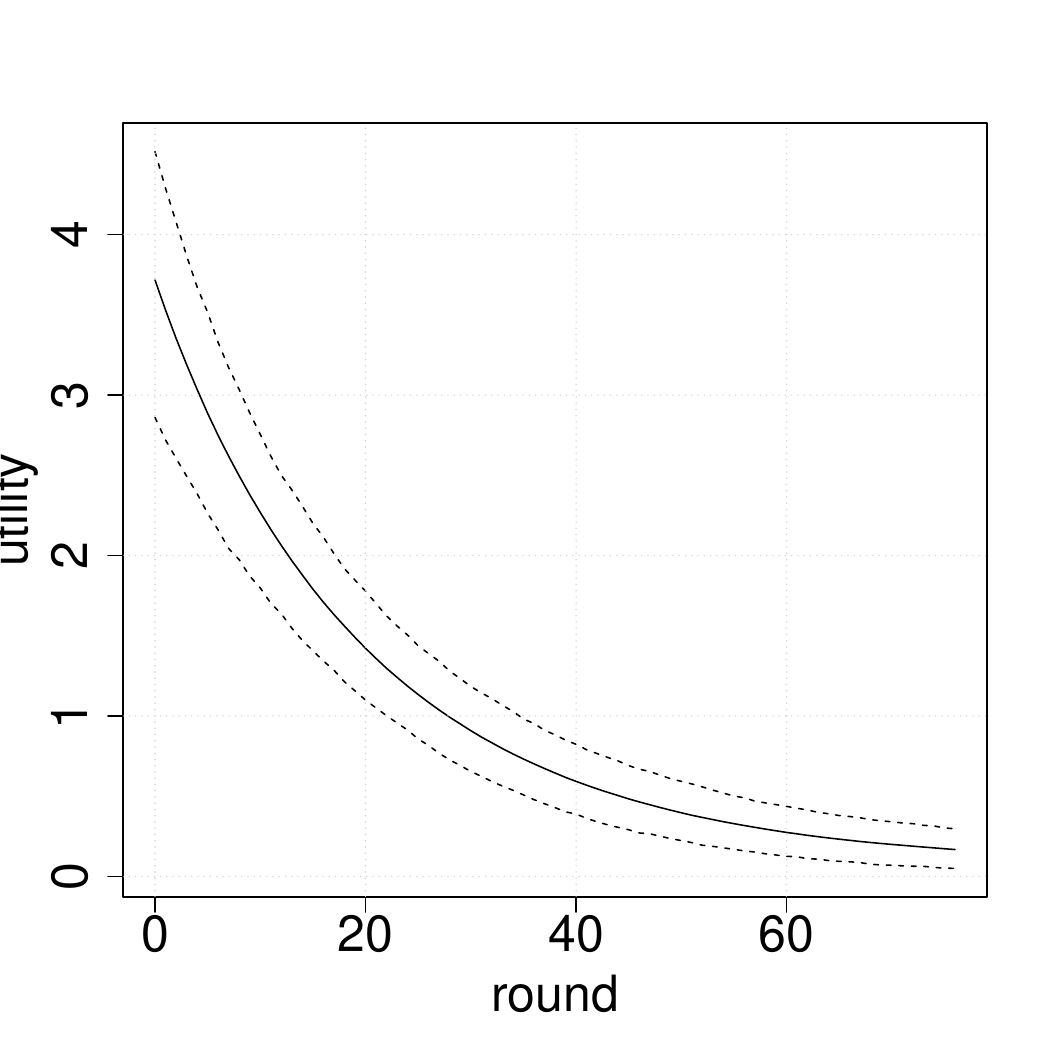}
		\caption{Total}
	\end{subfigure}
	\begin{subfigure}[b]{0.35\textwidth}
		\centering
		\includegraphics[width=\textwidth]{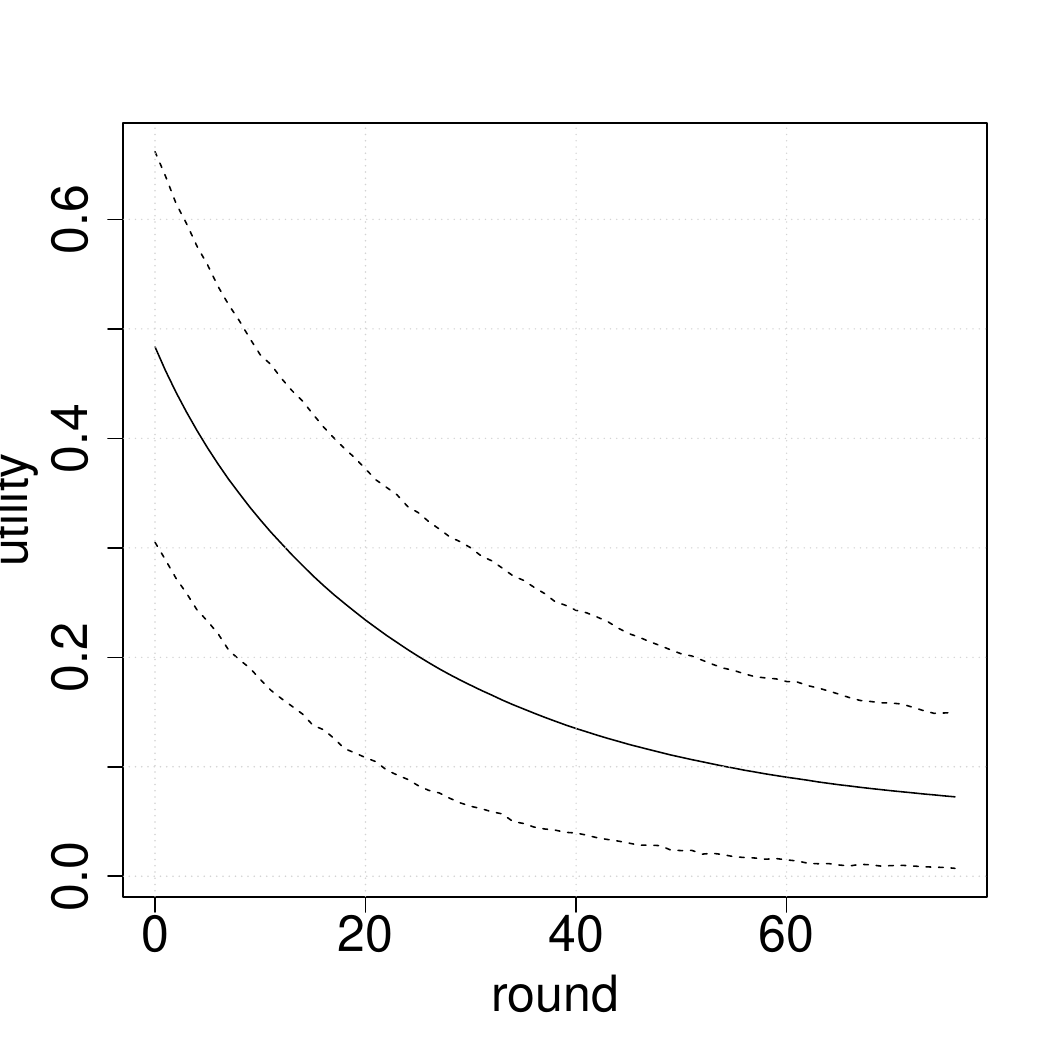}
		\caption{Direct}
	\end{subfigure}\\
	\vspace{-0.6em}
	\begin{subfigure}[b]{0.35\textwidth}
		\centering
		\includegraphics[width=\textwidth]{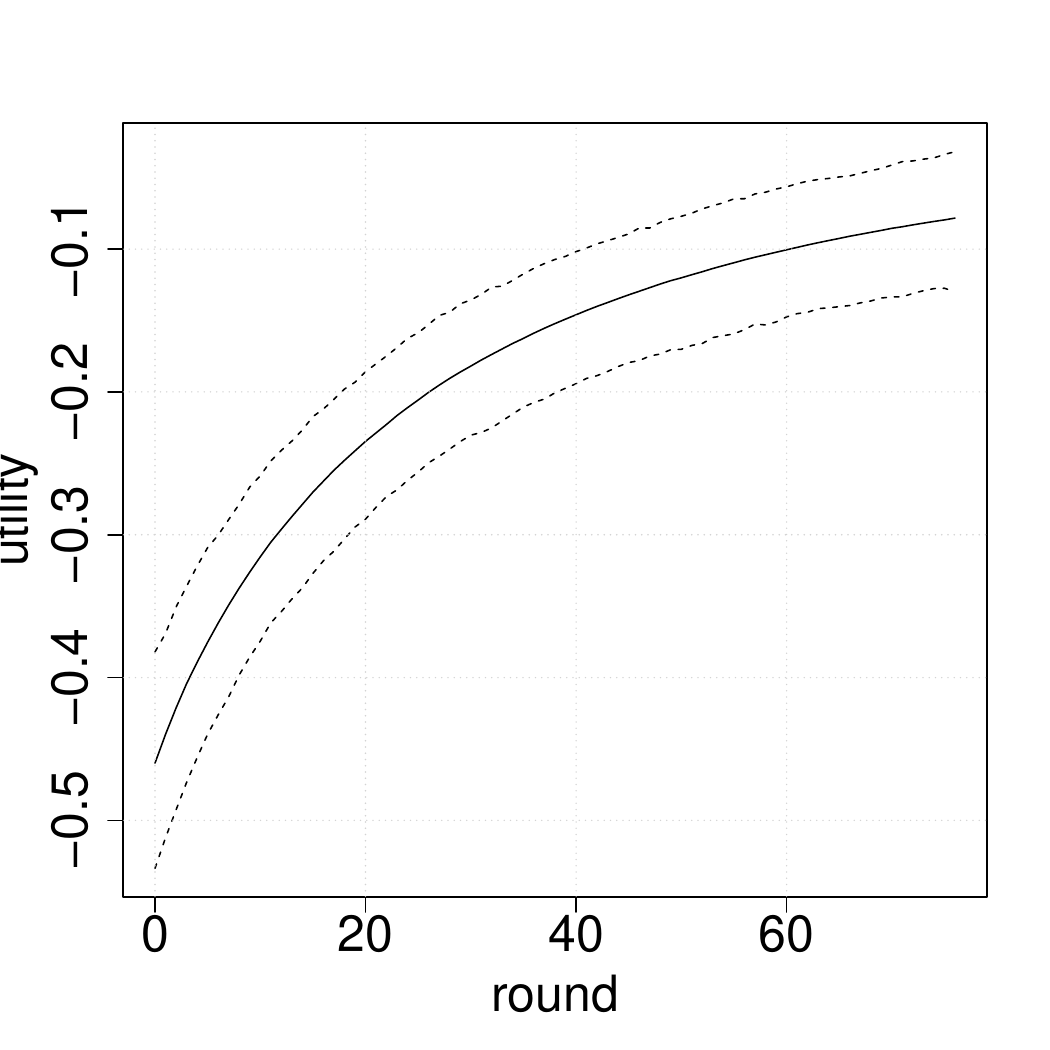}
		\caption{Mutual}
	\end{subfigure}
	\begin{subfigure}[b]{0.35\textwidth}
		\centering
		\includegraphics[width=\textwidth]{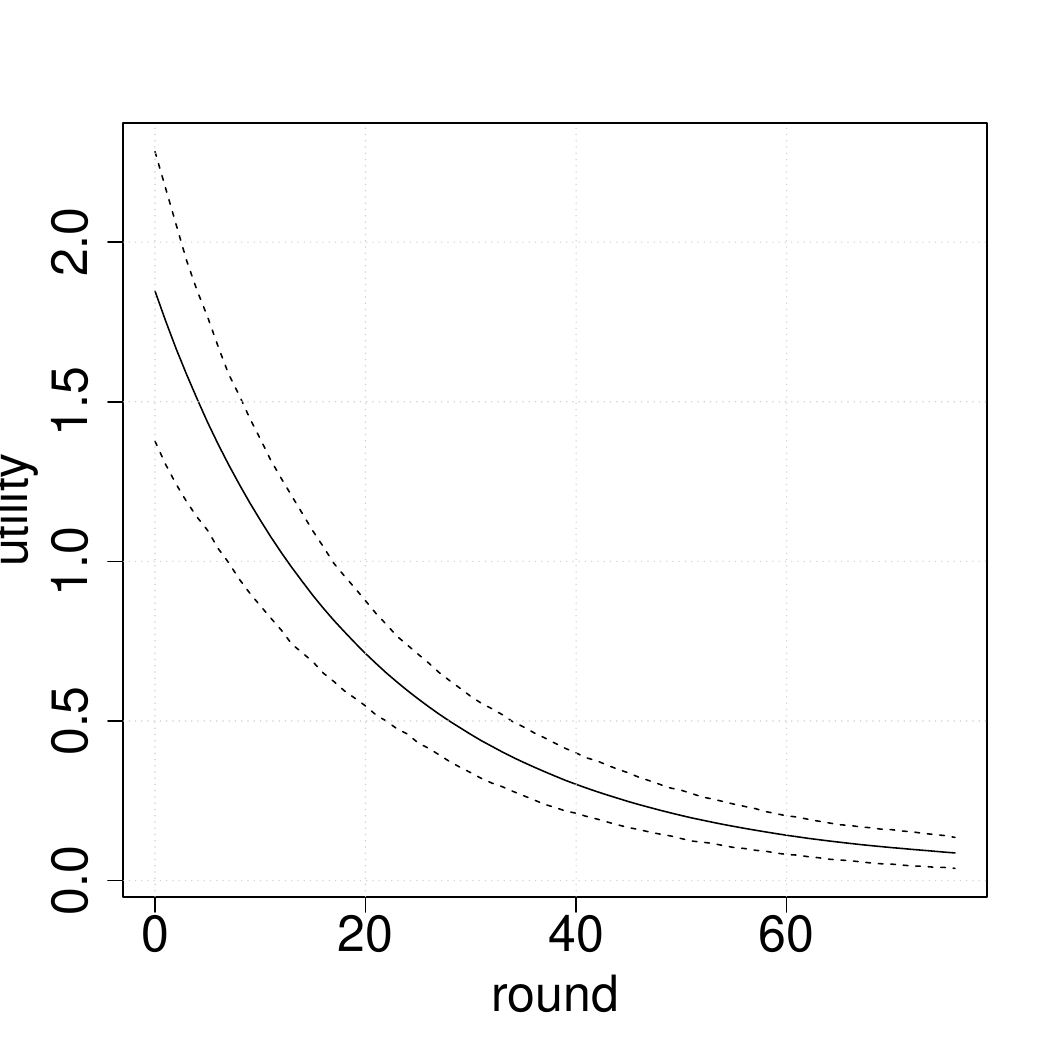}
		\caption{Indirect/Popularity}
	\end{subfigure}

\end{figure}

\begin{figure}[H]
	\centering
	\caption{Random friendships vs. Base case}
	\label{utilities_counterfactual_random_friendships}
	\begin{subfigure}[b]{0.35\textwidth}
		\centering
		\includegraphics[width=\textwidth]{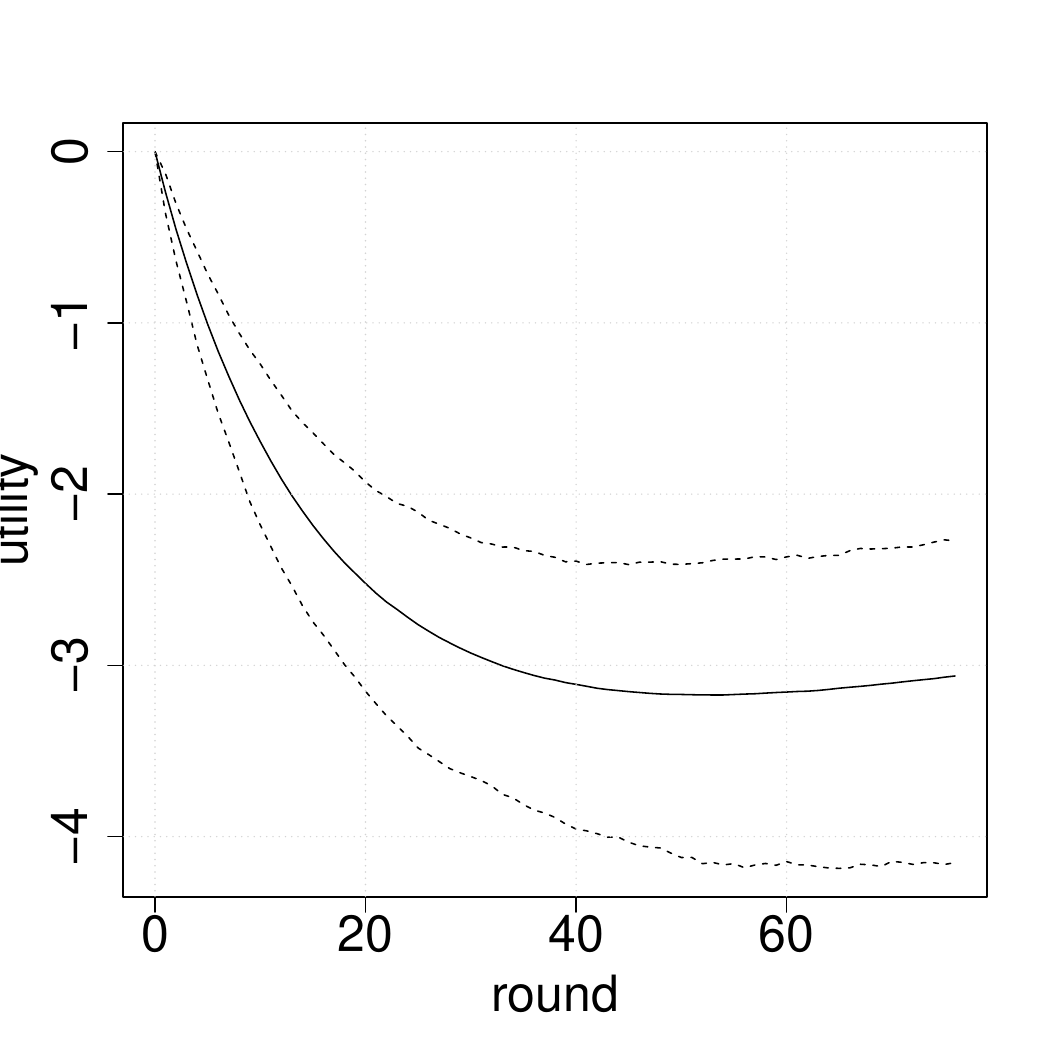}
		\caption{Total}
	\end{subfigure}
	\begin{subfigure}[b]{0.35\textwidth}
		\centering
		\includegraphics[width=\textwidth]{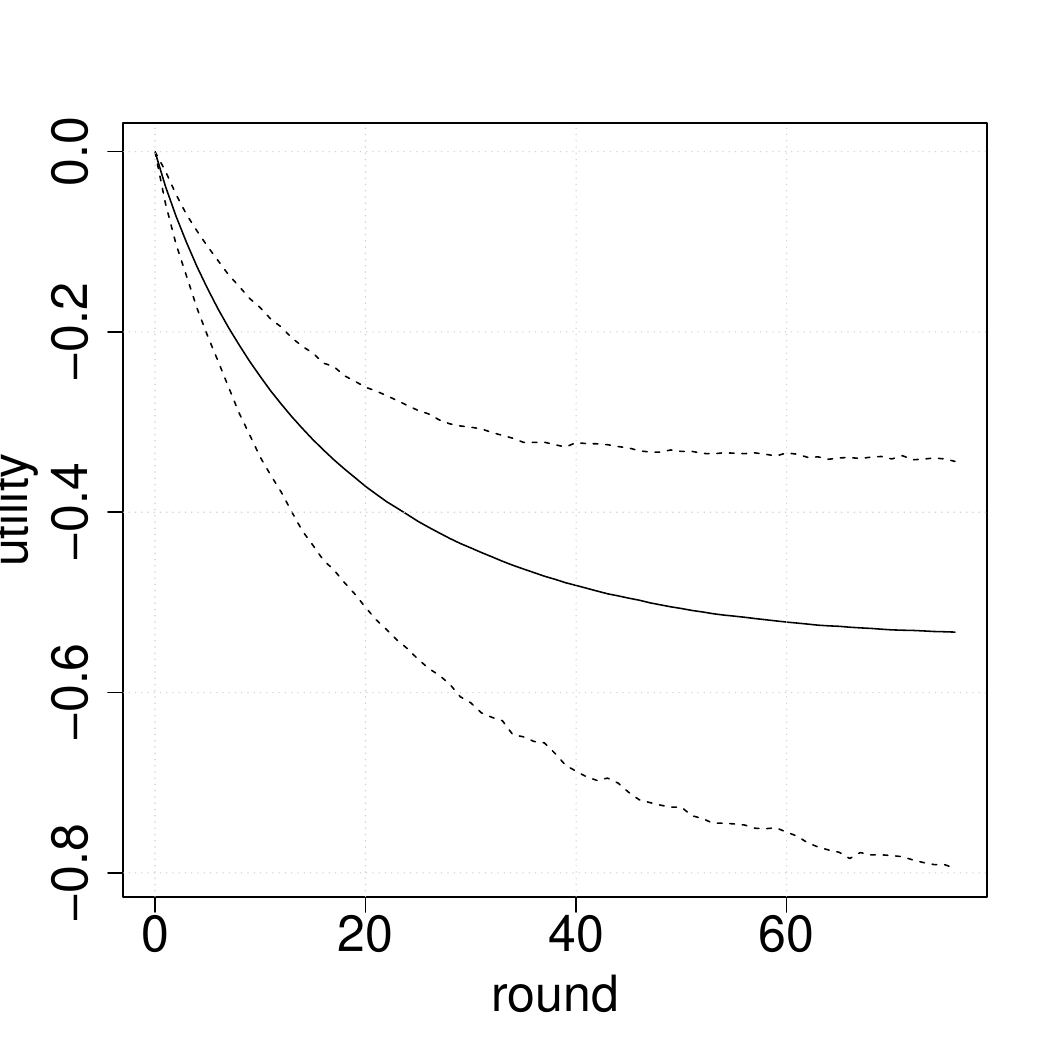}
		\caption{Direct}
	\end{subfigure}\\
	\vspace{-0.6em}
	\begin{subfigure}[b]{0.35\textwidth}
		\centering
		\includegraphics[width=\textwidth]{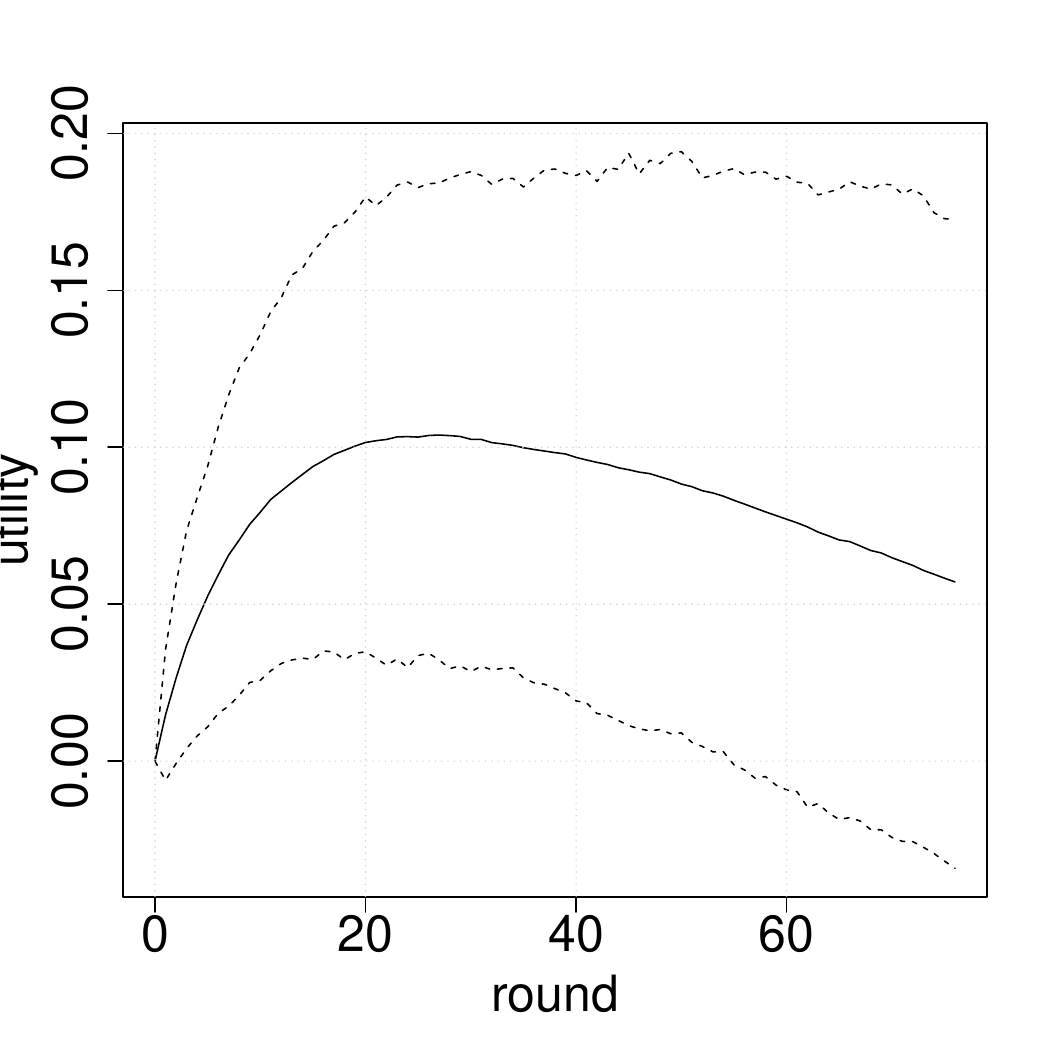}
		\caption{Mutual}
	\end{subfigure}
	\begin{subfigure}[b]{0.35\textwidth}
		\centering
		\includegraphics[width=\textwidth]{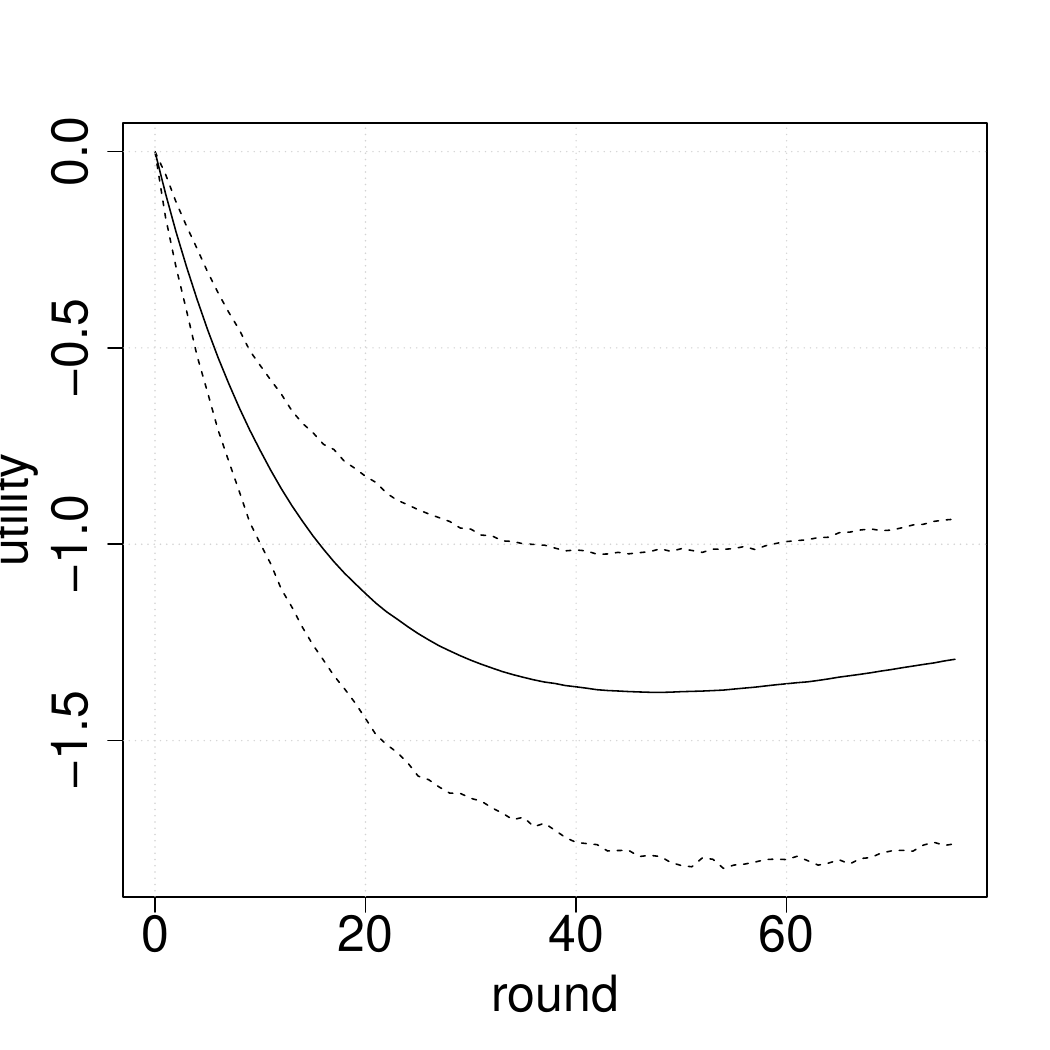}
		\caption{Indirect/Popularity}
	\end{subfigure}
\end{figure}

Figures \ref{utilities_counterfactual_random_matching}, \ref{utilities_counterfactual_tracking}, and \ref{utilities_counterfactual_random_friendships} compare the evolution of aggregate utility in the base case with each of our counterfactual scenarios. We plot posterior means and 95\% credible intervals of the aggregate utility index, $\sum_{c=1}^C \sum_{i=1}^{N_c} u_i(g^t_c, X_c)$, in the counterfactual scenario minus the same index in the base case, for each round from the baseline to the follow-up period. We normalize the difference in indices by the number of students in our sample multiplied by minus the posterior mean estimate of the distance-in-gender direct utility parameter; so results can be interpreted as the number of direct same-sex links each student should receive in the base case so that they are indifferent between policies (without taking into account spillovers on the remaining components of utility).\footnote{Since we only report the difference in aggregate utility between alternative policy scenarios, our welfare analyses are invariant to the normalization adopted to recover the \emph{levels} of utilities (observe that parametrization \eqref{eq_formulation} adopts the normalization that the utility of an empty network is zero). Computing welfare relatively to a baseline scenario thus helps us avoid the typical concern in counterfactual analyses in discrete choice models that require correctly identifying the \emph{levels} of pay-offs (see \citet{kalouptsidi2021counterfactual} for related discussion).}

Imposing random matching leads to a lower trajectory in aggregate utility over the school semester. As expected from the previous analysis, random friendship formation leads to even lower welfare than random matching. The decrease in welfare in both cases comes from a decrease in two components of the utility function: direct links and popularity. We also see that tracking leads to an improvement in welfare, though this benefit diminishes as time passes on. This indicates that a tracking policy in these schools has a positive effect on welfare in the short run, but this effect decreases over time, getting close to zero at the end of 76 rounds of iteration. The impact of tracking policies depends on the network structure, as pointed out by  \cite{Jackson2021}. Our example shows that homophily due to preferences is stronger than homophily due to opportunities. Then, when we change the meeting opportunities and place individuals in homogeneous classrooms, their welfare increases substantially in the first rounds of the game. However, as soon as they make their new connections, the relative effect of the tracking policy starts to decrease \textit{vis-à-vis} the base case. At the end of the game, the relative impact of the policy is close to zero.

In Supplemental Appendix \ref{Supplement-model_peer}, we show how our model can be used to assess the effects of counterfactual policies on measures of productivity and inequality in cognitive skills by coupling a peer effects model to our network formation algorithm. Our counterfactual exercises indicate that shutting down the preference channel in network formation may lead to higher average cognitive skills, though possibly at the expense of increased within-classroom inequality. All in all, these results suggest that policies aimed at changing the determinants of network formation (see \cite{Chetty2022social} for examples) may have nonnegligible impacts on students' outcomes.

\section{Conclusion}
\label{conclusion}

In this article, we studied the identification and estimation of a network formation model that distinguishes between homophily due to preferences and homophily due to meeting opportunities. The model builds upon the algorithm of \cite{Mele2017} by allowing for general classes of utilities and meeting processes. It is also well-grounded in the theoretical literature of network formation \citep{Jackson2002, Jackson2010social}. We provided researchers with a menu of identification assumptions that leverage different types of exclusion restriction to point identify both preference and meeting parameters. We also discussed a Bayesian estimation procedure that bypasses direct evaluation of the model likelihood -- a task that can be computationally unfeasible even for a moderate number of rounds of the network algorithm. All in all, our approach enables users to estimate the counterfactual effects of changes in the meeting technology between agents across time, something previous work could not do.

In the applied section of our article, we studied network formation in elementary schools in Northeastern Brazil. Our results suggest that tracking students according to their cognitive skills improves welfare, though the benefits reduce over time. The effect of this policy can be associated with the structure of the networks. In these classroom networks, homophily due to preferences seems more salient than homophily due to meeting opportunities. This can explain the large positive short-run effect of the tracking policies but the almost zero long-run impact.

{The identification strategy developed in this paper relies on a structural feature common to many models of decentralized interaction: the separation between a meeting process, which determines which agents interact, and a choice process, which governs the actions taken conditional on a meeting. In the model studied here, the meeting process determines which pair of agents obtains the opportunity to revise their relationship status, while agents’ preferences determine whether links are formed or maintained.

This structure is not specific to network formation models. Similar sequential protocols arise in a wide range of economic environments, including search models of financial and product markets \citep{Lagos2017,Wright2021}. In these settings, outcomes are generated by a combination of arrival processes (which determine who meets whom) and choice behavior conditional on interaction opportunities.

The identification arguments developed here suggest that, when observations are available on the evolution of outcomes across time and across many independent environments, it may be possible to recover both the meeting technology and the underlying preferences governing agents’ decisions. Exploring our identification strategies in these other contexts constitutes a promising direction for future research.}

\bibliography{bibliography}

\begin{thebibliography}{}

\bibitem[\protect\citeauthoryear{Aguirregabiria and Mira}{Aguirregabiria and
  Mira}{2010}]{Aguirregabiria2010}
Aguirregabiria, V. and P.~Mira (2010).
\newblock Dynamic discrete choice structural models: A survey.
\newblock {\em Journal of Econometrics\/}~{\em 156\/}(1), 38 -- 67.
\newblock Structural Models of Optimization Behavior in Labor, Aging, and
  Health.

\bibitem[\protect\citeauthoryear{Arnold, Dobbie, and Hull}{Arnold
  et~al.}{2022}]{Hull2022}
Arnold, D., W.~Dobbie, and P.~Hull (2022, September).
\newblock Measuring racial discrimination in bail decisions.
\newblock {\em American Economic Review\/}~{\em 112\/}(9), 2992--3038.

\bibitem[\protect\citeauthoryear{Auerbach}{Auerbach}{2022}]{Auerbach2022}
Auerbach, E. (2022).
\newblock Testing for differences in stochastic network structure.
\newblock {\em Econometrica\/}~{\em 90\/}(3), 1205--1223.

\bibitem[\protect\citeauthoryear{Badev}{Badev}{2021}]{Badev2017}
Badev, A. (2021).
\newblock Nash equilibria on (un)stable networks.
\newblock {\em Econometrica\/}~{\em 89\/}(3), 1179--1206.

\bibitem[\protect\citeauthoryear{Bajari, Hong, and Ryan}{Bajari
  et~al.}{2010}]{Bajari2010}
Bajari, P., H.~Hong, and S.~P. Ryan (2010).
\newblock Identification and estimation of a discrete game of complete
  information.
\newblock {\em Econometrica\/}~{\em 78\/}(5), 1529--1568.

\bibitem[\protect\citeauthoryear{Baron, Doyle, Emanuel, Hull, and Ryan}{Baron
  et~al.}{2024}]{Baron2024}
Baron, E.~J., J.~Doyle, Joseph~J, N.~Emanuel, P.~Hull, and J.~Ryan (2024, 02).
\newblock {Discrimination in Multiphase Systems: Evidence from Child
  Protection*}.
\newblock {\em The Quarterly Journal of Economics\/}~{\em 139\/}(3),
  1611--1664.

\bibitem[\protect\citeauthoryear{Barthelm{\'e}, Chopin, and
  Cottet}{Barthelm{\'e} et~al.}{2018}]{Barthelme2018}
Barthelm{\'e}, S., N.~Chopin, and V.~Cottet (2018).
\newblock Divide and conquer in {ABC}: Expectation-propagation algorithms for
  likelihood-free inference.
\newblock {\em Handbook of Approximate Bayesian Computation\/}, 415--34.

\bibitem[\protect\citeauthoryear{Barthelmé and Chopin}{Barthelmé and
  Chopin}{2014}]{Barthelme2014}
Barthelmé, S. and N.~Chopin (2014).
\newblock Expectation propagation for likelihood-free inference.
\newblock {\em Journal of the American Statistical Association\/}~{\em
  109\/}(505), 315--333.

\bibitem[\protect\citeauthoryear{Battaglini, Patacchini, and
  Rainone}{Battaglini et~al.}{2021}]{Battaglini2021}
Battaglini, M., E.~Patacchini, and E.~Rainone (2021).
\newblock Endogenous social interactions with unobserved networks.
\newblock {\em The Review of Economic Studies\/}.

\bibitem[\protect\citeauthoryear{Belloni and Chernozhukov}{Belloni and
  Chernozhukov}{2013}]{Belloni2013}
Belloni, A. and V.~Chernozhukov (2013).
\newblock {Least squares after model selection in high-dimensional sparse
  models}.
\newblock {\em Bernoulli\/}~{\em 19\/}(2), 521 -- 547.

\bibitem[\protect\citeauthoryear{Blum}{Blum}{2010}]{Blum2010}
Blum, M. G.~B. (2010).
\newblock Approximate bayesian computation: A nonparametric perspective.
\newblock {\em Journal of the American Statistical Association\/}~{\em
  105\/}(491), 1178--1187.

\bibitem[\protect\citeauthoryear{Bramoull{\'{e}}, Djebbari, and
  Fortin}{Bramoull{\'{e}} et~al.}{2009}]{Bramoulle2009}
Bramoull{\'{e}}, Y., H.~Djebbari, and B.~Fortin (2009).
\newblock Identification of peer effects through social networks.
\newblock {\em Journal of Econometrics\/}~{\em 150\/}(1), 41 -- 55.

\bibitem[\protect\citeauthoryear{Bramoullé, Currarini, Jackson, Pin, and
  Rogers}{Bramoullé et~al.}{2012}]{Bramoulle2012}
Bramoullé, Y., S.~Currarini, M.~O. Jackson, P.~Pin, and B.~W. Rogers (2012).
\newblock Homophily and long-run integration in social networks.
\newblock {\em Journal of Economic Theory\/}~{\em 147\/}(5), 1754 -- 1786.

\bibitem[\protect\citeauthoryear{Casanellas, Fern{\'a}ndez-S{\'a}nchez, and
  Roca-Lacostena}{Casanellas et~al.}{2023}]{Casanellas2023}
Casanellas, M., J.~Fern{\'a}ndez-S{\'a}nchez, and J.~Roca-Lacostena (2023).
\newblock {THE EMBEDDING PROBLEM FOR MARKOV MATRICES}.
\newblock {\em Publicacions Matemàtiques\/}~{\em 67\/}(1), 411 -- 445.

\bibitem[\protect\citeauthoryear{Centola}{Centola}{2011}]{Centola2011}
Centola, D. (2011).
\newblock An experimental study of homophily in the adoption of health
  behavior.
\newblock {\em Science\/}~{\em 334\/}(6060), 1269--1272.

\bibitem[\protect\citeauthoryear{Chandrasekhar}{Chandrasekhar}{2016}]{chandrasekhar2016econometrics}
Chandrasekhar, A. (2016).
\newblock Econometrics of network formation.
\newblock In {\em The Oxford Handbook of the Economics of Networks}, pp.\
  303--357. Oxford University Press.

\bibitem[\protect\citeauthoryear{Chandrasekhar and Jackson}{Chandrasekhar and
  Jackson}{2021}]{Jackson2016}
Chandrasekhar, A.~G. and M.~O. Jackson (2021).
\newblock A network formation model based on subgraphs.

\bibitem[\protect\citeauthoryear{Chetty, Jackson, Kuchler, Stroebel, Hendren,
  Fluegge, Gong, Gonzalez, Grondin, Jacob, et~al.}{Chetty
  et~al.}{2022a}]{Chetty2022socioeconomic}
Chetty, R., M.~O. Jackson, T.~Kuchler, J.~Stroebel, N.~Hendren, R.~B. Fluegge,
  S.~Gong, F.~Gonzalez, A.~Grondin, M.~Jacob, et~al. (2022a).
\newblock Social capital i: measurement and associations with economic
  mobility.
\newblock {\em Nature\/}~{\em 608\/}(7921), 108--121.

\bibitem[\protect\citeauthoryear{Chetty, Jackson, Kuchler, Stroebel, Hendren,
  Fluegge, Gong, Gonzalez, Grondin, Jacob, et~al.}{Chetty
  et~al.}{2022b}]{Chetty2022social}
Chetty, R., M.~O. Jackson, T.~Kuchler, J.~Stroebel, N.~Hendren, R.~B. Fluegge,
  S.~Gong, F.~Gonzalez, A.~Grondin, M.~Jacob, et~al. (2022b).
\newblock Social capital ii: determinants of economic connectedness.
\newblock {\em Nature\/}~{\em 608\/}(7921), 122--134.

\bibitem[\protect\citeauthoryear{Christakis, Fowler, Imbens, and
  Kalyanaraman}{Christakis et~al.}{2020}]{Christakis2010}
Christakis, N., J.~Fowler, G.~W. Imbens, and K.~Kalyanaraman (2020).
\newblock An empirical model for strategic network formation.
\newblock In B.~Graham and A.~de~Paula (Eds.), {\em The Econometric Analysis of
  Network Data}, Chapter Chapter 6, pp.\  123--148. Academic Press.

\bibitem[\protect\citeauthoryear{Colas and Morehouse}{Colas and
  Morehouse}{2022}]{Colas2020}
Colas, M. and J.~M. Morehouse (2022).
\newblock The environmental cost of land-use restrictions.
\newblock {\em Quantitative Economics\/}~{\em 13\/}(1), 179--223.

\bibitem[\protect\citeauthoryear{Currarini, Jackson, and Pin}{Currarini
  et~al.}{2009}]{Currarini2009}
Currarini, S., M.~O. Jackson, and P.~Pin (2009).
\newblock An economic model of friendship: Homophily,
  minorities, and segregation.
\newblock {\em Econometrica\/}~{\em 77\/}(4), 1003--1045.

\bibitem[\protect\citeauthoryear{Currarini, Jackson, and Pin}{Currarini
  et~al.}{2010}]{Currarini2010}
Currarini, S., M.~O. Jackson, and P.~Pin (2010).
\newblock Identifying the roles of race-based choice and chance in high school
  friendship network formation.
\newblock {\em Proceedings of the National Academy of Sciences\/}~{\em
  107\/}(11), 4857--4861.

\bibitem[\protect\citeauthoryear{Davezies, D'Haultfoeuille, and
  Fougère}{Davezies et~al.}{2009}]{Davezies2009}
Davezies, L., X.~D'Haultfoeuille, and D.~Fougère (2009).
\newblock Identification of peer effects using group size variation.
\newblock {\em The Econometrics Journal\/}~{\em 12\/}(3), 397--413.

\bibitem[\protect\citeauthoryear{de~Paula}{de~Paula}{2017}]{dePaula2016}
de~Paula, A. (2017).
\newblock Econometrics of network models.
\newblock In B.~Honor\'{e}, A.~Pakes, M.~Piazzesi, and L.~Samuelson (Eds.),
  {\em Advances in Economics and Econometrics: Eleventh World Congress},
  Econometric Society Monographs, pp.\  268–323. Cambridge University Press.

\bibitem[\protect\citeauthoryear{de~Paula, Richards‐Shubik, and
  Tamer}{de~Paula et~al.}{2018}]{dePaula2018a}
de~Paula, {\'{Á}}., S.~Richards‐Shubik, and E.~Tamer (2018).
\newblock Identifying preferences in networks with bounded degree.
\newblock {\em Econometrica\/}~{\em 86\/}(1), 263--288.

\bibitem[\protect\citeauthoryear{Engle, Hendry, and Richard}{Engle
  et~al.}{1983}]{Engle1983}
Engle, R.~F., D.~F. Hendry, and J.-F. Richard (1983).
\newblock Exogeneity.
\newblock {\em Econometrica\/}~{\em 51\/}(2), 277--304.

\bibitem[\protect\citeauthoryear{Fearnhead and Prangle}{Fearnhead and
  Prangle}{2012}]{Fearnhead2012}
Fearnhead, P. and D.~Prangle (2012).
\newblock Constructing summary statistics for approximate bayesian computation:
  semi-automatic approximate bayesian computation.
\newblock {\em Journal of the Royal Statistical Society: Series B (Statistical
  Methodology)\/}~{\em 74\/}(3), 419--474.

\bibitem[\protect\citeauthoryear{Frazier, Martin, Robert, and Rousseau}{Frazier
  et~al.}{2018}]{Frazier2018}
Frazier, D.~T., G.~M. Martin, C.~P. Robert, and J.~Rousseau (2018).
\newblock Asymptotic properties of approximate {B}ayesian computation.
\newblock {\em Biometrika\/}~{\em 105\/}(3), 593--607.

\bibitem[\protect\citeauthoryear{Gaonkar and Mele}{Gaonkar and
  Mele}{2023}]{gaonkar2023model}
Gaonkar, S. and A.~Mele (2023).
\newblock A model of inter-organizational network formation.
\newblock {\em Journal of Economic Behavior \& Organization\/}~{\em 214},
  82--104.

\bibitem[\protect\citeauthoryear{Goldsmith-Pinkham and
  Imbens}{Goldsmith-Pinkham and Imbens}{2013}]{Goldsmith2013}
Goldsmith-Pinkham, P. and G.~W. Imbens (2013).
\newblock Social networks and the identification of peer effects.
\newblock {\em Journal of Business \& Economic Statistics\/}~{\em 31\/}(3),
  253--264.

\bibitem[\protect\citeauthoryear{Graham}{Graham}{2008}]{Graham2009}
Graham, B.~S. (2008).
\newblock Identifying social interactions through conditional variance
  restrictions.
\newblock {\em Econometrica\/}~{\em 76\/}(3), 643--660.

\bibitem[\protect\citeauthoryear{Graham}{Graham}{2016}]{Graham2016}
Graham, B.~S. (2016).
\newblock Homophily and transitivity in dynamic network formation.
\newblock Working Paper 22186, National Bureau of Economic Research.

\bibitem[\protect\citeauthoryear{Graham}{Graham}{2017}]{Graham2017}
Graham, B.~S. (2017).
\newblock An econometric model of network formation with degree heterogeneity.
\newblock {\em Econometrica\/}~{\em 85\/}(4), 1033--1063.

\bibitem[\protect\citeauthoryear{Hobert}{Hobert}{2011}]{Hobert2011}
Hobert, J.~P. (2011).
\newblock Chapter 10 the data augmentation algorithm: Theory and methodology.
\newblock In {\em Handbook of Markov Chain Monte Carlo}, pp.\  253--294. CRC
  Press.

\bibitem[\protect\citeauthoryear{Horn and Johnson}{Horn and
  Johnson}{2012}]{Horn2012}
Horn, R.~A. and C.~R. Johnson (2012).
\newblock {\em Matrix Analysis}.
\newblock Cambridge University Press.

\bibitem[\protect\citeauthoryear{Hsieh and Lee}{Hsieh and
  Lee}{2016}]{Hsieh2016}
Hsieh, C.-S. and L.~F. Lee (2016).
\newblock A social interactions model with endogenous friendship formation and
  selectivity.
\newblock {\em Journal of Applied Econometrics\/}~{\em 31\/}(2), 301--319.

\bibitem[\protect\citeauthoryear{Jackson}{Jackson}{2010}]{Jackson2010social}
Jackson, M. (2010).
\newblock {\em Social and Economic Networks}.
\newblock Princeton University Press.

\bibitem[\protect\citeauthoryear{Jackson}{Jackson}{2021}]{Jackson2021}
Jackson, M.~O. (2021).
\newblock Inequality's economic and social roots: The role of social networks
  and homophily.
\newblock {\em {SSRN} Electronic Journal\/}.

\bibitem[\protect\citeauthoryear{Jackson and Rogers}{Jackson and
  Rogers}{2007}]{Jackson2007}
Jackson, M.~O. and B.~W. Rogers (2007, June).
\newblock Meeting strangers and friends of friends: How random are social
  networks?
\newblock {\em American Economic Review\/}~{\em 97\/}(3), 890--915.

\bibitem[\protect\citeauthoryear{Jackson and Watts}{Jackson and
  Watts}{2002}]{Jackson2002}
Jackson, M.~O. and A.~Watts (2002).
\newblock The evolution of social and economic networks.
\newblock {\em Journal of Economic Theory\/}~{\em 106\/}(2), 265 -- 295.

\bibitem[\protect\citeauthoryear{Jochmans}{Jochmans}{2023}]{Jochmans2023}
Jochmans, K. (2023).
\newblock Peer effects and endogenous social interactions.
\newblock {\em Journal of Econometrics\/}~{\em 235\/}(2), 1203--1214.

\bibitem[\protect\citeauthoryear{Kadelka and McCombs}{Kadelka and
  McCombs}{2021}]{Kadelka2021}
Kadelka, C. and A.~McCombs (2021).
\newblock Effect of homophily and correlation of beliefs on {COVID}-19 and
  general infectious disease outbreaks.
\newblock {\em {PLOS} {ONE}\/}~{\em 16\/}(12).

\bibitem[\protect\citeauthoryear{Kalouptsidi, Kitamura, Lima, and
  Souza-Rodrigues}{Kalouptsidi et~al.}{2021}]{kalouptsidi2021counterfactual}
Kalouptsidi, M., Y.~Kitamura, L.~Lima, and E.~Souza-Rodrigues (2021).
\newblock Counterfactual analysis for structural dynamic discrete choice
  models.
\newblock {\em NBER Working Paper Series,(26761)\/}.

\bibitem[\protect\citeauthoryear{Lagos, Rocheteau, and Wright}{Lagos
  et~al.}{2017}]{Lagos2017}
Lagos, R., G.~Rocheteau, and R.~Wright (2017, June).
\newblock Liquidity: A new monetarist perspective.
\newblock {\em Journal of Economic Literature\/}~{\em 55\/}(2), 371–440.

\bibitem[\protect\citeauthoryear{Lee}{Lee}{2007}]{Lee2007}
Lee, L. (2007).
\newblock Identification and estimation of econometric models with group
  interactions, contextual factors and fixed effects.
\newblock {\em Journal of Econometrics\/}~{\em 140\/}(2), 333--374.

\bibitem[\protect\citeauthoryear{Li and Racine}{Li and
  Racine}{2006}]{LiRacine2006}
Li, Q. and J.~S. Racine (2006).
\newblock {\em Nonparametric Econometrics: Theory and Practice}.
\newblock Princeton University Press.

\bibitem[\protect\citeauthoryear{Li and Fearnhead}{Li and
  Fearnhead}{2018}]{Wentao2018a}
Li, W. and P.~Fearnhead (2018).
\newblock On the asymptotic efficiency of approximate {B}ayesian computation
  estimators.
\newblock {\em Biometrika\/}~{\em 105\/}(2), 285--299.

\bibitem[\protect\citeauthoryear{McFadden}{McFadden}{1973}]{McFadden1974}
McFadden, D. (1973).
\newblock Conditional logit analysis of qualitative choice behaviour.
\newblock In P.~Zarembka (Ed.), {\em Frontiers in Econometrics}, pp.\
  105--142. New York, NY, USA: Academic Press New York.

\bibitem[\protect\citeauthoryear{Mele}{Mele}{2017}]{Mele2017}
Mele, A. (2017).
\newblock A structural model of dense network formation.
\newblock {\em Econometrica\/}~{\em 85\/}(3), 825--850.

\bibitem[\protect\citeauthoryear{Mele}{Mele}{2020}]{Mele2020}
Mele, A. (2020, May).
\newblock Does school desegregation promote diverse interactions? an
  equilibrium model of segregation within schools.
\newblock {\em American Economic Journal: Economic Policy\/}~{\em 12\/}(2),
  228–57.

\bibitem[\protect\citeauthoryear{Newey and McFadden}{Newey and
  McFadden}{1994}]{Newey1994}
Newey, W.~K. and D.~McFadden (1994).
\newblock Large sample estimation and hypothesis testing.
\newblock In {\em Handbook of Econometrics}, Volume~4, Chapter~36, pp.\  2111
  -- 2245. Elsevier.

\bibitem[\protect\citeauthoryear{Newey and Smith}{Newey and
  Smith}{2004}]{NeweySmith2004}
Newey, W.~K. and R.~J. Smith (2004).
\newblock Higher order properties of gmm and generalized empirical likelihood
  estimators.
\newblock {\em Econometrica\/}~{\em 72\/}(1), 219--255.

\bibitem[\protect\citeauthoryear{Norris}{Norris}{1997}]{Norris1997}
Norris, J.~R. (1997).
\newblock {\em Markov Chains}.
\newblock Cambridge Series in Statistical and Probabilistic Mathematics.
  Cambridge University Press.

\bibitem[\protect\citeauthoryear{Pinto and Ponczek}{Pinto and
  Ponczek}{2023}]{Pinto2017}
Pinto, C. C. d.~X. and V.~P. Ponczek (2023).
\newblock The building blocks of skill development.
\newblock {Unpublished}.

\bibitem[\protect\citeauthoryear{Rogerson, Shimer, and Wright}{Rogerson
  et~al.}{2005}]{Rogerson2005}
Rogerson, R., R.~Shimer, and R.~Wright (2005, December).
\newblock Search-theoretic models of the labor market: A survey.
\newblock {\em Journal of Economic Literature\/}~{\em 43\/}(4), 959–988.

\bibitem[\protect\citeauthoryear{Rothenberg}{Rothenberg}{1971}]{Rothenberg1971}
Rothenberg, T.~J. (1971).
\newblock Identification in parametric models.
\newblock {\em Econometrica\/}~{\em 39\/}(3), 577--591.

\bibitem[\protect\citeauthoryear{Seabrook and Wiskott}{Seabrook and
  Wiskott}{2023}]{Seabrook_2023}
Seabrook, E. and L.~Wiskott (2023, October).
\newblock A tutorial on the spectral theory of markov chains.
\newblock {\em Neural Computation\/}~{\em 35\/}(11), 1713–1796.

\bibitem[\protect\citeauthoryear{Sisson and Fan}{Sisson and
  Fan}{2011}]{Sisson2011}
Sisson, S.~A. and Y.~Fan (2011).
\newblock Chapter 12 likelihood-free mcmc.
\newblock In {\em Handbook of Markov Chain Monte Carlo}, pp.\  313--338. CRC
  Press.

\bibitem[\protect\citeauthoryear{Tamer}{Tamer}{2003}]{Tamer2003}
Tamer, E. (2003).
\newblock Incomplete simultaneous discrete response model with multiple
  equilibria.
\newblock {\em The Review of Economic Studies\/}~{\em 70\/}(1), 147--165.

\bibitem[\protect\citeauthoryear{Vaart}{Vaart}{1998}]{Vaart1998}
Vaart, A. W. v.~d. (1998).
\newblock {\em Asymptotic Statistics}.
\newblock Cambridge Series in Statistical and Probabilistic Mathematics.
  Cambridge University Press.

\bibitem[\protect\citeauthoryear{Wright, Kircher, Julien, and Guerrieri}{Wright
  et~al.}{2021}]{Wright2021}
Wright, R., P.~Kircher, B.~Julien, and V.~Guerrieri (2021, March).
\newblock Directed search and competitive search equilibrium: A guided tour.
\newblock {\em Journal of Economic Literature\/}~{\em 59\/}(1), 90–148.

\bibitem[\protect\citeauthoryear{Zeng and Xie}{Zeng and Xie}{2008}]{Zeng2008}
Zeng, Z. and Y.~Xie (2008).
\newblock A preference‐opportunity‐choice framework with applications to
  intergroup friendship.
\newblock {\em American Journal of Sociology\/}~{\em 114\/}(3), 615--648.

\end{thebibliography}

\appendix
\section{Proofs of the main results}
In the following, write $N^k(g)$, $k \in \mathbb{N}$ and $g \in \mathcal{G}$, for the set of networks that differ from $g$ in \emph{exactly} $k$ edges. Throughout, we also use extensively the following characterization of the matrix $\Pi^2(x)$:

{\tiny
	\begin{equation*}
		\label{eq.pi.square}
		(\Pi(x)^2)_{gw} = \begin{cases}
		 \rho_{ij}(g|x^m(g)) F_{\epsilon(g_{ij})-\epsilon(h_{ij})}(\Delta u_{ij}(g|x_{ij}(g))) \rho_{kl}(h|x^m(h)) F_{\epsilon(h_{kl})-\epsilon(w_{kl})}(\Delta u_{kl}(h|x_{kl}(h))) + &  \text{if } w \in N^2(g), w_{ij} \neq g_{ij}, w_{kl} \neq g_{kl} \\
		\rho_{kl}(g|x^m(g)) F_{\epsilon(g_{kl})-\epsilon(s_{kl})}(\Delta u_{kl}(g|x_{kl}(g))) \rho_{ij}(s|x^m(s)) F_{\epsilon(s_{ij})-\epsilon(w_{ij})}(\Delta u_{ij}(s|x_{ij}(s)))  & \text{where } h, s \in N(g), h_{ij}\neq g_{ij}, s_{kl} \neq  g_{kl} \\
		\\
			\Pi_{gg}(x) \rho_{ij}(g|x^m(g))  F_{\epsilon(g_{ij})-\epsilon(w_{ij})}(\Delta u_{ij}(g|x_{ij}(g)))  + &  \text{if } w \in N(g), w_{ij} \neq g_{ij}  \\
			\rho_{ij}(g|x^m(g))   F_{\epsilon(g_{ij})-\epsilon(w_{ij})}(\Delta u_{ij}(g|x_{ij}(g)))  \Pi_{ww}(x) &\\
			\\
			\Pi_{gg}(x)\Pi_{gg}(x) + \sum_{s \in N(g)}\Pi_{gs}(x)\Pi_{sg}(x) & \text{if } g = w \\
			\\
			0 & \text{otherwise}
		\end{cases} \, .
	\end{equation*}
}

\subsection{Proof of \Cref{claim_bound}}
\label{proof_bound}

We divide the proof in steps.

\vspace{0.5em}
\noindent \textit{Step 1: for every $t \in \mathbb{N}$, the number and position of the strictly positive entries in $\Pi(x;\theta)^t$ does not depend on $(x,\theta) \in \mathcal{X}\times \Theta$.}

\noindent The case $t = 1$ is immediate given \Cref{ass.matching}, \Cref{ass.shocks}  and \eqref{eq.trans.matrix}. Fix $t \in \mathbb{N}$. Suppose that the number and position of the strictly positive entries in $\Pi(x;\theta)^t$ does not depend on $(x;\theta)$. Observe that, since $\Pi(x;\theta)^t$ and $\Pi(x;\theta)$ are both transition matrices, they only admit nonnegative entries. It thus follows  that, for any $g, w \in \mathcal{G}$, $(\Pi(x;\theta)^{t+1})_{gw} > 0$ if, and only if, $(\Pi(x;\theta)^{t})_{gs} > 0$ and $(\Pi(x;\theta))_{sw} > 0$ for some $s \in \mathcal{G}$. Since the number and position of strictly positive entries in both $\Pi(x;\theta)^{t}$ and $\Pi(x;\theta)$ do not depend on $(x,\theta)$, we conclude the number and position of strictly positive entries of $\Pi(x;\theta)^{t+1}$ does not depend on $(x,\theta)$. By induction, we have the desired result.

\vspace{0.5em}
\noindent \textit{Step 2: the number of strictly positive entries of $\Pi(x;\theta)^t$ is nondecreasing in $t$.}

\noindent This follows from the fact that $\Pi(x;\theta)_{gg} > 0$ for every $g \in \mathcal{G}$. Indeed, observe that, for any $g, w \in \mathcal{G}$, $(\Pi(x;\theta)^t)_{gw} > 0 \implies (\Pi(x;\theta)^t)_{gw} \Pi(x;\theta^t)_{ww} > 0 \implies (\Pi(x;\theta)^{t+1})_{gw} > 0$.

\vspace{0.5em}
\noindent \textit{Step 3: the number of strictly positive entries of $\Pi(x;\theta)^t$ is \emph{strictly increasing} in $t$, for $t \in \{1,2,\ldots, N(N-1)\}$.}

\noindent Fix $t < N(N-1)$. Observe that, for any $g \in \mathcal{G}$ and $w \in N^{t+1}(w)$, $(\Pi(x;\theta)^{t})_{gw} = 0$, as it is impossible for a network to transition to another network that differs from it in $t+1$ edges in $t$ rounds. However, we note that $(\Pi(x;\theta)^{t+1})_{gw} > 0$. Indeed, one can define a sequence $s_1, s_2, \ldots, s_{t+1} \in \mathcal{G}$, $s_1=g$ and $s_{t+1}=w$, with the property that $s_{n} \in N(s_{n-1})$ for every $n \in \{2, \ldots, t+1\}$. It then follows  that $\Pi(x;\theta)_{s_{n-1} s_n} > 0$ for all $n \in \{2, \ldots, t+1\}$, and by the Markovian property of the model,  $(\Pi(x;\theta)^{t+1})_{gw} > 0$.

\vspace{0.5em}
\noindent \textit{Step 4: proof of identification}

\noindent Define the map $\phi: \{1,\ldots, N(N-1)\} \times \mathcal{X} \times \Theta \mapsto \mathbb{N}$, $\phi(t, x, \theta) = \# \{(\Pi(x;\theta)^t)_{gw} > 0: g,w \in \mathcal{G}\}$. By Step 1, $\phi(t, x, \theta)$ does not depend on $(x,\theta)$. It thus follows that we can define a map $\psi : \{1,2,\ldots, N(N-1)\} \mapsto \mathbb{N}$, $\psi(t)\coloneqq \phi(t,x,\theta)$ for some $(x,\theta)$. By Steps 2 and 3, $\psi$ is invertible in $t$, from which follows the desired conclusion.´

\subsection{Proof of Proposition \ref{prop_tau2_pref}}
\label{app_proof_tau2_pref}
Notice that the probability of the network transitioning from $g$ to $w$ in two rounds is given by:

\begin{equation}
	\label{eq_trans_gw}
	\begin{aligned}
		\Pi^2_{gw}(X) = \rho_{ij}(g|X^m(g)) F_{\epsilon(g_{ij})-\epsilon(h_{ij})}(\Delta u_{ij}(g|X_{ij}(g))) \rho_{kl}(h|X^m(h)) F_{\epsilon(h_{kl})-\epsilon(w_{kl})}(\Delta u_{kl}(h|X_{kl}(h))) + \\
		\rho_{kl}(g|X^m(g)) F_{\epsilon(g_{kl})-\epsilon(s_{kl})}(\Delta u_{kl}(g|X_{kl}(g))) \rho_{ij}(s|X^m(s)) F_{\epsilon(s_{ij})-\epsilon(w_{ij})}(\Delta u_{ij}(s|X_{ij}(s)))
	\end{aligned}\, ,
\end{equation}

Under Assumption \ref{ass_tau2}, $\Delta u_{ij}(g|x_{ij}(g))$ (and $\Delta u_{ij}(s|x_{ij}(s))$) are identified, for every support point $x_{ij}(g)$ of $X_{ij}(g)$, by the limit:

$$\frac{\Pi_{gw}^2(\{\tilde{x}^m(v)\}_{v \in \{g,h,s\}},x_{ij}(g),\tilde{x}_{kl}(g))}{\lim_{n \to \infty} \Pi_{gw}^2(\{\tilde{x}^m(v)\}_{v \in \{g,h,s\}},\tilde{z}_n,\tilde{x}_{kl}(g))} = F_{\epsilon(g_{ij})-\epsilon(h_{ij})}(\Delta u_{ij}(g|x_{ij}(g))) \, ,$$

and the fact that $u \mapsto F_{\epsilon(g_{ij})-\epsilon(h_{ij})}(u)$ is strictly increasing by Assumption \ref{ass.shocks}. Symmetrically, we establish identification of $\Delta u_{kl}(g|x_{kl}(g))$ (and $\Delta u_{kl}(h|x_{kl}(h))$) from the limit:

$$\frac{\Pi_{gw}^2(\{\check{x}^m(v)\}_{v \in \{g,h,s\}},\check{x}_{ij}(g),{x}_{kl}(g))}{\lim_{n \to \infty} \Pi_{gw}^2(\{\check{x}^m(v)\}_{v \in \{g,h,s\}},\check{x}_{ij}(g),\check{z}_n)} = F_{\epsilon(g_{kl})-\epsilon(s_{kl})}(\Delta u_{kl}(g|x_{kl}(g))) \, .$$

Next, we observe that:

\begin{equation}
	\label{eq_transititon_two_by_two}
	\footnotesize 
	\begin{aligned}		
		\Pi_{gh}^2(X) = \Pi_{gg}(X)\rho_{ij}(g|X^m(g))F_{\epsilon(g_{ij})-\epsilon(h_{ij})}(\Delta u_{ij}(g|X_{ij}(g))) +\rho_{ij}(g|X^m(g)) F_{\epsilon(g_{ij})-\epsilon(h_{ij})}(\Delta u_{ij}(g|X_{ij}(g))) \Pi_{hh}(X)\, , \\
				\Pi_{hg}^2(X) = \Pi_{hh}(X)\rho_{ij}(h|X^m(h))F_{\epsilon(h_{ij})-\epsilon(g_{ij})}( \Delta u_{ij}(h|X_{ij}(h))) + \rho_{ij}(h|X^m(h))F_{\epsilon(h_{ij})-\epsilon(g_{ij})}(\Delta u_{ij}(h|X_{ij}(h))) \Pi_{gg}(X)\, 
	\end{aligned} \, ,
\end{equation}
and from $\Delta u_{ij}(g|X_{ij}(g)) = -\Delta u_{ij}(h|X_{ij}(h))$, it follows that:

$$\frac{\Pi_{gh}^2(X)}{	\Pi_{hg}^2(X) } = \frac{\rho_{ij}(g|X^m(g))}{\rho_{ij}(h|X^m(h))}\frac{F_{\epsilon(g_{ij})-\epsilon(h_{ij})}(\Delta u_{ij}(g|X_{ij}(g))) }{1- F_{\epsilon(g_{ij})-\epsilon(h_{ij})}(\Delta u_{ij}(g|X_{ij}(g))) }\, ,$$
thus establishing identification of the ratio $\frac{\rho_{ij}(g|x^m(g))}{\rho_{ij}(h|x^m(h))}$  for every support point of $(X^m(g),X^m(h))$. A similar argument establishes identification of the ratios $\frac{\rho_{ij}(w|x^m(w))}{\rho_{ij}(s|x^m(s))}$, $\frac{\rho_{kl}(g|x^m(g))}{\rho_{kl}(s|x^m(s))}$ and $\frac{\rho_{kl}(w|x^m(w))}{\rho_{kl}(h|x^m(h))}$. Moreover, by \eqref{eq_trans_wg}, we identify 
$\rho_{ij}(g|X^m(g))  \rho_{kl}(h|X^m(h)) +	\rho_{kl}(g|X^m(g)) \rho_{ij}(s|X^m(s))$. Similarly, noting that:

\begin{equation}
	\label{eq_trans_wg}
	\begin{aligned}
		\Pi^2_{wg}(X) = \rho_{ij}(w|X^m(w)) F_{\epsilon(w_{ij})-\epsilon(s_{ij})}(\Delta u_{ij}(w|X_{ij}(w))) \rho_{kl}(s|X^m(s)) F_{\epsilon(s_{kl})-\epsilon(g_{kl})}(\Delta u_{kl}(s|X_{kl}(s))) + \\
		\rho_{kl}(w|X^m(w)) F_{\epsilon(w_{kl})-\epsilon(h_{kl})}(\Delta u_{kl}(w|X_{ij}(w))) \rho_{ij}(h|X^m(h)) F_{\epsilon(h_{ij})-\epsilon(g_{ij})}(\Delta u_{ij}(h|X_{ij}(h))) 
	\end{aligned} \, ,
\end{equation}
we identify $\rho_{ij}(w|X^m(w))  \rho_{kl}(s|X^m(s)) +	\rho_{kl}(w|X^m(w)) \rho_{ij}(h|X^m(h)) $. Under restriction \eqref{eq_homoge}, $\rho_{ij}(w|X^m(w))  \rho_{kl}(s|X^m(s)) +	\rho_{kl}(w|X^m(w)) \rho_{ij}(h|X^m(h))  =  \rho_{ij}(w|X^m(w)) \delta \rho_{ij}(w|X^m(w)) \rho_{ij}(g|X^m(g)) $, where $\delta$ is identified. Combining the six identifying relations with the two restrictions in \eqref{eq_homoge}, we obtain a log-linear system of 7 restrictions and 8 variables, given by:

$$  		\underbrace{\begin{bmatrix}
	1 & -1 & 0 & 0 &0 & 0 & 0 & 0 \\
	0 & 0 & 1 & -1 & 0 & 0 & 0 & 0 \\
	0 & 0 & 0 & 0 & 1 & 0 & -1 & 0 \\
	0 & 0 & 0 & 0 & 0 & 1 & 0 & -1 \\
	1 & 0 & 0 & 0 &  1 & 0 & 0& 0  \\
	1 & 0 & 0 & 0 & -1 & 0 & 0 &  0 \\
	0& 0 & 0 & 1 & 0 & 0 & 0 & -1 \\
\end{bmatrix}  }_{=A}\underbrace{\begin{bmatrix}
	\log \rho_{ij}(g|X^m(g))\\ 
	\log \rho_{ij}(h|X^m(h)) \\
	\log \rho_{ij}(s|X^m(s))\\
	\log \rho_{ij}(w|X^m(w)) \\
	\log \rho_{kl}(g|X^m(g))\\ 
	\log \rho_{kl}(h|X^m(h)) \\
	\log \rho_{kl}(s|X^m(s))\\
	\log \rho_{kl}(w|X^m(w)) \\
\end{bmatrix}}_{=\psi}  = \begin{bmatrix}
\boldsymbol{\Delta}_{5 \times 1} \\
0 \\
0
\end{bmatrix}\, , $$
where $\boldsymbol{\Delta}$ is identified. The identification region of the log-meeting probabilities is thus a subset of $\psi + \operatorname{nul}(A) $. Since $A$ rank 7, it follows from the rank nulity theorem that $\psi + \operatorname{nul}(A) $ is an affine subspace of dimension one, which establishes the second part of Proposition \ref{prop_tau2_pref}. 

\subsection{Proof of Proposition \ref{prop_exclude_meeting}}
\label{proof_prop_exclude_meeting}

\begin{enumerate}
	\item[a] The identification of relative pay-offs is established by the fact that, under the assumption in the statement of the proposition, we have, for any $g \in \mathcal{G}$ and $h \in N(g)$ with $h_{ij} \neq g_{ij}$
	
	$$\frac{\Pi_{gh}(X)}{\Pi_{hg}(X)} = \frac{F_{\epsilon(g_{ij})-\epsilon(h_{ij})}(\Delta u_{ij}(g|X_{ij}(g))) }{1- F_{\epsilon(g_{ij})-\epsilon(h_{ij})}(\Delta u_{ij}(g|X_{ij}(g))) }\,.$$
	Since preferences are identified for any pair and network, meeting probabilities are then identified by:
	
	$$\rho_{ij}(g|X^m(g)) = \frac{\Pi_{gw}(X)}{F_{\epsilon(g_{ij})-\epsilon(h_{ij})}(\Delta u_{ij}(g|X_{ij}(g))) } \, . $$
	
	\item[b] The identification of relative pay-offs is established by the fact that, under the assumption in the statement of the proposition, it follows by \eqref{eq_transititon_two_by_two} that, for any $g \in \mathcal{G}$ and $h \in N(g)$ with $h_{ij} \neq g_{ij}$:
	
	$$\frac{\Pi_{gh}^2(X)}{	\Pi_{hg}^2(X) } = \frac{F_{\epsilon(g_{ij})-\epsilon(h_{ij})}(\Delta u_{ij}(g|X_{ij}(g))) }{1- F_{\epsilon(g_{ij})-\epsilon(h_{ij})}(\Delta u_{ij}(g|X_{ij}(g))) }\, ,$$
	
	Now, if we restrict our attention to pairs  $(i,j)\neq (k,l)$, networks $g \in \mathcal{G}$, and support points satisfying:
	
	$$\rho_{ij}(g|x^m(g)) = \rho_{kl}(g|x^m(g)) = \rho_{ij}(w|x^m(w)) = \rho_{kl}(w|x^m(w))\, ,$$
	we have that:
	
	\begin{equation}
		\begin{aligned}
			\Pi^2_{gw}(x) = \rho_{ij}(g|x^m(g)) F_{\epsilon(g_{ij})-\epsilon(h_{ij})}(\Delta u_{ij}(g|x_{ij}(g))) \rho_{kl}(h|x^m(h)) F_{\epsilon(h_{kl})-\epsilon(w_{kl})}(\Delta u_{kl}(h|x_{kl}(h))) + \\
			\rho_{kl}(g|x^m(g)) F_{\epsilon(g_{kl})-\epsilon(s_{kl})}(\Delta u_{kl}(g|x_{kl}(g))) \rho_{ij}(s|x^m(s)) F_{\epsilon(s_{ij})-\epsilon(w_{ij})}(\Delta u_{ij}(s|x_{ij}(s))) =\\
		 \rho_{ij}(g|x^m(g))^2 [F_{\epsilon(g_{ij})-\epsilon(h_{ij})}(\Delta u_{ij}(g|x_{ij}(g))) \rho_{kl}(h|x^m(h))  F_{\epsilon(h_{kl})-\epsilon(w_{kl})}(\Delta u_{kl}(h|x_{kl}(h))) + \\
	F_{\epsilon(g_{kl})-\epsilon(s_{kl})}(\Delta u_{kl}(g|x_{kl}(g))) ) F_{\epsilon(s_{ij})-\epsilon(w_{ij})}(\Delta u_{ij}(s|x_{ij}(s))) ] \, .
		\end{aligned}\, ,
	\end{equation}
	
	Since marginal utilities are identified, we thus identify $\rho_{ij}(g|x^m(g))$.
\end{enumerate}

\subsection{Proof of Proposition \ref{prop_meet_excl}}
\label{app_meet_excl}
\begin{enumerate}
	\item[a] Marginal utilities $\Delta u_{ij}(g|x_{ij}(g))$ are identified, for every support point $x_{ij}(g)$ of $X_{ij}(g)$ as:
	
	$$\lim_{n \to \infty}\Pi(\{x_{ij}(g), z_n\};\theta_0) = F_{\epsilon(g_{ij})-\epsilon(w_{ij})}(\Delta u_{ij}(g|X_{ij}(g)))\, .$$
	
	Meeting probabilities are then identified by:
	
	$$\rho_{ij}(g|X^m(g)) =\frac{\Pi(X;\theta_0)}{F_{\epsilon(g_{ij})-\epsilon(w_{ij})}(\Delta u_{ij}(g|X_{ij}(g)))}\, ,$$
		$$\rho_{ij}(w|X^m(w)) =\frac{\Pi(X;\theta_0)}{1 - F_{\epsilon(g_{ij})-\epsilon(w_{ij})}(\Delta u_{ij}(g|X_{ij}(g)))}\, .$$
	
	\item[b] Relative utilities are identified, for every support point $x_{ij}(g)$ of $X_{ij}(g)$, by:
	
	$$\lim_{n\to \infty}\frac{ \Pi^2_{gw}(\{a_n, x_{ij}(g), x^{m}(w)\})}{ \Pi^2_{wg}(\{b_n, x_{ij}(g), x^{m}(w)\})} =\frac{F_{\epsilon(g_{ij})-\epsilon(w_{ij})}(\Delta u_{ij}(g|x_{ij}(g)))}{1-F_{\epsilon(g_{ij})-\epsilon(w_{ij})}(\Delta u_{ij}(g|x_{ij}(g)))}$$
	
	Next, we observe that:
	$$\Pi^2_{gg}(X;\theta_0) = \sum_{s \in N(g)\cup\{g\}}\Pi_{gs}(X;\theta_0)\Pi_{sg}(X;\theta_0)\ , .$$
	
	Consequently, for support points $(x_{ij}(g), x^{m}(w))$ of $(X_{ij}(g),X^{m}(w))$, we have that:
	\begin{equation*}
		\begin{aligned}
			\lim_{n\to \infty} \Pi^2_{gg}(\{\tilde{x}_n, z_n, x_{ij}(g), x^{m}(w)\}) =\\  (1-F_{\epsilon(g_{ij})-\epsilon(w_{ij})}(\Delta u_{ij}(g|x_{ij}(g))))^2  + \\ 
			 \rho_{ij}(w|x^m(w)) F_{\epsilon(g_{ij})-\epsilon(w_{ij})}(\Delta u_{ij}(g|x_{ij}(g)))(1-F_{\epsilon(g_{ij})-\epsilon(w_{ij})}(\Delta u_{ij}(g|x_{ij}(g)))) \, ,
		\end{aligned}
	\end{equation*}	
	where $\tilde{x_n}_{n \in \mathbb{N}}$ is a sequence of values in the support of the other covariates entering the decisions relevant to $\Pi^2_{gg}$. Since relative utilities are identified, $ \rho_{ij}(w|x^m(w))$ is identified. It then follows that $\rho_{ij}(g|x^m(g))$ is identified by the ratio $\Pi^2_{gw}/\Pi^2_{wg}$.
\end{enumerate}

\subsection{Proof of Proposition \ref{prop_spectral}}
\label{proof_prop_spectral}
To establish the result in Proposition \ref{prop_spectral}, we begin with the following result:

\begin{lemma}
	Under Assumptions \ref{ass.matching}, \ref{ass.shocks}, \ref{ass_distribution}, \ref{ass_pot_pref} and \ref{ass_pot_meet}, and for each $x \in \mathcal{X}$, the Markov chain $\Pi(x;\tau_0)$ admits a unique conditional stationary distribution $\pi(\cdot|x)$ given by:
	$$\pi(g|x)\propto \exp(Q(g,x)- M(g,x))\, .$$
	
	The stationary distribution satisfies the flow balancedness condition:
	
	\begin{equation}
		\label{eq_flow}
		\pi(g|x) \Pi_{gw}(x;\theta_0)  =  \pi(w|x)	\Pi_{wg}(x;\theta_0) , \quad \forall g,w\in \mathcal{G}
	\end{equation}
	\begin{proof}
		That the Markov chain admits a unique stationary distribution follows from Remark \ref{rmk.existence}. It then suffices to verify that the candidate distribution satisfies the flow balancedness condition \eqref{eq_flow} to assert that it is the stationary distribution. Notice that, for any $g \in \mathcal{G}$, \eqref{eq_flow} is trivially satisfied for $w \notin N(g)$. Consider now that $w \in N(g)$, with $w_{ij} \neq g_{ij}$. In this case, we have that:
		
		\begin{equation*}
			\begin{aligned}
				\frac{\pi(w|x)}{\pi(g|x)} = \exp(Q(w,x)-Q(g,x) - (M(w,x) - M(g,x)) ) = \exp(\Delta u_{ij}(g|x_{ij}(g)))\frac{\rho_{ij}(g|x^m(g))}{\rho_{ij}(w|x^m(w))}  = \\
				\frac{\frac{\exp(\Delta u_{ij}(g|x_{ij}(g)))}{1+\exp(\Delta u_{ij}(g|x_{ij}(g))) }}{\frac{1}{1+\exp(\Delta u_{ij}(g|x_{ij}(g))) }}\frac{\rho_{ij}(g|x^m(g))}{\rho_{ij}(w|x^m(w))} = \frac{\Pi_{gw}(x;\theta_0)}{\Pi_{wg}(x;\theta_0)}\, .
			\end{aligned}
		\end{equation*}
		
	\end{proof}

\end{lemma}

Fix $x \in \mathcal{X}$. Since the Markov chain $\Pi(x;\theta_0)$ is irreducible (Remark \ref{rmk_irr}), it is \emph{recurrent} \citep[Proposition 2]{Seabrook_2023}. Since, by the above lemma, the unique stationary distribution satisfies the flow-balancedness condition \eqref{eq_flow}, it follows by Theorem 4 of \cite{Seabrook_2023} that the chain is \emph{reversible}. But then, applying Theorem 14 of \cite{Seabrook_2023},  we conclude that $\Pi(x;\theta_0)$ is diagonalizable with real eigenvalues, meaning that:

\begin{equation}
	\label{eq_diag_val}
	\Pi(x) = Y(x)\operatorname{diag}(\lambda_1(x), \lambda_2(x),\ldots,\lambda_{|\mathcal{G}|}(x))Y(x)^{-1}\, ,
\end{equation}
 where $\lambda_j(x) \in \mathbb{R}$ is the $j$-th eigenvalue of $\Pi(x)$, arranged in decreasing order, and $Y(x)$ is a matrix with linearly independent columns, with $j$-column equal to an eigenvector associated with $\lambda_j(x)$. The above representation implies that, to identify $\Pi(x)$, it suffices to identify its eigenvalues and the eigenspace associated with each eigenvalue.\footnote{\textit{Proof:} Since $\Pi(x;\theta_0)$ is diagonalizable, the algebraic and geometric mulitplicities of eigenvalues coincide. Consequently, letting $V_\lambda$ denote a matrix whose columns correspond to the $n_\lambda$ distinct eigenvalues associated with the eigenvalue $\lambda$ and entering \eqref{eq_diag_val}, we have that, for any $|\mathcal{G}|\times n_\lambda$ matrix $B_\lambda$  storing in its columns  a basis of the eigenspace of $\lambda$, we have that: $$B_\lambda = V_\lambda A_\lambda \, ,$$ where $A_\lambda$ is a $n_\lambda \times n_\lambda$ invertible matrix. Consequently, having identified the $s$ distinct eigenvalues $\lambda_1 \geq \lambda_2 \geq \ldots \geq \lambda_s$ of $\Pi(x;\theta_0)$ and corresponding eigenbasis $B_{\lambda_1}, \ldots, B_{\lambda_s}$, we have that:
 \begin{equation*}
 	\begin{aligned}
 		 \begin{bmatrix}
 			B_{\lambda_1} & B_{\lambda_2} & \ldots & B_{\lambda_s}
 		\end{bmatrix} \operatorname{diag}(\underbrace{\lambda_1}_{n_{\lambda_1} \ \text{times}},\underbrace{\lambda_2}_{n_{\lambda_2} \ \text{times}} \ldots \underbrace{\lambda_s}_{n_{\lambda_s} \ \text{times}} ) \begin{bmatrix}
 			B_{\lambda_1} & B_{\lambda_2} & \ldots & B_{\lambda_s}
 		\end{bmatrix}^{-1} = \\ Y(x) A  \operatorname{diag}(\underbrace{\lambda_1}_{n_{\lambda_1} \ \text{times}},\underbrace{\lambda_2}_{n_{\lambda_2} \ \text{times}} \ldots \underbrace{\lambda_s}_{n_{\lambda_s} \ \text{times}} ) A^{-1} Y(x) = Y(x)   \operatorname{diag}(\underbrace{\lambda_1}_{n_{\lambda_1} \ \text{times}},\underbrace{\lambda_2}_{n_{\lambda_2} \ \text{times}} \ldots \underbrace{\lambda_s}_{n_{\lambda_s} \ \text{times}} ) Y(x) = \Pi(x;\theta_0)\, , 
 	\end{aligned}
 \end{equation*}

 where $A$ is a block-diagonal matrix with $s$ blocks and $j$-th block given by $A_{\lambda_j}$. Since its inverse is also a block-diagonal matrix, with $j$-th block given by $A_{\lambda_j}^{-1}$, it follows that $A  \operatorname{diag}(\underbrace{\lambda_1}_{n_{\lambda_1} \ \text{times}},\underbrace{\lambda_2}_{n_{\lambda_2} \ \text{times}} \ldots \underbrace{\lambda_s}_{n_{\lambda_s} \ \text{times}} ) A^{-1} = \operatorname{diag}(\underbrace{\lambda_1}_{n_{\lambda_1} \ \text{times}},\underbrace{\lambda_2}_{n_{\lambda_2} \ \text{times}} \ldots \underbrace{\lambda_s}_{n_{\lambda_s} \ \text{times}} )$, which establishes identifiability of $\Pi(x;\theta_0)$ as given by the expression above. 
   } Now, recalling that, if $v$ is an eigenvector associated with an eigenvalue $\lambda$ of a matrix $A$, then $v$ is an eigenvector of an eigenvalue associated with eigenvalue $\lambda^n$ of  matrix $A^n$, we conclude that, if $\tau_0$ is odd, it is possible to recover the eigenvalues and eigenspace of $\Pi(x;\theta_0)$ from the eigenvalues and eigenspace of $\Pi(x;\theta_0)^{\tau_0}$. If $\tau_0$ is even, then this is not the case, since the sign of eigenvalues is not identified. However, it is possible to identify the eigenvalues and associated eigenspaces of $\Pi(x;\theta_0)^2$. Since $\Pi(x;\theta_0)^2$ is also an irreducible transition matrix and has the same stationary distribution as $\Pi(x;\theta_0)$, we conclude that  $\Pi(x;\theta_0)^2$ is also diagonalizable and thus it is possible to identify $\Pi(x;\theta_0)^2$ from $\Pi(x;\theta_0)^{\tau_0}$.
   
   The previous paragraph established item (a) and the first part of item (b) in the Proposition. As for the final part of item (b), the result is an immediate consequence of Corollary 3.3 of \cite{Casanellas2023}.

\subsection{Proof of Proposition \ref{proposition_parametric}}
\label{app_parametric}
We first begin by noticing that, since we assumed that the potential function approximation \eqref{approx_meeting} holds, the conditions of Proposition \ref{prop_spectral} are satisfied. In this case, it suffices to establish identification from $\Pi(X;\theta_0)$, if $\tau_0$ is odd; or identification from $\Pi(X;\theta_0)^2$, if $\tau_0$ is even. Let $\boldsymbol{b}_{1}$ denote the first entry of $\beta_{\text{ud}}$, and $\boldsymbol{b}_{-1}$ the subvector consisting of the remaining entries of  $\beta_{\text{ud}}$. We split the proof in three parts

	\paragraph{Part 1} Let $s_0 =1$ if $\tau_0$ is odd; and $s_0 = 2$ if $\tau_0$ is even. Let $\boldsymbol{O}$ denote an empty network, and $\boldsymbol{I}^{ij} = [1,\boldsymbol{O}_{-ij}]$.  Observe that, by the potential function approximation:
	
	$$\log(\Pi_{\boldsymbol{O}\boldsymbol{I}^{ij}}(X;\theta_0)^{s_0}) - \log(\Pi_{\boldsymbol{I}^{ij}\boldsymbol{O}}(X;\theta_0)^{s_0}) = \beta'_{\text{ud}}\begin{bmatrix}
		1 \\ W_{ij}
	\end{bmatrix}  - \delta  - \psi' Z_{ij}\, ,$$
	which, under the rank condition in item (i) in the statement of the proposition, establishes identification of $\psi$, $\boldsymbol{b}_{-1}$ and $\boldsymbol{b}_{1}+\delta$. Next, we identify the reciprocal component of preferences. Let $\tilde{\boldsymbol{I}}^{ij}_{-jl} =  [1, \boldsymbol{I}^{ij}_{-jl}]$ and $\tilde{\boldsymbol{O}}_{-jl}= [1, \boldsymbol{O}_{-jl}]$, for $d \in \{0,1\}$. We obtain that:
	
	\begin{align*}
		\small
		\log(\Pi_{\tilde{\boldsymbol{O}}_{-jl}\tilde{\boldsymbol{I}}^{ij}_{-jl}}(X;\theta_0)^{s_0}) - \log(\Pi_{\tilde{\boldsymbol{I}}^{ij}_{-jl}\tilde{\boldsymbol{O}}_{-jl}}(X;\theta_0)^{s_0})  - \left(\log(\Pi_{\boldsymbol{O}\boldsymbol{I}^{ij}}(X;\theta_0)^{s_0}) - \log(\Pi_{\boldsymbol{I}^{ij}\boldsymbol{O}}(X;\theta_0)^{s_0})\right)  = \\  \beta'_{\text{ur}}\begin{bmatrix}
			1 \\ W_{ij}
		\end{bmatrix} ,
	\end{align*}
	thus establishing identification of the $\beta_{\text{ur}}$ under the rank condition. Finally, for identification of the ``indirect'' component of preferences, consider an agent $k \in \mathcal{I}\setminus\{i,j\}$. We have that
	
		\begin{align*}
			\small
		\log(\Pi_{\tilde{\boldsymbol{O}}_{-kj}\tilde{\boldsymbol{I}}^{ik}_{-kj}}(X;\theta_0)^{s_0}) - \log(\Pi_{\tilde{\boldsymbol{I}}^{ik}_{-kj}\tilde{\boldsymbol{O}}_{-kj}}(X;\theta_0)^{s_0}) - \left(\log(\Pi_{\boldsymbol{O}\boldsymbol{I}^{ik}}(X;\theta_0)^{s_0}) - \log(\Pi_{\boldsymbol{I}^{ik}\boldsymbol{O}}(X;\theta_0)^{s_0})\right)  = \\  \beta'_{\text{un}}\begin{bmatrix}
			1 \\ W_{ij}
		\end{bmatrix} ,
	\end{align*}
		thus establishing identification of the $\beta_{\text{un}}$ under the rank condition.
	
	\paragraph{Part 2} We now consider the case where $\tau_0$ is odd, and $\tau_0$ is even; separately:
	
	\begin{enumerate}
		\item $\tau_0$ is odd: By the assumption in the theorem, we have that:
		
		\begin{align*}
		\log(\Pi_{\boldsymbol{O}\boldsymbol{I}^{ab}}(X))-\log(\Pi_{\boldsymbol{O}\boldsymbol{I}^{cd}}(X)) =  \gamma' (W_{a,b} - W_{c,d}) + \Delta u_{ab}(\boldsymbol{O}|X) - \Delta u_{cd}(\boldsymbol{O}|X) \\- \left(\log(1+\exp(\Delta u_{ab}(\boldsymbol{O}|X))) - \log(1+\Delta u_{cd}(\boldsymbol{O}|X))\right)
		\end{align*}
		
		Given the result in the first part, the unknown quantities in the right-hand side of the above expression are $\boldsymbol{b}_1$ and $\gamma$. Consider the identified moments:
		
		$	\mathbb{E}\left[(\log(\Pi_{\boldsymbol{O}\boldsymbol{I}^{ab}}(X))-\log(\Pi_{\boldsymbol{O}\boldsymbol{I}^{cd}}(X)))  \begin{bmatrix}
			(W_{a,b}-W_{c,d}) \\
			1
		\end{bmatrix}\right] \eqqcolon\mu(\boldsymbol{b}_1, \gamma) $
		
		The Jacobian with respect to the unknown parameters, evaluated at values $\tilde{\boldsymbol{b}}_1, \tilde{\gamma}$, is given by:
		
	$$	J(\tilde{\boldsymbol{b}}, \tilde{\gamma}) =	\begin{bmatrix}
		\mathbb{E}[	\Delta \tilde{W} \Delta \tilde{W}']  & \mathbb{B}[\Delta \tilde{p}\Delta \tilde{W}] \\
		\mathbb{E}[\Delta \tilde{W}'] & \mathbb{E}[\Delta \tilde{p}]
		\end{bmatrix}$$
where $\Delta \tilde{p} = \operatorname{Logit}(\tilde{\boldsymbol{b}}_1+ \boldsymbol{b}_{-1}'W_{a,b}) - \operatorname{Logit}(\tilde{\boldsymbol{b}}_1 +  \boldsymbol{b}_{-1}'W_{c,d})$. By the rank conditions, $J(\tilde{\boldsymbol{b}}, \tilde{\gamma}) $ has full rank, for every value $\tilde{\boldsymbol{b}}_1, \tilde{\gamma}$.\footnote{Clearly, by the first rank condition, the upper-left block $\mathbb{E}[	\Delta \tilde{W} \Delta \tilde{W}']  $ has full rank. Consequently, for the whole matrix to have full rank, it is sufficient that the last column of  $J(\tilde{\boldsymbol{b}}, \tilde{\gamma}) $ is not in the linear span of the remaining columns. But the second rank condition precisely implies that. } It thus follows from Theorem 6 of \cite{Rothenberg1971} that $\boldsymbol{b}_1$ and $\gamma$ are locally identified. Since $\gamma$ enters linearly in the moment function $\mu$, it is also globally identified \citep[Theorem 5]{Rothenberg1971}. Finally, we may globally identify $\boldsymbol{b}_1$ from $\Pi_{\boldsymbol{O}\boldsymbol{I}^{ab}}(X))$, since this identified quantity now only depends on a single unknown ($\boldsymbol{b}_1$), and one may invert this mapping to obtain $\boldsymbol{b}_1$. Since $\boldsymbol{b}_1$ is now identified, and given that $\boldsymbol{b}_1+\delta$ was identified in Part 1, we identify $\delta$.

\item $\tau_0$ is even:  Consider the transition in two rounds from $\boldsymbol{O}$ to the network $\boldsymbol{I}_{(a,b)(e,f)}$ where edges $(a,b)$ and $(e,f)$ are turned on. By the continuous variation assumption, the directional derivative $\partial_{ \overrightarrow{t} } \Pi^2_{\boldsymbol{O}\boldsymbol{I}_{(a,b)(e,f)}}$ at $z_{e,f}$ is identifed, for $\mathbb{P}_X$-almost every value of the $\{W_m:m \in \mathcal{M}\}$. By the relevance condition, this derivative is non-zero. Similarly, considering the transition in two rounds from $\boldsymbol{O}$ to the network $\boldsymbol{I}_{(c,d)(e,f)}$ where edges $(c,d)$ and $(e,f)$ are turned on, we have that $\partial_{ \overrightarrow{t} } \Pi^2_{\boldsymbol{O}\boldsymbol{I}_{(c,d)(e,f)}}$at $z_{e,f}$ is nonzero and identified, for $\mathbb{P}_X$-almost every value of the $\{W_m:m \in \mathcal{M}\}$. Consequently, we obtain that, $\mathbb{P}_X$-almost surely:

\begin{align*}
	\log(\partial_{ \overrightarrow{t} } \Pi^2_{\boldsymbol{O}\boldsymbol{I}_{(a,b)(e,f)}})-\log(\partial_{ \overrightarrow{t} } \Pi^2_{\boldsymbol{O}\boldsymbol{I}_{(c,d)(e,f)}}) =  \gamma' (W_{a,b} - W_{c,d}) + \Delta u_{ab}(\boldsymbol{I}^{e,f}|X) - \Delta u_{cd}(\boldsymbol{I}^{e,f}||X) \\- \left(\log(1+\exp(\Delta u_{ab}(\boldsymbol{I}^{e,f}|X))) - \log(1+\Delta u_{cd}(\boldsymbol{I}^{e,f}|X))\right) \, ,
\end{align*}
which establishes global identification of $\boldsymbol{\gamma}$ and local identification of $\boldsymbol{b}_1$, by an analogous reasoning underlying part one, since $\Delta u_{ab}(\boldsymbol{I}^{e,f}|X)=\Delta u_{ab}(\boldsymbol{O}|X)$ a $\Delta u_{cd}(\boldsymbol{I}^{e,f}|X)=\Delta u_{abcd}(\boldsymbol{O}|X)$ by our assumption on the identity of agents $e,f$ and the fact that the rank conditions in the statement of the Proposition now hold in the limit where $\epsilon$ is arbitrarily small. We can then globally identify $\boldsymbol{b}_1$ by inverting $\partial_{ \overrightarrow{t} } \Pi^2_{\boldsymbol{O}\boldsymbol{I}_{(a,b)(e,f)}}$.
	\end{enumerate}
	
	\begin{remark}
		Even though the identification argument in the proof of the proposition above considered transitions that started from an empty adjacency matrix, it is clear from the structure of the argument that it is also possible to consider alternative starting networks, provided that the formulae relating the identified transition probabilities to structural parameters remain the same. For example, one could consider transitions starting from ``locally''  empty networks around the pairs upon which the identification argument relies, or more generally any transition starting from an initial network that does not alter the relative pay-offs of agents across the relevant transitions.
	\end{remark}
	
\subsection{Proof of \Cref{lemma.est.tau0}}
\label{tau_proof}
\begin{proof}
 Consider the event $\{\hat{\tau} \to \tau_0\}^\complement $. Observe that $\{\hat{\tau} \to \tau_0\}^\complement = \cap_{c \in \mathbb{N}} \{\lVert G^{T_1}_c - G^{T_0}_c\rVert_{F} < \tau_0\}$. Fix $k \in \mathbb{N}$ and notice that
 $$\prob\left[\cap_{c=1}^k \{\lVert G^{T_1}_c - G^{T_0}_c\rVert_{F} < \tau_0\}\right] = \phi^k \, ,$$
where $\phi$ is the ex-ante (unconditional) probability  that $G^{T_1}_c$ differs from $G^{T_0}_c$ in strictly less than $\tau_0$ edges. Since $\prob[\lVert G^{T_1}_c - G^{T_0}_c\rVert_{F} = \tau_0] \in (0,1)$ (which follows from \ref{ass.matching}, \ref{ass.shocks}, and \ref{ass.upperbound.tau0}), we have $\phi \in (0,1)$. Passing $k$ to the limit and using continuity of $\prob$ from above, we conclude the desired result.
\end{proof}

It is apparent from the proof above that it is possible extend the result of \Cref{lemma.est.tau0} in order to allow for a sequence of \emph{independent} observations stemming from a game with common $\tau_0$, but allowing utilities, the meeting process, the distribution of covariates and the number of players to vary between networks. In this case, we must restrict the distribution of covariates, utilities, the meeting process and the number of players to not shift excessively to regions where the probability of the network changing by strictly less than $\tau_0$ edges is high. Formally, the proof would change as the ex-ante probability would now depend on $c$, i.e. we would have $\prod_{c=1}^k \phi_c$ in the formula. If $\limsup_{c \to \infty} \phi_c < 1$, we would get the same result. Observe this implies $\limsup_{c \to \infty} \mathbb{P}[N_c (N_c-1) \geq \tau_0] > 0$.

\pagebreak
\pagestyle{plain}
 \setcounter{page}{1}
\setcounter{footnote}{0} 

\section*{\centering Supplemental Appendix to ``Homophily in preferences or meetings? Identifying and estimating an iterative network formation model''}
\subsection*{Contents}
\startcontents[appendix]
\printcontents[appendix]{ }{1}{ }

\section{Identification results in the nonparametric model}
\label{app_np}
This Appendix studies the identification of utilities and meeting processes in the nonparametric setting defined in \Cref{setup}. Section \ref{Supplement-identification_stationary} shows that, in a setting where one period of data from the model's stationary distribution is available, nonparametric identification is generally impossible. Section \ref{Supplement-non_two} shows that, even when two periods of data are available, nonparametric identification is generally impossible absent further restrictions on meetings or preferences. In light of this result, Appendices \ref{identification_covariates} and \ref{Supplement-ident.exclusion.matching}  present identification results that rely on the existence of large-support covariates entering preferences (meetings) but that are ``locally'' exluded from the meeting process (preferences). We show that, under such exclusion restrictions, preference and meeting parameters may be recovered through  identification-at-infinity arguments. Finally, Supplemental Appendices \ref{Supplement-proof.claim.N2} and \ref{proposition3.1} present the proofs of results omitted in earlier subsections.
\subsection{Non-identification with one period of data from the stationary distribution}
\label{Supplement-identification_stationary}

In this section, we explore identification in a context where we have access to a sample of $C$ networks stemming from the stationary distribution of the game described in \Cref{setup} of the main text. In particular, we have access to a sample $\{X_c , G_c\}_{c=1}^C$.  It follows from \Cref{rmk.existence} in the main text that one may interpret the stationary distribution as a long-run distribution. If one assumes that the observed network data ($G_c$) was drawn from the (conditional) stationary distribution, then the (conditional) stationary distribution $\pi(X)$ is identified. In this context, identification of the transition matrix $\Pi(X)$ is a \emph{necessary} condition for identification of $((u_i)_{i=1}^N,\rho)$, the objects of interest. We thus propose to analyze the identification of $\Pi(X)$. In particular, we explore the identification of $\Pi(X)$ without imposing further restrictions. In light of this, and without loss of generality, we may essentially view $X$ as nonstochastic throughout the remainder of this section and suppress dependence of $\Pi(X)$ on $X$ by writing $\Pi$.

In our setting, the identification problem (of $\Pi$) reduces to providing conditions under which no other transition matrix $\tilde{\Pi} \in \mathcal{S}$ is observationally equivalent to $\Pi$; where $\mathcal{S}$ is admissible (by the model) set of Markov chains. In other words, $\Pi$ is identified if

\begin{equation}
	\begin{aligned}
		\forall \  \tilde{\Pi} \in \mathcal{S} \quad  (I - \tilde{\Pi}')\pi =0 \implies \tilde{\Pi} = \Pi \, .
	\end{aligned}
\end{equation}

If $\mathcal{S}$ were the set of all row-stochastic matrices, $\Pi$ would clearly not be identified, as $\mathbb{I}_{2^{N(N-1)} \times 2^{N(N-1)}}$ is observationally equivalent. But $\mathcal{S}$ is not the set of all Markov matrices. Indeed, the model imposes restrictions on the set of admissible Markov matrices. As we do not impose further restrictions on utilities and the matching function, \emph{$\mathcal{S}$ is the set of all  $2^{N(N-1)} \times 2^{N(N-1)}$ row-stochastic matrices with strictly positive entries $\Pi_{gw}$ for all $g \in \mathcal{G}$, $w \in N(g)\cup \{g\}$,  and 0 otherwise}.

Do the restrictions implied by the model identify $\Pi$? The following lemma is a negative result.

\begin{lemma}[Non-identification]
	\label{Supplement-thm.non}
	Under Assumptions \ref{ass.matching} and \ref{ass.shocks}, and if $F_\epsilon (e_0, e_1) = \exp[-\exp(e_0) - \exp(e_1)] $ (i.e. $(\epsilon(0), \epsilon(1))$ are independent EV type 1), then the model is not identified.
	\begin{proof}
		Fix $\Pi_0 \in \mathcal{S}$ and let $\pi_0$ be the (unique) solution to $(I - \Pi_0')\pi_0 = 0$, $\pi_0'\iota=1$. The proof presents a family of observationally equivalent versions of the model of \cite{Mele2017}. Consider a family of utilities $(u_i)_{i=1}^N$ where $u_i(g,X) = \ln(\pi_0(g))$ for all $i \in \mathcal{I}$ and all $g \in \mathcal{G}$. These utilities are well defined, as $\pi_0 >> 0$ from \Cref{rmk.pos.distr} in the main text. Moreover, fix some arbitrary $\rho$ satisfying (i) \Cref{ass.matching}; and (ii) $\rho((i,j), [0, g_{-ij}], X) = \rho((i,j), [1, g_{-ij}], X)$ for all $(i,j) \in \mathcal{M}$, $g \in \mathcal{G}$ (the matching probability does not depend on the existence of a link between $i$ and $j$). The pair $((u_i)_i, \rho)$ satisfies Assumptions \ref{ass.matching} and \ref{ass.shocks} in the main text. Therefore, there exists a unique stationary distribution $\tilde{\pi}$ associated with the chain $\tilde{\Pi}$ from this game. Further notice that the family of utilities admits a \emph{potential function} $Q: \mathcal{G} \times \mathcal{X} \mapsto \mathbb{R}$ satisfying $Q([1, g_{-ij}], X) - Q([0, g_{-ij}, X]) = u_i([1, g_{-ij}], X) - u_i([0, g_{-ij}], X)$ for all $(i,j) \in \mathcal{M}$, $g \in \mathcal{G}$. Indeed, $Q(g,X) = \ln(\pi_0(g))$ is a potential function for this class of utilities.  But then, this distribution has a closed form expression: $\tilde{\pi}(g) = \frac{\exp(Q(g,X))}{\sum_{w \in \mathcal{G}} \exp(Q(w,X))} = \pi_0(g)$. To see this, one can follow the argument in Theorem 1 of \cite{Mele2017} and verify that this distribution satisfies the flow balancedness condition $\tilde{\pi}_g \tilde{\Pi}_{gw}= \tilde{\pi}_w \tilde{\Pi}_{wg}$ for all $g, w \in \mathcal{G}$.\footnote{See Online Appendix A of \cite{Mele2017}.} But then the stationary distribution does not depend on the matching probabilities, so the Markov matrix is not identified, as it is always possible to choose $\rho$ satisfying $(i)-(ii)$ s.t. $\tilde{\Pi} \neq \Pi_0$ (and $\pi_0 = \tilde{\pi}$ follows, as we saw).
	\end{proof}
\end{lemma}

We interpret this result as suggesting the need to impose further restrictions to identify $\Pi$. Note that verification of the flow balancedness condition relies crucially on the functional form of the distribution function for the error term. One approach would then be to restrict the analysis to different distributions. However, we do not recommend this approach, as it is not grounded in knowledge regarding the social interactions being analyzed (identification by functional form). Moreover, and as it will become clearer later sections, the exclusion restriction approaches to identification in settings where two periods of data are available that we outline in Appendices \ref{identification_covariates} and \ref{Supplement-ident.exclusion.matching} do not bring additional identifying power in the setting of \Cref{Supplement-thm.non} without further restrictions on the matching function and/or utilities. In particular, one would require restricting $((u_i)_{i=1}^N,\rho)$ to classes where the assumptions of \cite{Mele2017} regarding the matching function and/or utilities do \textbf{not} hold. Since this would restrict the generality and applicability of the model, we refrain from further analyzing identification with one period of data from the stationary distribution.

\subsection{(Non)-identification with two periods of data and no further restrictions}
\label{Supplement-non_two}
To illustrate the difficulty of identification without imposing further restrictions, let us briefly analyze the identification of $\theta_0$ from $\Pi(X;\theta_0)$, which is a necessary condition for identification of $\theta_0$ from $\Pi(X;\theta_0)^{\tau_0}$. Indeed, knowledge of $\Pi(X;\theta_0)$ implies knowledge of $\Pi(X;\theta_0)^{\tau_0}$. Thus, in a sense, our analysis in this subsection provides a ``best-case'' scenario for achieving identification without additional restrictions.

As we do not impose further restrictions in the model, we essentially view $X$ as nonstochastic throughout the remainder of this subsection and suppress dependence of $\Pi(X;\theta)$ on $X$ by writing $\Pi(\theta)$. To make the identification problem clearer, define, for all $g \in \mathcal{G}$ and $w \in N(g)$ with $g_{ij} \neq w_{ij}$, $F_{ij}(g,w) \coloneqq F_{\epsilon(g_{ij}) - \epsilon(w_{ij})}(u_i(w, X) - u_i(g, X))$. Observe that $F_{ij}(g,w) + F_{ij}(w,g) = 1$. Write $\rho_{ij}(g)$ for $\rho((i,j),g,X)$. Observe that $\sum_{(i,j) \in \mathcal{M} }\rho_{ij}(g)=1$. Let $\gamma \coloneqq (\left(\rho_{ij}(g)\right)_{g \in \mathcal{G}, (i,j) \in \mathcal{M}}, \left(F_{ij}(g,w)\right)_{g \in \mathcal{G}, w \in N(g), g_{ij}\neq w_{ij}} )$ be a parameter vector, and $\gamma_0$ the ``true'' parameter. Observe that, under Assumptions \ref{ass.matching} and \ref{ass.shocks}, the parameter space, which we denote by $\Gamma$, is a subset of $\mathbb{R}_{++}^{\dim \gamma_0}$, an open set. Put another way, $\Gamma = \{\gamma \in \mathbb{R}_{++}^{\dim \gamma_0}:  F_{ij}(g,w) + F_{ij}(w,g) = 1,  \sum_{(k,l) \in \mathcal{M} }\rho_{k,l}(g) = 1 \text{ for all } g \in \mathcal{G}, w \in N(g) \text{ with } g_{ij} \neq w_{ij} \}$. Identification from the transition matrix thus requires us to show that, for all $ \gamma' \in \Gamma$, $\gamma' \neq \gamma_0 \implies \Pi(\gamma') \neq \Pi(\gamma_0)$, where $\Pi(\gamma)$ is the matrix in \eqref{eq.trans.matrix} constructed under $\gamma$. If we can uniquely recover $\gamma_0$ from $\Pi(\gamma_0)$, then we can recover differences in utilities, $u_i(g,X) - u_i(w,X)$, for all $i \in \mathcal{I}$ and $g,w \in \mathcal{G}$, as  $F_{\epsilon(1) - \epsilon(0)}$ is invertible under \Cref{ass.shocks}. Levels (and thus $\theta_0$) are then identified under a location normalization on pay-offs (e.g., $u_i(g_0, X) = 0$ for all $i$ and some $g_0$).

Observe that $\dim \gamma =  N(N-1)2^{N(N-1)} + N(N-1) 2^{N(N-1)} $. The first summand is the dimension of $\left(p_{ij}(g): g \in \mathcal{G}, (i,j) \in \mathcal{M}\right)$ and the second term is the dimension of $(F_{ij}(g,w): (i,j) \in \mathcal{M}, g \in \mathcal{G}, w \in N(g), w_{ij} \neq g_{ij})$. Matrix $\Pi(\gamma)$ has $2^{N(N-1)}(N(N-1))$ freely-varying entries. The parameter space imposes $2^{N(N-1)}$ restrictions of the type $\sum_{(i,j) \in \mathcal{M}} \rho_{i,j}(g) = 1$ and $N(N-1)2^{N(N-1)}/2$ restrictions of the type $F_{ij}(g,w) + F_{ij}(w,g) = 1$. A simple order condition would thus require

\begin{equation}
	\label{eq.order}
	2^{N(N-1)}\left[2 N(N-1)\right] \leq 2^{N(N-1)}\left[N(N-1) + 1 + N(N-1)/2 \right]\,,
\end{equation} which implies that $N$ should be less than or equal to $2$. The point is that the map $\gamma \mapsto \Pi(\gamma)$ is nonlinear, so the order condition is neither necessary nor sufficient for identification. Nonetheless, we can show that the model is identified when $N = 2$.

\begin{claim}
	\label{claim.N2}
	If $N = 2$, then $\gamma_0$ is identified from $\Pi(\gamma_0)$ under Assumptions \ref{ass.matching}, \ref{ass.shocks}, and \ref{ass.full.support.X}.
	\begin{proof}
		See Supplemental Appendix \ref{Supplement-proof.claim.N2}.
	\end{proof}
\end{claim}

Extending such a direct argument to $N>2$ is not feasible, as $\Pi(\gamma)$ is a $2^{N(N-1)} \times 2^{N(N-1)}$ matrix. Notice that \eqref{eq.order} implies that the rank condition of \cite{Rothenberg1971} for local identification is not satisfied. The problem is that, for this failure of the order condition to be sufficient for nonidentification, the Jacobian of $\gamma \mapsto \Pi(\gamma)$ must be rank-regular (i.e., it must have constant rank in a neighborhood of $\gamma_0$), which is not trivial to show. Of course, if that were the case, we would know the model is nonidentified for $N>2$.

Given the difficulty of establishing identification without imposing further restrictions, even when $\Pi(X;\theta_0)$ is known (or $\tau_0=1$), in the following subsections, we explore the identifying power of restrictions on (i) how covariates affect utilities and the matching function; and (ii) how the network structure affects pay-offs and meetings. To make both the exposition and proofs clearer, in what follows, we maintain the notation introduced in this section, and the dependence of objects on covariates remains implicit whenever it does not cause confusion.

\subsection{Identification with exclusion restriction on meeting process}
\label{identification_covariates}
In this subsection, we explore the identifying power of restrictions on covariates. We work in an environment where we observe two data periods and assume $\tau_0$ is identified or known \textit{a priori}. We follow an identification at infinity approach to identify $\theta_0$ \citep{Tamer2003, Bajari2010, Colas2020}. The main idea is to use information from pairs that would almost certainly accept a relationship if selected by the meeting process to infer about the matching technology. We require a large support instrument that affects preferences and is excluded from the matching function. In Supplemental Appendix \ref{Supplement-ident.exclusion.matching}, we show a similar argument is valid if a large support instrument enters the meeting process but is excluded from preferences. In this case, information from pairs almost certainly selected by the algorithm is used to infer preferences. Importantly, we provide identification results directly from $\Pi(X;\theta_0)^{\tau_0}$. Therefore, differently from the results in the main text, we do not need to assume that Assumptions \ref{ass_distribution}, \ref{ass_pot_pref} and \ref{ass_pot_meet}, as we need not infer $\Pi(X;\theta_0)$ or $\Pi(X;\theta_0)^2$ from  $\Pi(X;\theta_0)^{\tau_0}$ to establish identification.

To make the dependency in covariates explicit, we follow the notation introduced in Section \ref{sec_ident_pref_meeting} of the main text. Specifically,  we use $X^m(g)$ for the covariates that enter the matching function under network $g$, i.e. we write $\rho((i,j),g,X) = \rho_{ij}(g|X^m(g))$ for all $(i,j) \in \mathcal{M}$ and $g \in \mathcal{G}$. We also write $X_{ij}(g)$ for the transformation of $X$ with the covariates relevant in the marginal gain of $i$ moving from $g$ to $[1-g_{ij}, g_{-ij}]$. In addition, we write $\Pi(X)^{\tau_0} = \Pi(X;\theta_0)^{\tau_0}$ for the observed transition matrix. Finally, we use the notation $A \setminus B$ for the subvector of $A$ such that, up to permutations, $A = [A \setminus B, B]$.

The following assumption states our exclusion restriction.

\begin{assumption}[Large support exclusion restriction]
	\label{ass.exclusion.strong}
 Fix $g \in \mathcal{G}$ and $w \in N(g)$ with $g_{ij} \neq w_{ij}$. We assume that:
 \begin{enumerate}
 	
 			 	\item[a] For $\mathbb{P}_X$-almost every support point $(x^m(g),x^m(w), x_{ij}(g))$ ot $(X^m(g),X^m(w),X_{ij}(g))$, there exists a known sequence $\{(z_{kl,\downarrow}^n(g),z_{kl,\downarrow}^n(w))_{(k,l)\neq (i,j)}\}_{n \in \mathbb{N}}$ in the support of $$(X_{kl}(g),X_{kl}(w))_{(k,l)\neq (i,j)}|X^m(g)=x^m(g),X^m(w)=x^m(w), X_{ij}(g)=x_{ij}(g)\, ,$$ such that:  	$$\lim_{n\to \infty} \Delta u_{kl}(g|z_{kl,\downarrow}^n(g)) = \lim_{n\to \infty} \Delta u_{kl}(w|z_{kl,\downarrow}^n(w)) = - \infty\, ,$$ for every $(k,l) \neq (i,j)$.

 \item[b]    For $\mathbb{P}_X$-almost every support point $x^m(g)$ ot $X^m(g)$, there exists a known sequence $\{z_{ij,\uparrow}^n(g), (z_{kl,\downarrow}^n(g),z_{kl,\downarrow}^n(w))_{(k,l)\neq (i,j)}\}_{n \in \mathbb{N}}$ in the support of $$X_{ij}(g),(X_{kl}(g),X_{kl}(w))_{(k,l)\neq (i,j)}|X^m(g)=x^m(g)\,,$$ such that:
 $$\lim_{n\to \infty} \Delta u_{ij}(g|z_{ij,\uparrow}^n(g)) = \infty\,,$$ and 	$$\lim_{n\to \infty} \Delta u_{kl}(g|z_{kl,\downarrow}^n(g)) = \lim_{n\to \infty} \Delta u_{kl}(w|z_{kl,\downarrow}^n(w)) = - \infty\,,$$ for every $(k,l) \neq (i,j)$.

 	\item[c]   For $\mathbb{P}_X$-almost every support point $x^m(w)$ ot $X^m(w)$, there exists a known sequence $\{z_{ij,\downarrow}^n(g), (z_{kl,\downarrow}^n(g),z_{kl,\downarrow}^n(w))_{(k,l)\neq (i,j)}\}_{n \in \mathbb{N}}$ in the support of $$X_{ij}(g),(X_{kl}(g),X_{kl}(w))_{(k,l)\neq (i,j)}|X^m(w)=x^m(w	)\, ,$$ such that:
 	$$\lim_{n\to \infty} \Delta u_{ij}(g|z_{ij,\downarrow}^n(g)) = - \infty\,,$$ and 	$$\lim_{n\to \infty} \Delta u_{kl}(g|z_{kl,\downarrow}^n(g)) = \lim_{n\to \infty} \Delta u_{kl}(w|z_{kl,\downarrow}^n(w)) = - \infty\,,$$ for every $(k,l) \neq (i,j)$.

 \end{enumerate}

\end{assumption}

Part (a) of \Cref{ass.exclusion.strong} requires that, for a given pair of networks $g \in \mathcal{G}$ and  $w \in \mathcal{N}(g)$, $g_{ij} \neq w_{ij}$, it is possible to find large support covariates entering the decision of pairs $(k,l)\neq (i,j)$ of moving from $g$ and $w$, but excluded from the relative pay-off of  $(i,j)$ of moving to and from $g$, and also from the set of covariates entering meeting probabilities under $g$ and $w$. These covariates should admit sufficiently extreme realisations, such that the probability of a pair $(k,l)$ transitioning from $g$ or $w$ upon being selected by the meeting process can be made arbitrarily close to zero. Parts (b) and (c) further require the existence of a large support covariate entering the decision of pair $(i,j)$ transitioning from $g$, but excluded from the meeting process, such that the probability of pair $(i,j)$ moving from network $g$ upon meeting can be made arbitrarily close to unity or zero. 

The previous restrictions imply the following result:

\begin{proposition}
	\label{thm.ident.infinity.general}
	Suppose Assumptions \ref{ass.matching}, \ref{ass.shocks}, and \ref{ass.exclusion.strong} hold. If $\Pi(X)^{\tau_0}$ is identified, then $\rho_{ij}(g|X^m(g))$, $\rho_{ij}(w|X^m(w))$ and $\Delta u_{ij}(g|X_{ij}(g))$ are identified for any $\tau_0$ known or identified.
	\begin{proof}
		See \Cref{proposition3.1}.
	\end{proof}
\end{proposition}

The intuition behind the main identification result of this Appendix is that, when the instruments attain large values in their support, the probability of staying/returning to the same network after $\tau_0$ rounds depends solely on the matching process, which remains unchanged by the exclusion restriction. Similarly, large values of the instruments for pairs $ms \neq ij$ imply that the probability of, starting from $g$, arriving at network $w \in N(g)$, $w_{ij} \neq g_{ij}$ after $\tau_0$ rounds, depends only on the meeting process \textbf{and} on the decisions of pair $ij$. These limiting probabilities then allow us to identify the desired parameters.

\subsection{Identification with exclusion restriction on preferences}
\label{Supplement-ident.exclusion.matching}

In this section, we analyze how the identification result in the previous section would change if large support variables were included in the matching function ($\rho$) but not in utilities. Fix $g \in \mathcal{G}$, $w \in N(g)$, $g_{ij} \neq w_{ij}$. We start by establishing the following claim:

\begin{claim}
	Under a limit which drives $\rho_{ij}(g) \to 1$ and $\rho_{ij}(w) \to 1$; but leaves $F_{ij}(g,w)$ unaltered, we have
	
	$$\lim_{t^*} (\Pi^{\tau_0})_{gw} = F_{ij}(g,w)\, .$$
	\begin{proof}
		The case where $\tau_0 = 1$ is immediate, since $\lim_{t^*} (\Pi^{\tau_0})_{gw} = F_{ij}(g,w)$ follows directly from  equation \eqref{eq.trans.matrix} in the main paper. Suppose $\tau_0 > 1$. Recall that:
		
		$$(\Pi^{\tau_0})_{gw} = \sum_{m \in N(w) \cup \{w\}} (\Pi^{\tau_0-1})_{gm} \Pi_{mw} \, .$$
		
		Using a similar argument as in Proposition 3.1, we can show the limit in the statement of the Claim is such that $\lim_{t^*} (\Pi^{\tau_0-1})_{gm} \to 0$ for all $m \in N(w) \setminus \{g\}$.\footnote{If $m \in N(w) \setminus \{g\}$, then $m \in N^2(g)$. Fix a summand in $(\Pi^{\tau_0-1})_{gm}$. If a transition from $g$ to $m$ occurs at pair $(k,l) \neq (i,j)$, the limit in the statement of the Claim drives the summand to $0$; if all transitions occur at pair $(i,j)$, a transition from $w$ to some $z \in N(w) \setminus \{g\}$ must occur, for, if not, then either $m = g$ or $m = w$, which is not true. The limit in the statement of the Claim thus drives the summand to zero.} We are thus left with
		
		$$\lim_{t^*}(\Pi^{\tau_0})_{gw} = \lim_{t^*} (\Pi^{\tau_0-1})_{gw} \Pi_{ww} + \lim_{t^*} (\Pi^{\tau_0-1})_{gg} \Pi_{gw} =   (\lim_{t^*} (\Pi^{\tau_0-1})_{gw}  +  \lim_{t^*} (\Pi^{\tau_0-1})_{gg}) F_{ij}(g,w)\, . $$
		
		Next, under the limit in the statement of the claim
		
		$$\lim_{t^*}(\Pi^{\tau_0})_{gg} = (\lim_{t^*} (\Pi^{\tau_0-1})_{gw}  +  \lim_{t^*} (\Pi^{\tau_0-1})_{gg}) F_{ij}(w,g)\, , $$ which follows from $(\Pi^{\tau_0})_{gg} = \sum_{m \in N(g) \cup \{g\}} (\Pi^{\tau_0-1})_{gm} \Pi_{mg}$ and an argument similar to the previous one. We then get
		
		$$\lim_{t^*}(\Pi^{\tau_0})_{gw} + \lim_{t^*}(\Pi^{\tau_0})_{gg} = \lim_{t^*}(\Pi^{\tau_0-1})_{gw} + \lim_{t^*}(\Pi^{\tau_0 -1})_{gg}.$$
		
		Induction on $\lim_{t^*}(\Pi)_{gw} + \lim_{t^*}(\Pi)_{gg} = 1$ then yields the desired result.
	\end{proof}
	
\end{claim}

The previous argument suggests that if large support variables are included in $\rho_{ij}(g)$ and $\rho_{ij}(w)$ -- but excluded from $F_{ij}(g,w)$ -- may be shifted in a direction that (simultaneously) drives the probability of selecting pair $(i,j)$ under networks $w$ and $g$ to $1$; then marginal utilities are identified.

Identification of the matching process in this setting is more intricate, as we require the ``feasibility'' of a different limit. We consider the case where $\tau_0 \leq N (N-1)$ (\Cref{ass.upperbound.tau0} holds). Fix some $s \in N^{\tau_0}(g)$ such that $g_{ij} \neq s_{ij}$. Denote by $\mathcal{D} \subseteq \mathcal{M}$ be the set of pairs where $g$ and $s$ differ. We then have

\begin{equation*}
	\begin{aligned}
		(\Pi^{\tau_0})_{gs} = \\  \sum_{(a_1, a_2 \ldots a_{\tau_0}) \in P(D)} \rho_{a_1}(g) F_{a_1}(g, [1-g_{a_1}, g_{-a_1}])  \times \\
		\rho_{a_2}([1-g_{a_1}, g_{-a_1}]) F_{a_2}([1-g_{a_1}, g_{-a_1}], [1-g_{a_1}, 1 - g_{a_2}, g_{-a_1, -a_2}]) \times \ldots \\
		\times \rho_{a_{\tau_0}}([1- g_{a_1}, 1 - g_{a_2} \ldots 1 - g_{a_{\tau_0 - 1}}, g_{-a_1,-a_2 \ldots - a_{\tau_0-1}}]) \times \\
		F_{a_{\tau_0}}([1- g_{a_1}, 1 - g_{a_2} \ldots 1 - g_{a_{\tau_0 - 1}}, g_{-a_1,-a_2 \ldots - a_{\tau_0-1}}], s) \, ,
	\end{aligned}
\end{equation*} where $P(\mathcal{D})$ is the set of all vectors constructed from permutations of the elements in $\mathcal{D}$.  If there exists a limit that vanishes all summands not starting on $\rho_{ij}(g)$ (but leaves the latter unchanged), and if it is further feasible to simultaneously drive $\rho_{ij}(g)$ to 1, then a ratio of limits identifies matching probabilities.

\subsection{Proof of Claim \ref{claim.N2}}
\label{Supplement-proof.claim.N2}

Observe that $\Pi(\gamma)$ is written as

\resizebox{1 \textwidth}{!}{
	$
	\Pi(\gamma) =
	\begin{blockarray}{ccccc}
		g' = (0,0) & g' = (1,0) & g' = (0,1) & g' = (1,1) \\
		\begin{block}{(cccc)c}
			\rho_{12}(0,0) F_{12}((1,0), (0,0)) + \rho_{21}(0,0) F_{21}((0,1), (0,0)) & \rho_{12}(0,0) F_{12}((0,0), (1,0)) & \rho_{21}(0,0) F_{21}((0,0), (0,1))  & 0 & g = (0,0) \\
			\rho_{12}(1,0) F_{12}((1,0),(0,0)) & \rho_{12}(1,0) F_{12}((0,0), (1,0)) + \rho_{21}(1,0) F_{21}((1,1), (1,0)) & 0 & \rho_{21}(1,0) F_{21}((1,0),(1,1)) & g = (1,0) \\
			\rho_{21}(0,1)F_{21}((0,1),(0,0)) & 0 & \rho_{12}(0,1) F_{12}((1,1), (0,1)) + \rho_{21}(1,1) F_{21}((0,0), (0,1)) & \rho_{12}(0,1)F_{12}((0,1),(1,1))  & g = (0,1)\\
			0 & \rho_{21}(1,1)F_{21}((1,1),(1,0)) & \rho_{12}(1,1)F_{12}((1,1),(0,1))  & \rho_{12}(1,1) F_{12}((0,1), (1,1)) + \rho_{21}(1,1) F_{21}((1,0), (1,1)) & g = (1,1). \\
		\end{block}
	\end{blockarray}\ .
	$ 
}

Now suppose there exists $\gamma \neq \tilde{\gamma}$, $\Pi(\gamma) = \Pi(\tilde{\gamma})$. Suppose $\rho_{21}(0,0)> \widetilde{\rho_{21}(0,0)}$. The argument is symmetric for the remaining parameters. Since objects are observationally equivalent, it must be that $F_{21}((0,0),(0,1)) < \widetilde{F_{21}((0,0),(0,1))} $, which in its turn yields $F_{21}((0,1),(0,0)) > \widetilde{F_{21}((0,1),(0,0))} $. This implies $\rho_{21}(0,1) < \widetilde{\rho_{21}(0,1)}$; consequently, $\rho_{12}(0,1) > \widetilde{\rho_{12}(1,0)}$ and $F_{12}((0,1),(1,1)) < \widetilde{F_{12}((0,1), (1,1))}$ thereafter. But then  $F_{12}((1,1),(0,1)) < \widetilde{F_{12}((1,1),(0,1))}$, leading to $\rho_{12}(1,1)>\widetilde{\rho_{12}(1,1)}$, $\rho_{21}(1,1) <\widetilde{\rho_{21}(1,1)}$, $F_{21}((1,1),(1,0))> \widetilde{F_{21}((1,1),(1,0))}$. Proceeding analogously, we get $\rho_{21}(1,0)>\widetilde{\rho_{21}(0,1)}$, $F_{12}((0,0),(1,0)) < \widetilde{F_{12}((0,0),(1,0))}$ and finally $\rho_{12}((0,0)) > \widetilde{\rho_{12}(0,0)}$. But we also had $\rho_{21}(0,0)> \widetilde{\rho_{21}(0,0)}$, leading to $1 > 1$, a contradiction.

\subsection{Proof of \Cref{thm.ident.infinity.general}}
\label{proposition3.1}

\begin{proof}[Proof of \Cref{thm.ident.infinity.general}]
	Fix $g \in \mathcal{G}$, $w \in N(g)$, $g_{ij} \neq w_{ij}$. Observe that
	
	$$(\Pi^{\tau_0})_{gg} = \sum_{m \in N(g) \cup \{g\}} (\Pi^{\tau_0-1})_{gm} \Pi_{mg}\, .$$
	
	We first prove the following claim:
	
	\begin{claim.no}
		Under the limit in item (b) of Assumption \ref{ass.exclusion.strong}, which drives $F_{pq}(w,m) \to 0$ for all $m \in N(w)$, $m_{pq} \neq w_{pq}$, and  $F_{kl}(g,s) \to 0$ for all $s \in N(g) \setminus \{w\}$, $s_{kl} \neq g_{kl}$, we have $\lim_{t^*} (\Pi^{\tau_0})_{gg} = (1 - \rho_{ij}(g))^{\tau_0} $, where $\lim_{t^*}$ is shorthand for the appropriate limit.
		
		\begin{proof}
			The case $\tau_0 = 1$ is readily verified by driving $F_{kl}(g,s) \to 0$ for all $s \in N(g) \setminus \{w\}$, $s_{kl} \neq g_{kl}$ and $F_{ij}(g,w) \to 1$. For $\tau_0 > 1$, we begin by noticing that we may drive $(\Pi^{\tau_0-1})_{gm} \to 0$ for all $m \in N(g)\setminus \{w\}$. Since $m$ and $g$ differ in exactly one edge (say, $m_{pq} \neq g_{pq}$), a transition in edge $pq$ must appear in every summand in $(\Pi^{\tau_0 - 1})_{gm}$. Indeed, $(\Pi^{\tau_0 - 1})_{gm}$ sums over all possible transitions in edge $pq$ from value  $g_{pq}$ to $m_{pq}$ in $\tau_0 - 1$ rounds. Put another way, for each summand in $(\Pi^{\tau_0-1})_{gm}$, there exists $t \in \{0, 1 \ldots \tau_0 - 2\}$, $g^t_{pq} = g_{pq}$ and $g^{t+1}_{pq} = m_{pq}$.\footnote{Recall $g^t$ is the stochastic process on $\mathcal{G}$ induced by the game.} Fix a summand in $(\Pi^{\tau_0 - 1})_{gm}$. We analyze the following cases:
			
			\begin{enumerate}
				\item There exists $t \in \{0,1\ldots \tau_0-2\}$, $g^{t}_{pq} = g_{pq}$, $g^{t+1}_{pq} = m_{pq}$ \textbf{and} $g^t = g$. In this case, by taking $F_{pq}(g,m) \to 0$, we drive the summand to 0.
				\item For all $t \in \{0,1\ldots \tau_0 - 2\}$ such that $g^{t}_{pq} = g_{pq}$, $g^{t+1}_{pq} = m_{pq}$, we have $g^t \neq g$. Take $t^*$ to be the smallest $t$ satisfying the above. Observe that $t^*  > 0$, as $g^0 = g$ (we always start at $g$). Since $g^{t^*} \neq g$, there exists $t' < t^*$, $g^{t'} = g$ and $g^{t'+1} = z$, $z \in N(g)$. If there exists some $t'$ satisfying this property such that $z \neq w$ (with $g_{kl} \neq z_{kl}$), then driving $F_{kl}(g,z) \to 0$ vanishes the term. If, for all such $t'$, $z = w$, take $t^{**}$ to be the maximum of $t'$. Observe that $t^{**} < t^{*}$. If
				$g^{t^*} = w$, we may safely drive $F_{pq}(w,[m_{pq}, w_{-pq}]) \to 0$. If not, then $t^{**} + 1 < t^{*}$ and there exists a transition from $w$ to some element in $N(w)$ which we can safely drive to 0.
			\end{enumerate}
			
			Since the above argument holds, irrespective of the summand (the common limit will vanish all terms), we conclude $(\Pi^{\tau_0 - 1})_{gm} \to 0$. Since $F_{ij}(g,w) \to 1$, $\lim_{t^*}(\Pi^{\tau_0-1})_{gw} \Pi_{wg} = 0$.\footnote{In all previous arguments, we implicitly use the sandwich lemma to infer that, if one term of the product goes to zero, the whole product does. This is immediate, since we are working with products of probabilities.} The common limit in the statement of the claim thus leaves us with
			
			$$\lim_{t^*}(\Pi^{\tau_0})_{gg} = \lim_{t^*}(\Pi^{\tau_0 -1 })_{gg} \lim_{t^*}(\Pi)_{gg} = \lim_{t^*}(\Pi^{\tau_0 -1 })_{gg} (1 -  \rho_{ij}(g))  \, , $$
			
			Induction then yields the desired result.
		\end{proof}
	\end{claim.no}
	Since $\lim_{t^*}(\Pi^{\tau_0})_{gg} \in [0,1]$, we can uniquely solve for $\rho_{ij}(g)$, thus establishing identification. A symmetric argument that relies on the limit in item (c) of Assumption \ref{ass.exclusion.strong} then establishes identification of $\pi_{ij}(w)$.
	
	Next, we proceed to identification of $F_{ij}(g,w)$. Note that
	
	$$(\Pi^{\tau_0})_{gw} = \sum_{m \in N(w) \cup \{w\}} (\Pi^{\tau_0-1})_{gm} \Pi_{mw}\, .$$
	
	We then prove the following claim:
	
	\begin{claim.no}
		Under the limit in item (c) of Assumption \ref{ass.exclusion.strong}, which drives $F_{pq}(w,m) \to 0$ for all $m \in N(w) \setminus \{g\}$, $m_{pq} \neq w_{pq}$, and  $F_{kl}(g,s) \to 0$ for all $s \in N(g) \setminus \{w\}$, $s_{kl} \neq g_{kl}$:
		
		\begin{equation*}
			\begin{aligned}
				\lim_{t^{**}} (\Pi^{\tau_0})_{gw} = \lim_{t^{**}} (\Pi^{\tau_0-1})_{gg} \rho_{ij}(g) F_{ij}(g,w) + \lim_{t^{**}} (\Pi^{\tau_0-1})_{gw} (1 - \rho_{ij}(w) F_{ij}(w,g)) \, ,
			\end{aligned}
		\end{equation*}
		where $\lim_{t^{**}}$ is shorthand for the appropriate limit. We also have that, under such a limit,
		
		$$\lim_{t^{**}} (\Pi^{\tau_0})_{gw} + \lim_{t^{**}} (\Pi^{\tau_0})_{gg} = 1 \, .$$
		
		\begin{proof}
			For $\tau_0 = 1$, we have
			
			\begin{equation*}
				\begin{aligned}
					\lim_{t^{**}} (\Pi)_{gw} = \rho_{ij}(g)F_{ij}(g,w) \, , \\
					\lim_{t^{**}} (\Pi)_{gg} = (1 - \rho_{ij}(g)F_{ij}(g,w)) \, .
				\end{aligned} 
			\end{equation*}
			
			These expressions follow directly from the limit being taken and equation \eqref{eq.trans.matrix}.
			
			Consider next the case $\tau_0 >  1$. The limit in the statement of the lemma drives $(\Pi^{\tau_0-1})_{gm} \to 0$ for all $m \in N(w) \setminus \{g\}$. Indeed, if $m \in N(w) \setminus \{g\}$, then $m \in N^2(g)$. Recall that $(\Pi^{\tau_0-1})_{gm}$ sums over all possible transitions from $g$ to $m$ in $\tau_0-1$ rounds.  Fix a summand in $(\Pi^{\tau_0-1})_{gm}$.  If a transition from $g$ occurs at pair $(a,b) \in \mathcal{M}$, $(a,b) \neq (i,j)$, then the limit vanishes the term. If all transitions from $g$ occur at pair $(i,j)$ (i.e. $g$ only transitions to $w$), a transition from $w$ must occur, since $m \in N^2(g)$. If $w$ only transitions to $g$, then either $m = g$ or $m = w$, which is not true. Therefore, there exists a transition from $w$ to some $z \in N(w) \setminus \{g\}$, so we can vanish the summand. The limit in the statement thus drives the term $(\Pi^{\tau_0-1})_{gm}$ to zero.
			
			From the above discussion, we thus get
			
			\begin{equation*}
				\begin{aligned}
					\lim_{t^{**}} (\Pi^{\tau_0})_{gw} = \lim_{t^{**}} (\Pi^{\tau_0-1})_{gg} \Pi_{gw} + \lim_{t^{**}} (\Pi^{\tau_0-1})_{gw}\Pi_{ww} = \\ \lim_{t^{**}} (\Pi^{\tau_0-1})_{gg} \rho_{ij}(g) F_{ij}(g,w) + \lim_{t^{**}} (\Pi^{\tau_0-1})_{gw} (1 - \rho_{ij}(w) F_{ij}(w,g)) \, ,
				\end{aligned}
			\end{equation*}
			which establishes the first part of the claim.
			
			Next, under the limit in the statement of the claim
			
			\begin{equation*}
				\begin{aligned}
					\lim_{t^{**}}(\Pi^{\tau_0})_{gg} = \lim_{t^{**}} (\Pi^{\tau_0-1})_{gg} (1-\rho_{ij}(g)F_{ij}(g,w)) + \lim_{t^{**}} (\Pi^{\tau_0-1})_{gw} \rho_{ij}(w) F_{ij}(w,g) \, .
				\end{aligned}
			\end{equation*}
			
			This follows from observation that, in the proof of the \emph{previous} claim, we can still drive $(\Pi^{\tau_0-1})_{gm} \to 0$ for all $m \in N(g) \setminus \{w\}$ even though $F_{ij}(g,w)$ does not vanish.\footnote{Suppose $g$ only transitions to $w$. Since $m \neq w$, $w$ must transition to some other $z \in N(w)$. If $w$ only transitions to $g$, then either $m = g$ or $m = w$, which is not true. Therefore, we can always vanish a summand in $(\Pi^{\tau_0 -1})_{gm}$, even though $F_{ij}(g,w)$ does not vanish.} We are thus left with the terms related to staying in $g$ or transitioning to $w$ in $\tau_0-1$ rounds.
			
			Finally, the second part of the claim can be asserted by noticing that $\lim_{t^{**}} (\Pi)_{gw} + \lim_{t^{**}} (\Pi)_{gg}= 1$ and applying this fact inductively on the expression for $\lim_{t^{**}} (\Pi^{\tau_0})_{gw} + \lim_{t^{**}} (\Pi^{\tau_0})_{gg}= \lim_{t^{**}} (\Pi^{\tau_0-1})_{gw} + \lim_{t^{**}} (\Pi^{\tau_0-1})_{gg}$.
		\end{proof}
	\end{claim.no}
	To establish identification of $F_{ij}(g,w)$, we need to show that the expression for $\lim_{t^{**}} (\Pi^{\tau_0})_{gw}$ is strictly increasing in $(0,1)$ as a function of $F_{ij}(g,w)$. Denoting by $D_{F_{ij}(g,w)} \lim_{t^{**}} (\Pi^{\tau_0})_{gw}$ the derivative of $\lim_{t^{**}} (\Pi^{\tau_0})_{gw}$ as a function of $F_{ij}(g,w)$, we get
	
	\begin{equation*}
		\begin{aligned}D_{F_{ij}(g,w)} \lim_{t^{**}} (\Pi^{\tau_0})_{gw} = D_{F_{ij}(g,w)}  \lim_{t^{**}} (\Pi^{\tau_0-1})_{gw} (1 - \rho_{ij}(w)  + (\rho_{ij}(w) - \rho_{ij}(g)) F_{ij}(g,w) ) \\ + \rho_{ij}(g) ( 1 - \lim_{t^{**}} (\Pi^{\tau_0-1})_{gw} ) + \rho_{ij}(w) \lim_{t^{**}} (\Pi^{\tau_0-1})_{gw} \, ,
		\end{aligned}
	\end{equation*}
	where we use that  $\lim_{t^{**}} (\Pi^{\tau_0})_{gw} + \lim_{t^{**}} (\Pi^{\tau_0})_{gg} = 1$ and $F_{ij}(g,w) + F_{ij}(w,g)=1 $. By noticing $ \rho_{ij}(g) ( 1 - \lim_{t^{**}} (\Pi^{\tau_0-1})_{gw} ) + \rho_{ij}(w) \lim_{t^{**}} (\Pi^{\tau_0-1})_{gw} \geq \min \{\rho_{ij}(g) , \rho_{ij}(w)\} > 0$ and applying induction on the fact that $D_{F_{ij}(g,w)}  \lim_{t^{**}} (\Pi)_{gw} \geq 0$, we get that the derivative is strictly positive in $(0,1)$, thus showing that the map is invertible and establishing identification of $F_{ij}(g,w)$.
\end{proof}

\section{Approximation property of the Approximate Bayesian Computation (ABC) algorithm}
\label{Supplement-abc.proof}
In this Appendix, we show how our likelihood-free algorithm approximates moments of the posterior distribution. As in our main text, we observe a sample $\{G^{T_0}_c, G^{T_1}_c, X_c\}_{c=1}^C$ from the model. As our focus lies on the posterior distribution, we essentially view this sample as fixed (nonstochastic) throughout the remainder of this section. The model parameters are $(\beta, \tau) \in \mathbb{B}\times \mathbb{N}$. The prior density is $p_0$, and the model likelihood is $\prob(\cdot|\mathbf{X}_C; \beta, \tau)$, where $\mathbf{X}_C \coloneqq \{X_c, G^{T_0}_c\}_{c=1}^C$. Approximate Bayesian Computation requires that we draw samples from $\prob(\cdot | \mathbf{X}_C; \beta, \tau)$ and compare them with the data $\mathbf{Y}_C \coloneqq \{G^{T_1}_c\}_{c=1}^C$. In particular, we consider computing (approximations of) moments of the posterior distribution $\prob(\cdot | \mathbf{X}_C, \mathbf{Y}_C)$ according to \Cref{alg:abc_approx}, where $K(\tilde{\mathbf{Y}}_{Cs}, \mathbf{Y}_C; \epsilon)$ is a rescaled kernel and $q_0$ is a proposal density.

\begin{algorithm}[H]
	\caption{Approximating posterior moments}
	\label{alg:abc_approx}
	\begin{algorithmic}
		\State define some tolerance $\epsilon > 0$
		\State define a function $h : \mathbb{B} \times \mathbb{N} \mapsto \mathbb{R}^m$
		\For{$s \in \{1,2\ldots S\}$}
		\State draw $(\beta_s, \tau_s) \sim q_0$
		\State generate an artificial sample $\tilde{\mathbf{Y}}_{Cs} \sim \prob(\cdot|\mathbf{X}_C; \beta_s, \tau_s) $
		\State accept $(\beta_s, \tau_s)$ with probability $K(\tilde{\mathbf{Y}}_{Cs}, \mathbf{Y}_C; \epsilon)$
		\EndFor
		\State compute the approximation to the posterior mean of $h$ using the accepted draws according to $\hat{h} \coloneqq \sum_{s : \text{accepted}} h(\beta_s, \tau_s) w_s /\sum_{s : \text{accepted}} w_s $, where $w_s \coloneqq p_0(\beta_s, \tau_s) / q_0(\beta_s, \tau_s) $
	\end{algorithmic}
\end{algorithm}

The next proposition shows that, if we let $\epsilon \to 0$ as $S \to \infty$, our approximation will be consistent for the posterior mean $\expec_{\prob(\cdot | \mathbf{X}_C, \mathbf{Y}_C)}[h(\cdot)]$.

\begin{proposition}
	Suppose that (1) the prior distribution admits a density $p_0$ with respect to some measure $\mu$ on $\mathbb{B}\times \mathbb{N}$; (2) the proposal density $q_0$ is such that $\text{supp } p_0 \subseteq \text{supp } q_0$; and (3) the map $K: \mathcal{G}^C \times \mathcal{G}^C  \times \mathbb{R}_{+} \mapsto [0,1]$ is such that (3.i) $K(\mathbf{Y}_1,\mathbf{Y}_2, \cdot)$ is continuous at $0$ for all $(\mathbf{Y}_1, \mathbf{Y}_2) \in \mathcal{G}^C \times \mathcal{G}^C$; and (3.ii) $K(\mathbf{Y}_1,\mathbf{Y}_2, 0) = \mathbbm{1}\{\mathbf{Y}_1 = \mathbf{Y}_2\}$. Then, for any $h : \mathbb{B} \times \mathbb{N} \mapsto \mathbb{R}^m$ such that $\int \lVert h(\beta, \tau) \rVert p_0(\beta, \tau) d \mu < \infty$, the approximation $\hat{h}$  in \Cref{alg:abc_approx} is such that $\hat{h} \overset{p}{\to} \expec_{\prob(\cdot | \mathbf{X}_C, \mathbf{Y}_C)}[h(\cdot)]$ as $\epsilon \to 0$ and $S \to \infty$.
	
	\begin{proof}
		Observe that $\hat{h}$ may be written as
		
		\begin{equation}
			\hat{h} = \frac{S^{-1} \sum_{s=1}^S h(\beta_s, \tau_s) w_s \mathbbm{1}\{u_s \leq K(\tilde{\mathbf{Y}}_{Js}, \mathbf{Y}_C; \epsilon)\}}{S^{-1} \sum_{s=1}^S w_s \mathbbm{1}\{u_s \leq K(\tilde{\mathbf{Y}}_{Js}, \mathbf{Y}_C; \epsilon)\} } \, ,
		\end{equation} where $\{u_s\}_{s=1}^S$ are iid draws from a niform distribution; independent from $\{\beta_s, \tau_s, \tilde{\mathbf{Y}}_{Cs}\}_{s=1}^S$.
		
		We next note the random map $a(\epsilon) =  h(\beta_s, \tau_s) w_s \mathbbm{1}\{u_s \leq K(\tilde{\mathbf{Y}}_{Cs}, \mathbf{Y}_C; \epsilon)\}$ is almost surely continuous at 0, for $\prob[\{a \text{ is discontinuous at 0}\}] \leq \prob[u_s = 1] = 0$. Moreover, we have that $\expec[\sup_{\epsilon \geq 0} \lVert a(\epsilon) \rVert] \leq
		\expec[\lVert h(\beta_s, \tau_s) \rVert w_s] = \int \lVert h(\beta, \tau) \rVert p_0(\beta, \tau) d \mu < \infty $. Then, by applying Lemma 4.3 of \cite{Newey1994}, we have that, as $S \to \infty$ and $\epsilon \to 0$
		
		\begin{equation}
			S^{-1} \sum_{s=1}^S h(\beta_s, \tau_s) w_s \mathbbm{1}\{u_s \leq K(\tilde{\mathbf{Y}}_{Js}, \mathbf{Y}_C; \epsilon)\} \overset{p}{\to} \expec[h(\beta_s, \tau_s) w_s \mathbbm{1}\{u_s \leq K(\tilde{\mathbf{Y}}_{Js}, \mathbf{Y}_C; 0)\}].
		\end{equation}
		But we further have that:
		\begin{equation*}
			\begin{aligned}
				\expec[h(\beta_s, \tau_s) w_s \mathbbm{1}\{u_s \leq K(\tilde{\mathbf{Y}}_{Js}, \mathbf{Y}_C; 0)\}] = \expec[h(\beta_s, \tau_s) w_s \mathbbm{1}\{\tilde{\mathbf{Y}}_{Js} = \mathbf{Y}_C\}] = \\
				= \expec[h(\beta_s, \tau_s) w_s \expec[\mathbbm{1}\{\tilde{\mathbf{Y}}_{Js} = \mathbf{Y}_C\}|\beta_s, \tau_s]] = \expec[h(\beta_s, \tau_s) w_s  \prob[\mathbf{Y}_C|\mathbf{X}_C; \beta_s, \tau_s]] = \\= \int h(\beta_s, \tau_s) \prob[\mathbf{Y}_C|\mathbf{X}_C; \beta_s, \tau_s] p_0 (\beta_s, \tau_s) d \mu .
			\end{aligned}
		\end{equation*}
		An analogous argument establishes that the denominator converges in probability to $\quad \quad \quad \quad \quad \quad$ $\int \prob[\mathbf{Y}_C|\mathbf{X}_C; \beta_s, \tau_s] p_0 (\beta_s, \tau_s) d \mu$, which establishes the desired result.
	\end{proof}
\end{proposition}
Examples of maps that satisfy our required property are

$$K(\mathbf{Y}_1, \mathbf{Y}_2; \epsilon) = \mathbbm{1}\{\lVert \mathbf{Y}_1 - \mathbf{Y}_2\rVert \leq \epsilon \}\, ,$$
which corresponds to the ``sharp'' rejection rule in Algorithm 1 of the main text. A ``smooth'' alternative is:

$$K(\mathbf{Y}_1, \mathbf{Y}_2; \epsilon) = \begin{cases} \phi(\lVert \mathbf{Y}_1 - \mathbf{Y}_2\rVert / \epsilon)/\phi(0) & \epsilon > 0 \\
	\mathbbm{1}\{ \mathbf{Y}_1 = \mathbf{Y}_2 \} & \epsilon = 0
\end{cases}\,,$$
where $\phi$ is the standard normal pdf.

\section{Alternative Algorithm: expectation propagation with ``local'' summary statistics}
\label{Supplement-EPlocal}

This method replaces the vector of classroom edge indicators in the approach described in the main paper with a data-driven ``local'' summary statistic. The idea is to achieve further dimensionality reduction. We explore the result in \cite{Fearnhead2012}, according to whom, in the environment of \Cref{alg:abc_approx}, the optimal -- in terms of minimizing posterior risk under quadratic loss -- choice of summary statistic is the posterior mean:\footnote{As in the main text, we estimate the number of rounds in a first step.}

\begin{equation}
	T^*(\mathbf{g}) = \mathbb{E}[\beta | {\mathbf{G}_1} = \mathbf{g}, \mathbf{G}_0, \mathbf{X}] \, .
\end{equation}

The authors suggest estimating this function in a pre-step, where we first construct an artificial dataset by drawing multiple times from the prior; simulate an artificial dataset from each of these draws; and then use a flexible estimation method to approximate $T^*$.

We adapt the approach of \cite{Fearnhead2012} to our EP-ABC setting. In particular, we run a variant of EP-ABC, where given the local nature of the algorithm, we now aim to approximate the conditional expectation of the parameter given the data in a \textbf{specific classroom}. We do so by running the post-LASSO algorithm of \cite{Belloni2013} in an artificial dataset restricted to each classroom (see \Cref{alg:local_summary} for details). We thus end up with a sequence of summary functions $T^c$, $c=1,2\ldots C$, which are used to assess the quality of draws in each step of the EP algorithm.

\begin{algorithm}
	\caption{Constructing local summary statistics}
	\label{alg:local_summary}
	\begin{algorithmic}[1]
		\For{$r \in \{1,2\ldots R\}$}
		\State draw $\beta_r \sim F$
		\State generate an artificial sample $\widetilde{\mathbf{G}}_1^r\sim \mathbb{P}[\cdot|\beta_r, \mathbf{G}_0, \mathbf{X}]$
		\State store $\beta_r$ and $\widetilde{\mathbf{G}}_1^r$
		\EndFor
		\For{$c \in \{1,2,\ldots C\}$}
		\For{$l \in \{1,2\ldots k\}$}
		\State estimate $T^{c}_l(\mathbf{g})= \alpha_{l}^c  + \gamma_l^{c'}\operatorname{vec}(\mathbf{g})$, where $(\alpha_l^c,\gamma_l^c)$ is obtained from a Post-Lasso regression of $\beta_{r}(l)$, the $l$-th entry of $\beta_r$, on an intercept and $\operatorname{vec} \left(\left(\widetilde{\mathbf{G}_c^{T_1}}\right)_r\right)$,
		\EndFor
		\EndFor
	\end{algorithmic}
	
\end{algorithm}

\section{Application: Results with the EP local summary algorithm}
\label{Supplement-results_EPlocal}
This section presents the results of our empirical application using the algorithm outlined in the previous section. First, we set the number of draws in the construction of summary statistics (\Cref{alg:local_summary}) to $R=100,000$.
The results of the counterfactuals using the alternative method are qualitatively similar to those presented in \Cref{application} of the main text, though we
find some differences in the structural parameter estimates, especially in the matching process.

% latex table generated in R 4.0.3 by xtable 1.8-4 package
% Fri Apr  9 16:55:51 2021
\begin{table}[H]
\centering
\caption{Posterior estimates - Expectation propagation (summary statistic)}
\begin{tabular}{lllll}
  \hline
  & Mean & Q 0.025 & Q 0.975 & Prob $<$ 0 \\
  \hline \multicolumn{5}{c}{ Meeting process } \\ \hline  \hline
distance in age &  0.1031 & -0.0640 &  0.2702 &  0.1132 \\
  distance in gender & -0.1613 & -0.4138 &  0.0912 &  0.8947 \\
  distance in cognitive skills & -0.5771 & -1.7264 &  0.5723 &  0.8375 \\
  distance in conscientiousness & -0.0057 & -0.1913 &  0.1798 &  0.5241 \\
  distance in neuroticism & -0.0862 & -0.2410 &  0.0686 &  0.8624 \\
  g\_ij & -0.0346 & -0.3319 &  0.2628 &  0.5902 \\
  (1-g\_ij)*distance in class list & -0.0722 & -0.1308 & -0.0136 &  0.9921 \\
   \hline \multicolumn{5}{c}{ Utility -- Direct Links } \\ \hline intercept & -1.0931 & -1.7330 & -0.4532 &  0.9996 \\
  distance in age & -0.1402 & -0.5192 &  0.2387 &  0.7658 \\
  distance in gender & -3.8073 & -4.6269 & -2.9876 &  1.0000 \\
  distance in cognitive skills & -0.9604 & -2.7615 &  0.8406 &  0.8520 \\
  distance in conscientiousness & -1.6017 & -2.2789 & -0.9245 &  1.0000 \\
  distance in neuroticism &  0.3845 & -0.0817 &  0.8507 &  0.0530 \\
   \hline \multicolumn{5}{c}{ Utility -- Reciprocity Links } \\ \hline intercept &  2.7245 &  1.6589 &  3.7900 &  0.0000 \\
  distance in age & -0.5096 & -1.4204 &  0.4011 &  0.8636 \\
  distance in gender &  1.0101 & -0.8506 &  2.8708 &  0.1437 \\
  distance in cognitive skills & -0.1921 & -2.1391 &  1.7549 &  0.5767 \\
  distance in conscientiousness &  2.5556 &  0.8627 &  4.2485 &  0.0015 \\
  distance in neuroticism & -0.5859 & -1.7417 &  0.5698 &  0.8398 \\
   \hline \multicolumn{5}{c}{ Utility -- Indirect/Popularity } \\ \hline intercept &  0.3193 &  0.0444 &  0.5942 &  0.0114 \\
  distance in age &  0.0874 & -0.0877 &  0.2626 &  0.1640 \\
  distance in gender & -0.7141 & -1.1727 & -0.2555 &  0.9989 \\
  distance in cognitive skills & -0.3506 & -1.9198 &  1.2185 &  0.6693 \\
  distance in conscientiousness & -0.2376 & -0.6888 &  0.2137 &  0.8489 \\
  distance in neuroticism & -0.4962 & -0.7736 & -0.2187 &  0.9998 \\
   \hline
\end{tabular}
\end{table}

% latex table generated in R 4.0.3 by xtable 1.8-4 package
% Fri Apr  9 16:56:20 2021
\begin{table}[H]
\centering
\caption{Projection coefficients and edge statistics - Expectation propagation (summary statistic)}

\begin{adjustbox}{width=\textwidth}
\begin{tabular}{llllll}
  \hline
  & Data & Base case & Random matching & Tracking & Random friendship \\
  \hline \multicolumn{4}{c}{ Regression coefficients } \\ \hline  \hline
distance in class list & -0.0016 & -0.0018 & -0.0003 & -0.0037 & -0.0084 \\
    & [-0.0028;-0.0004] & [-0.0033;-0.0003] & [-0.0010; 0.0004] & [-0.0059;-0.0017] & [-0.0125;-0.0034] \\
  distance in age &  0.0018 & -0.0099 & -0.0070 & -0.0099 &  0.0041 \\
    & [-0.0048; 0.0083] & [-0.0199; 0.0018] & [-0.0164; 0.0031] & [-0.0181;-0.0003] & [-0.0062; 0.0147] \\
  distance in gender & -0.1864 & -0.1721 & -0.1980 & -0.1193 & -0.1066 \\
    & [-0.2072;-0.1657] & [-0.1942;-0.1540] & [-0.2224;-0.1760] & [-0.1354;-0.1049] & [-0.1227;-0.0899] \\
  distance in cognitive skills & -0.1275 & -0.1010 & -0.1288 &  0.0613 & -0.1475 \\
    & [-0.2172;-0.0378] & [-0.1805;-0.0171] & [-0.2077;-0.0428] & [-0.0377; 0.1693] & [-0.2360;-0.0618] \\
  distance in conscientiousness & -0.0193 & -0.0339 & -0.0370 & -0.0338 & -0.0083 \\
    & [-0.0332;-0.0053] & [-0.0452;-0.0229] & [-0.0498;-0.0256] & [-0.0460;-0.0219] & [-0.0252; 0.0065] \\
  distance in neuroticism & -0.0231 & -0.0131 & -0.0165 & -0.0076 & -0.0126 \\
    & [-0.0356;-0.0106] & [-0.0262;-0.0001] & [-0.0293;-0.0034] & [-0.0202; 0.0075] & [-0.0254; 0.0003] \\
   \hline \multicolumn{4}{c}{ Edge summary statistics (followup)} \\ \hline edge indicator mean & 0.1788 &  0.1312 &  0.1501 &  0.0846 &  0.2564 \\
    & [0.168;0.1896] & [0.1204;0.1445] & [0.1390;0.1633] & [0.0763;0.0935] & [0.2324;0.2841] \\
  edge indicator stdev & 0.3832 &  0.3375 &  0.3571 &  0.2781 &  0.4364 \\
    & [0.374;0.3921] & [0.3255;0.3516] & [0.3459;0.3696] & [0.2655;0.2911] & [0.4224;0.4510] \\
   \hline
\end{tabular}
\end{adjustbox}
\begin{minipage}{0.8\textwidth}
	\vspace{1em}
	\footnotesize
	\textit{Notes:} In the first column, the output from a frequentist regression of the edge indicator $g_{ij}$ on pair characteristics along with 95\% confidence intervals is reported. Summary statistics on the edge indicator are also reported. Confidence intervals assume a Gaussian approximation and are constructed using standard  errors clustered at the classroom level. Confidence intervals on the standard deviation of the edge indicator additionally use the Delta Method. Columns 2-4 report mean estimates and 95\% credible intervals on the projection coefficients and summary statistics. We obtain these from 1,000 simulations of network data from draws of the posterior distribution under base and counterfactual modifications in parameter values.
\end{minipage}
\end{table}

\begin{figure}[H]
	\centering
	\caption{Random matching vs. base case}
	\begin{subfigure}[b]{0.35\textwidth}
		\centering
		\includegraphics[width=\textwidth]{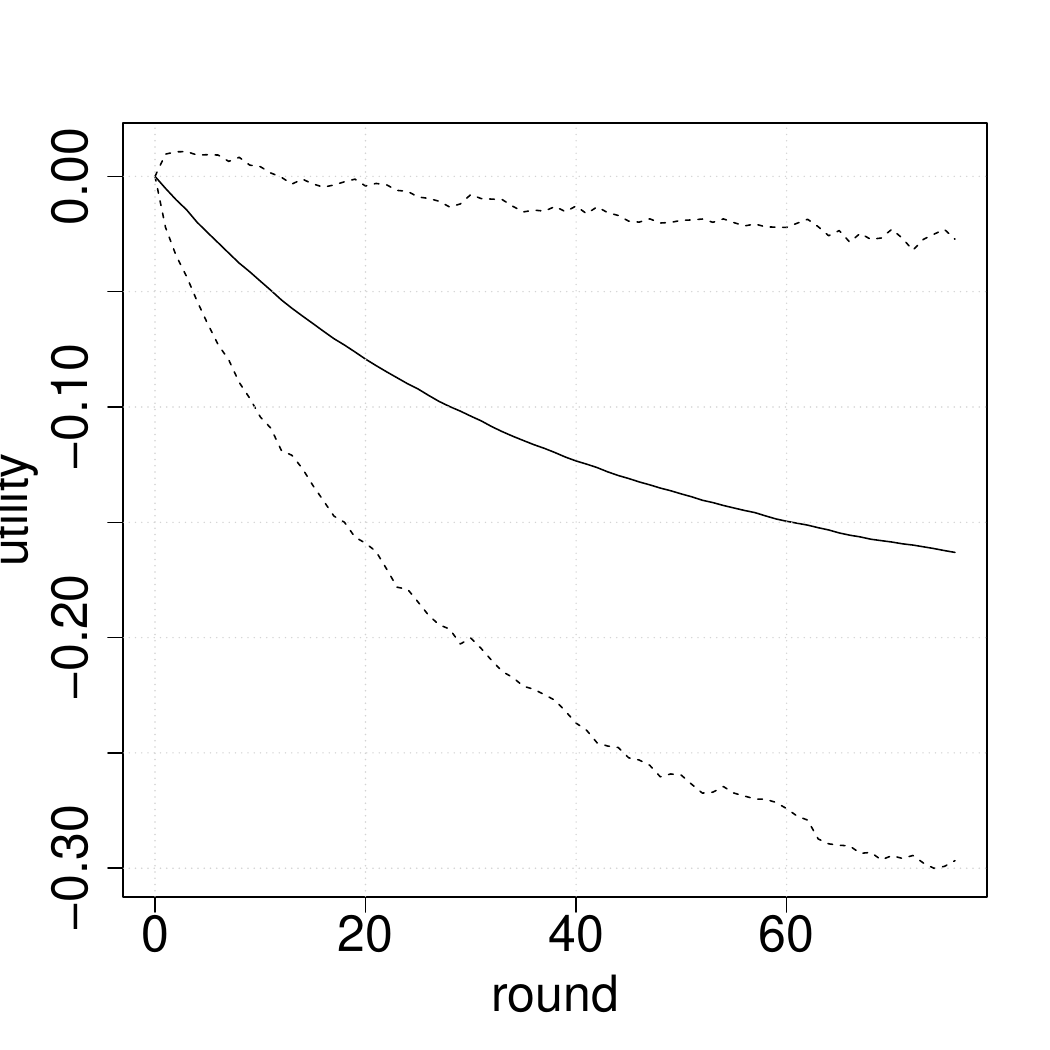}
		\caption{Total}
	\end{subfigure}
	\begin{subfigure}[b]{0.35\textwidth}
		\centering
		\includegraphics[width=\textwidth]{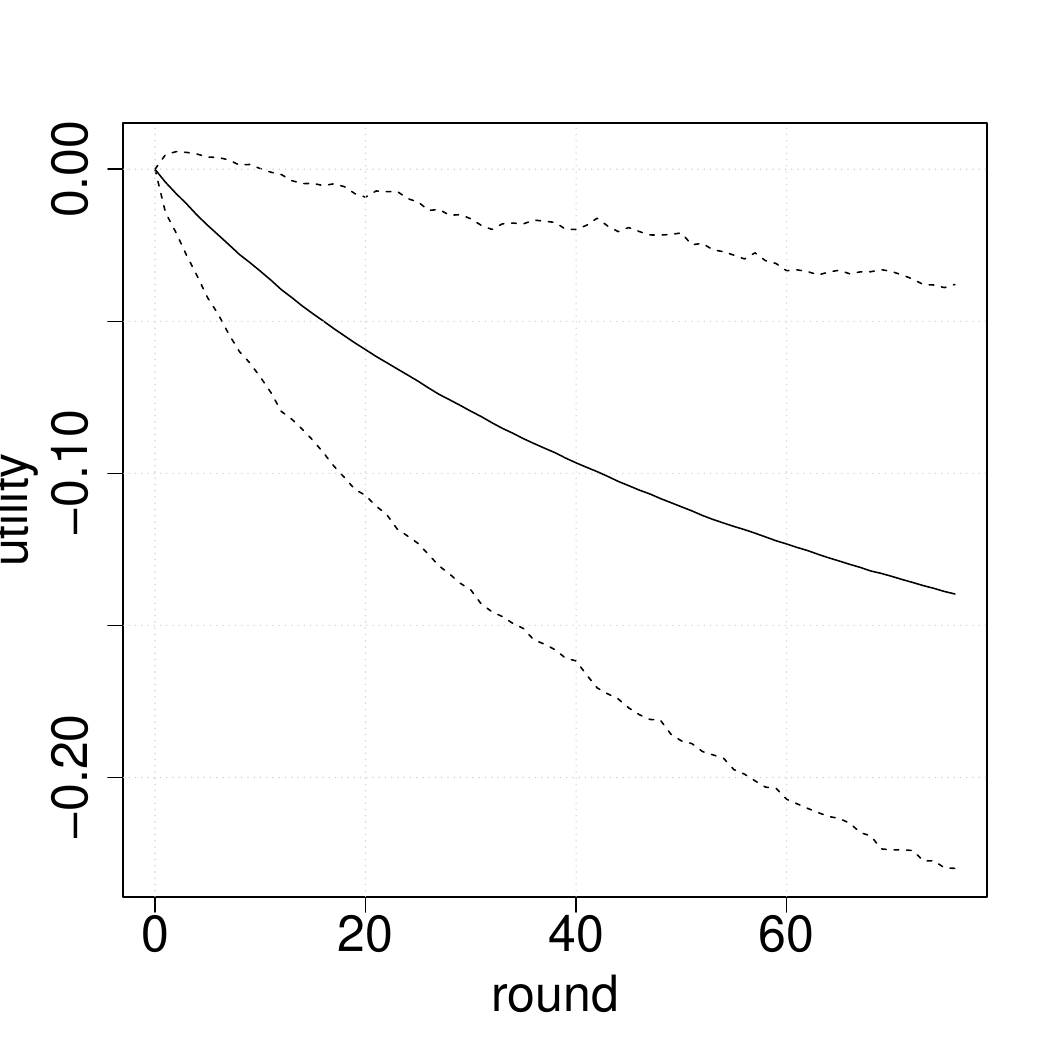}
		\caption{Direct}
	\end{subfigure}\\
	\vspace{-0.6em}
	\begin{subfigure}[b]{0.35\textwidth}
		\centering
		\includegraphics[width=\textwidth]{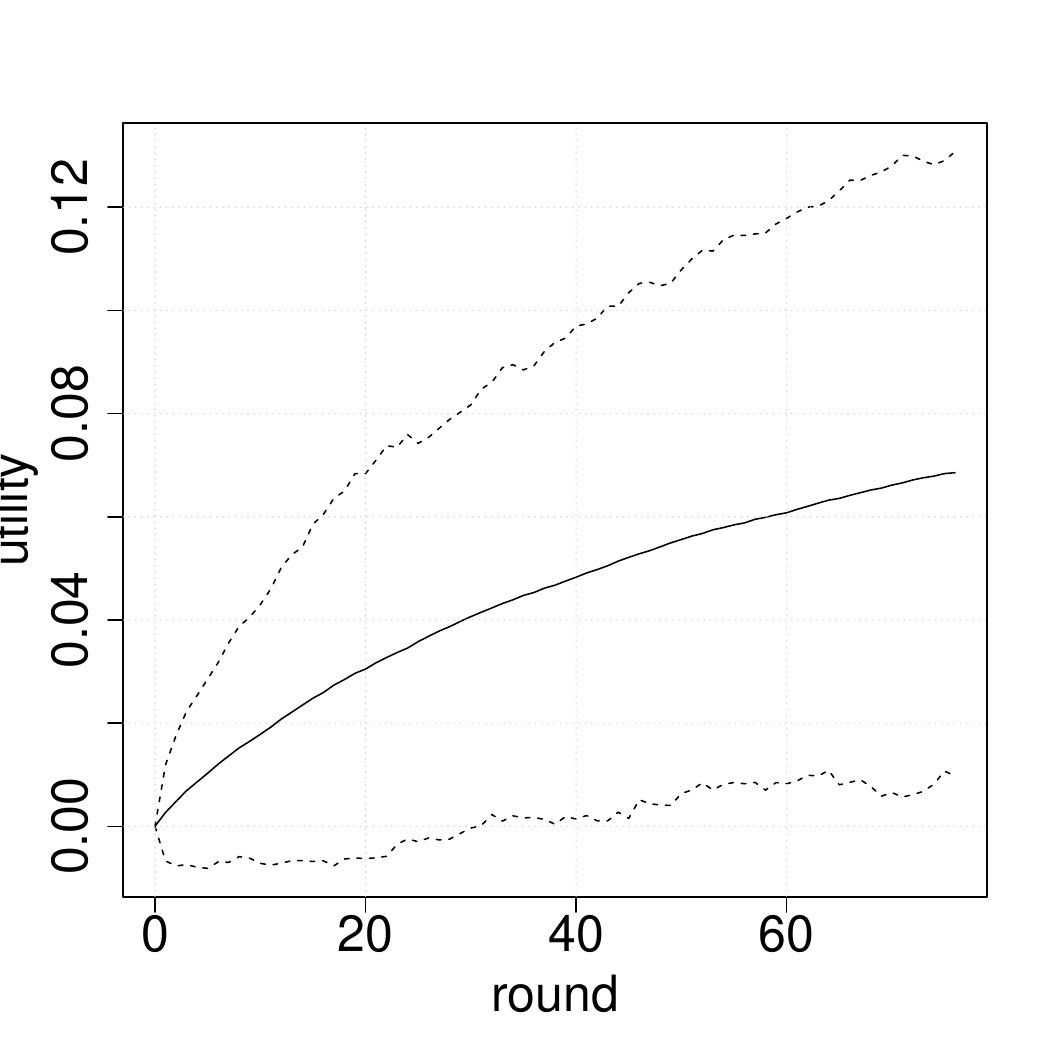}
		\caption{Mutual}
	\end{subfigure}
	\begin{subfigure}[b]{0.35\textwidth}
		\centering
		\includegraphics[width=\textwidth]{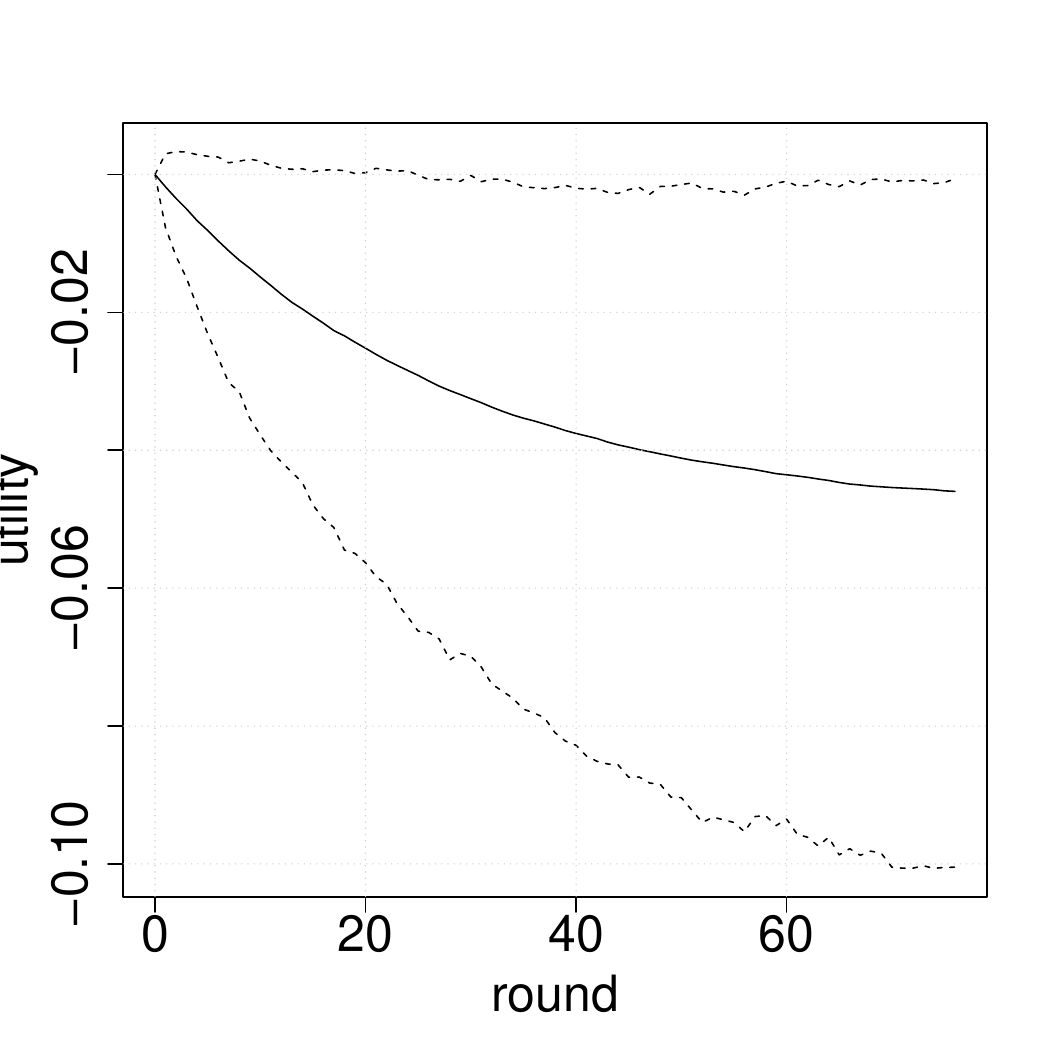}
		\caption{Indirect/Popularity}
	\end{subfigure}

\end{figure}

\begin{figure}[H]
	\centering
	\caption{Tracking vs. base case}
	\begin{subfigure}[b]{0.35\textwidth}
		\centering
		\includegraphics[width=\textwidth]{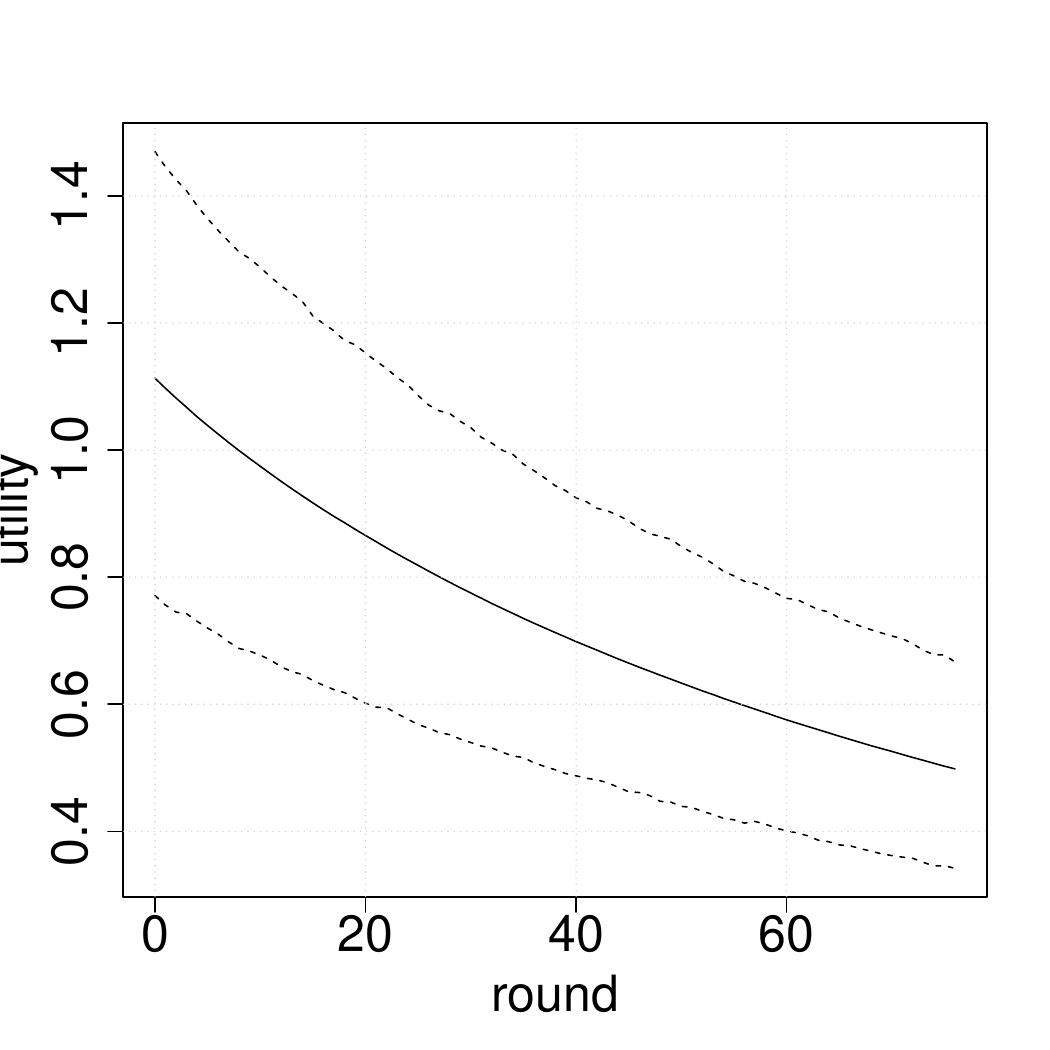}
		\caption{Total}
	\end{subfigure}
	\begin{subfigure}[b]{0.35\textwidth}
		\centering
		\includegraphics[width=\textwidth]{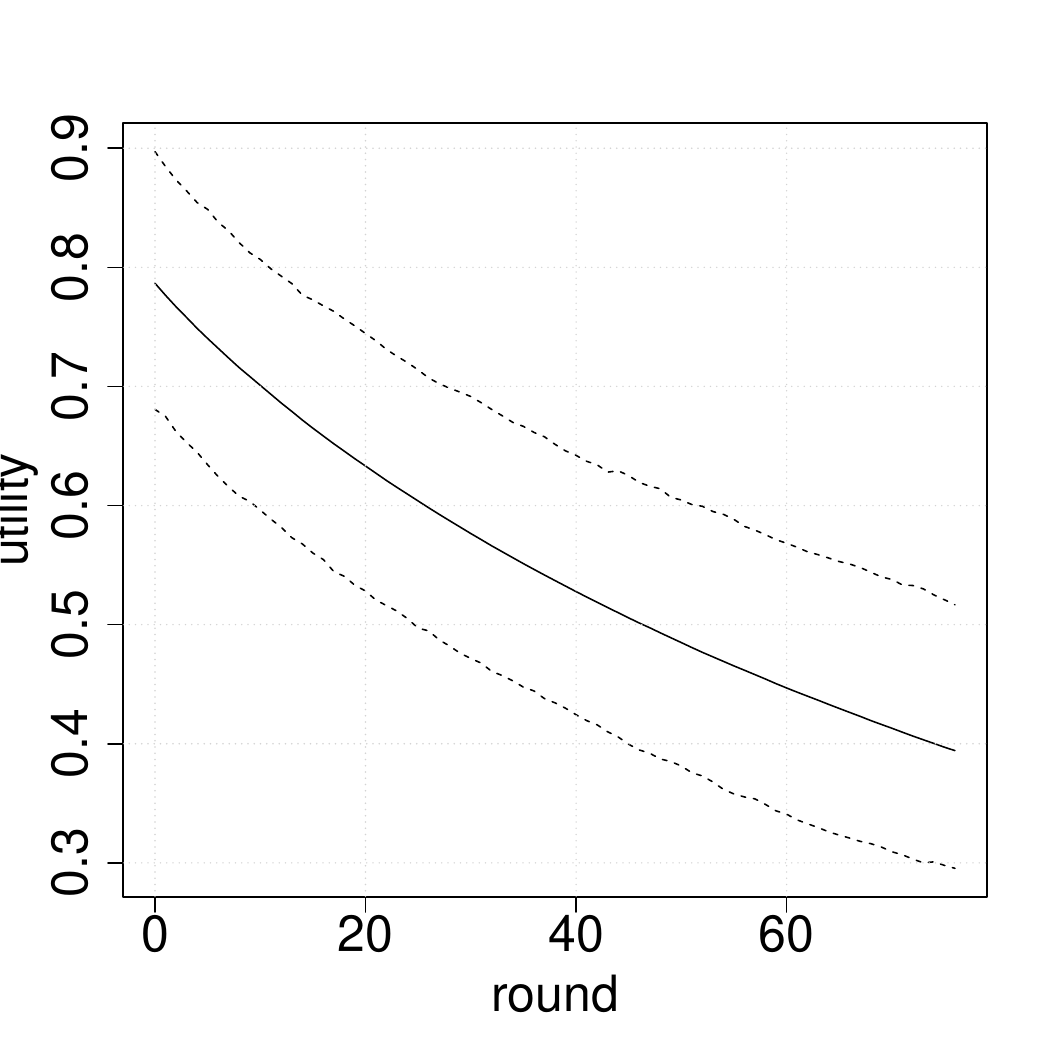}
		\caption{Direct}
	\end{subfigure}\\
	\vspace{-0.6em}
	\begin{subfigure}[b]{0.35\textwidth}
		\centering
		\includegraphics[width=\textwidth]{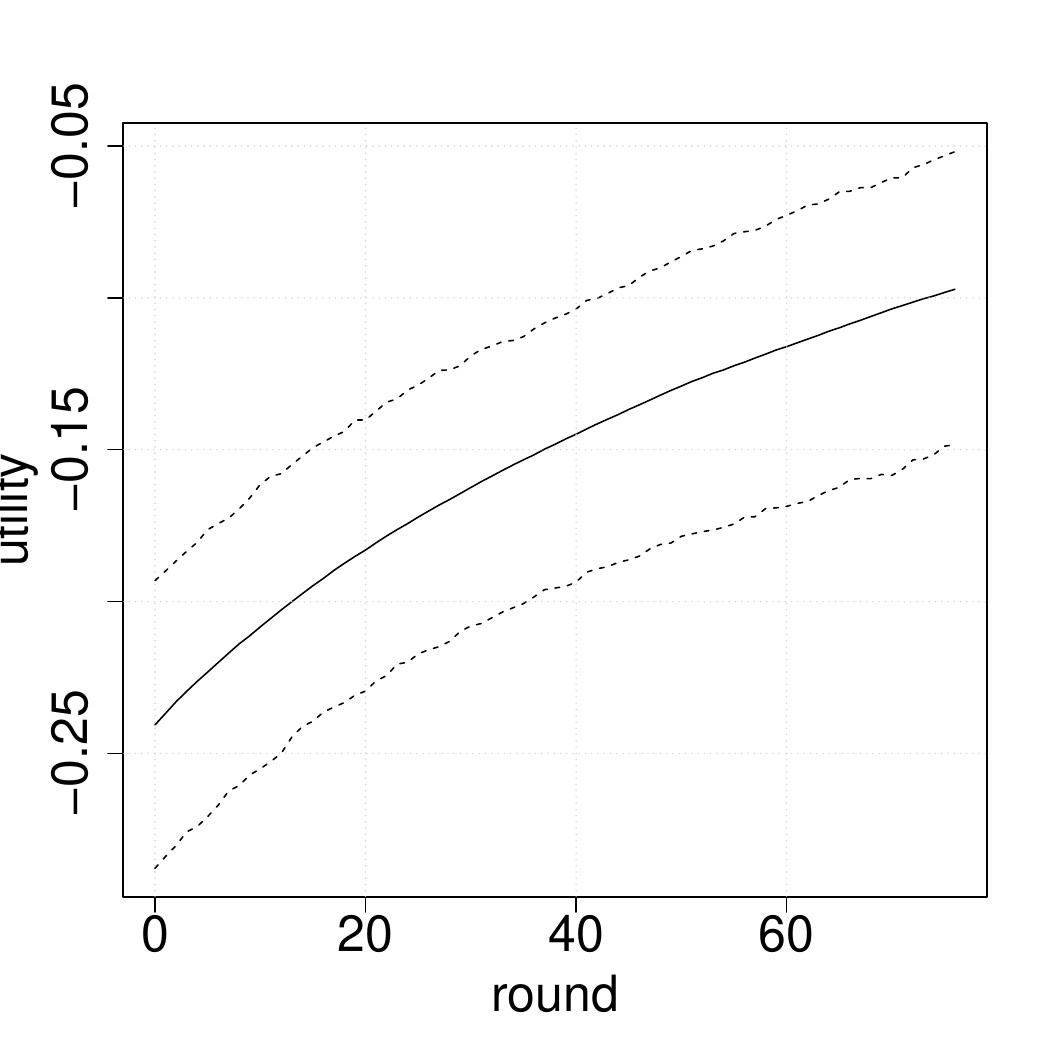}
		\caption{Mutual}
	\end{subfigure}
	\begin{subfigure}[b]{0.35\textwidth}
		\centering
		\includegraphics[width=\textwidth]{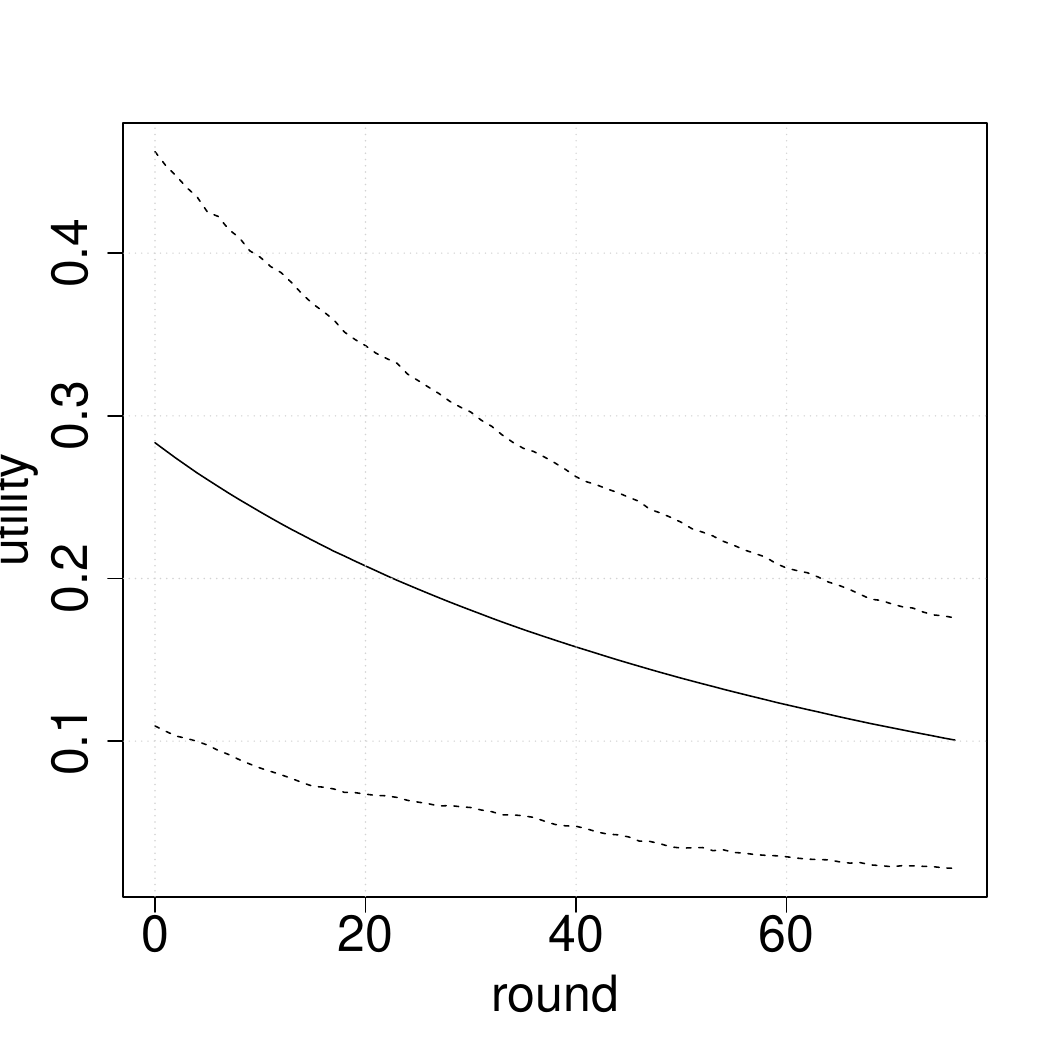}
		\caption{Indirect/Popularity}
	\end{subfigure}

\end{figure}

\begin{figure}[H]
	\centering
	\caption{Random friendships vs. base case}
	\begin{subfigure}[b]{0.35\textwidth}
		\centering
		\includegraphics[width=\textwidth]{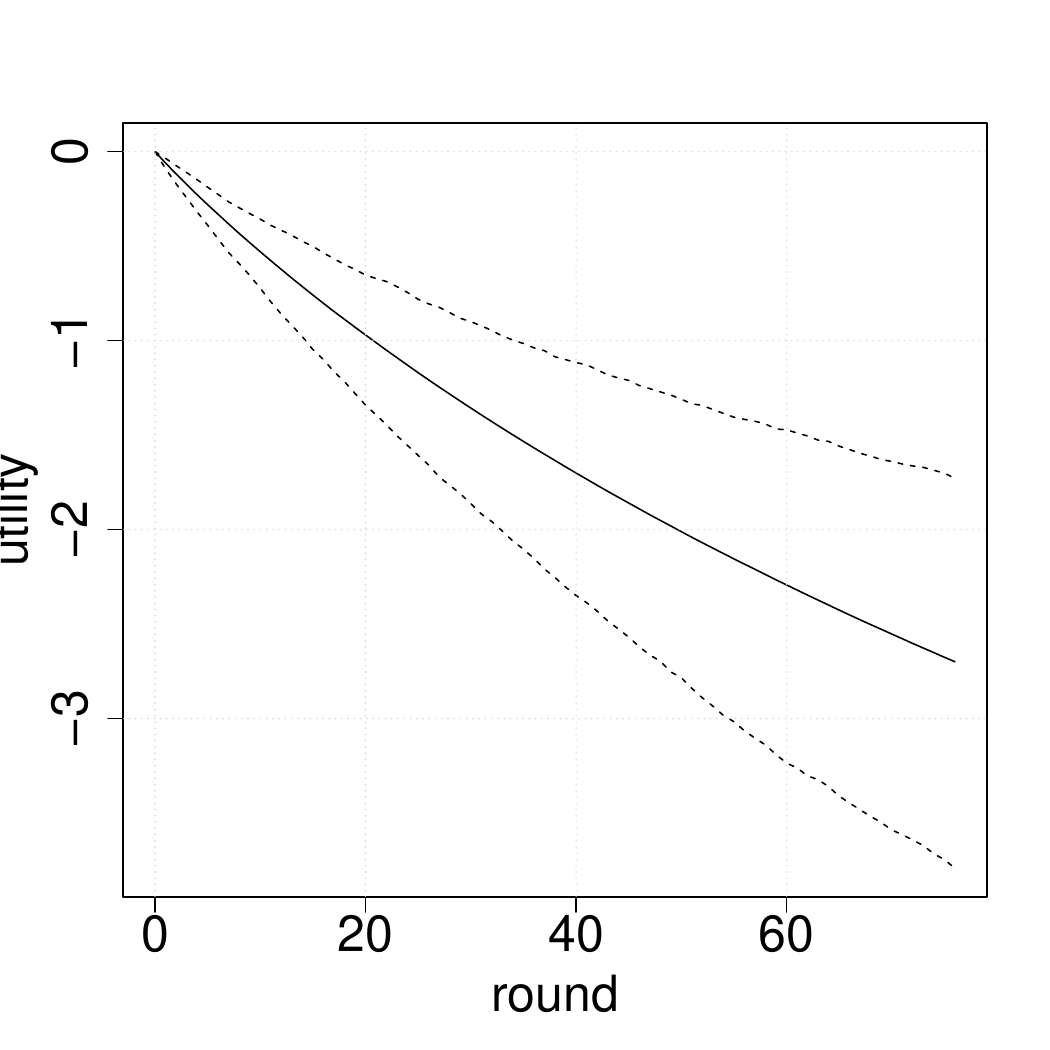}
		\caption{Total}
	\end{subfigure}
	\begin{subfigure}[b]{0.35\textwidth}
		\centering
		\includegraphics[width=\textwidth]{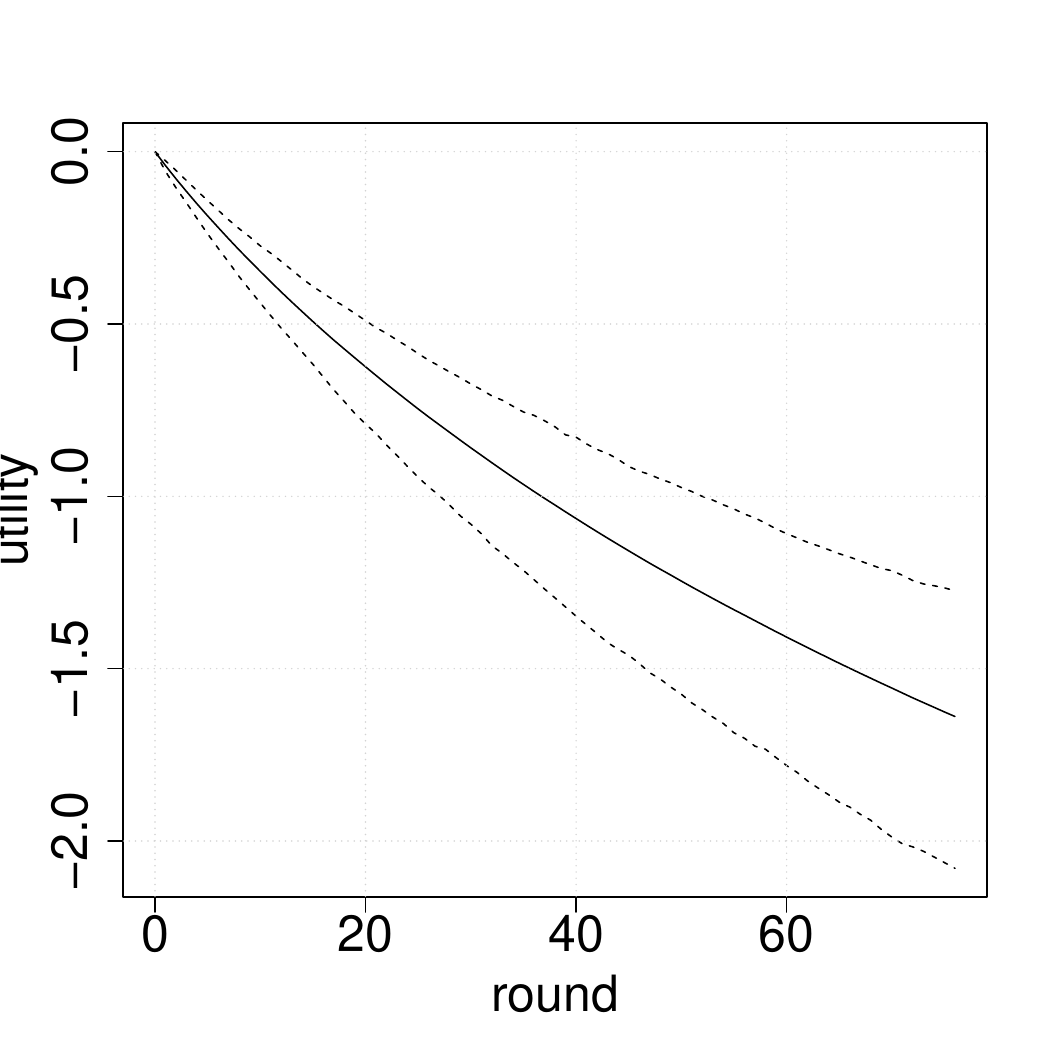}
		\caption{Direct}
	\end{subfigure}\\
	\vspace{-0.6em}
	\begin{subfigure}[b]{0.35\textwidth}
		\centering
		\includegraphics[width=\textwidth]{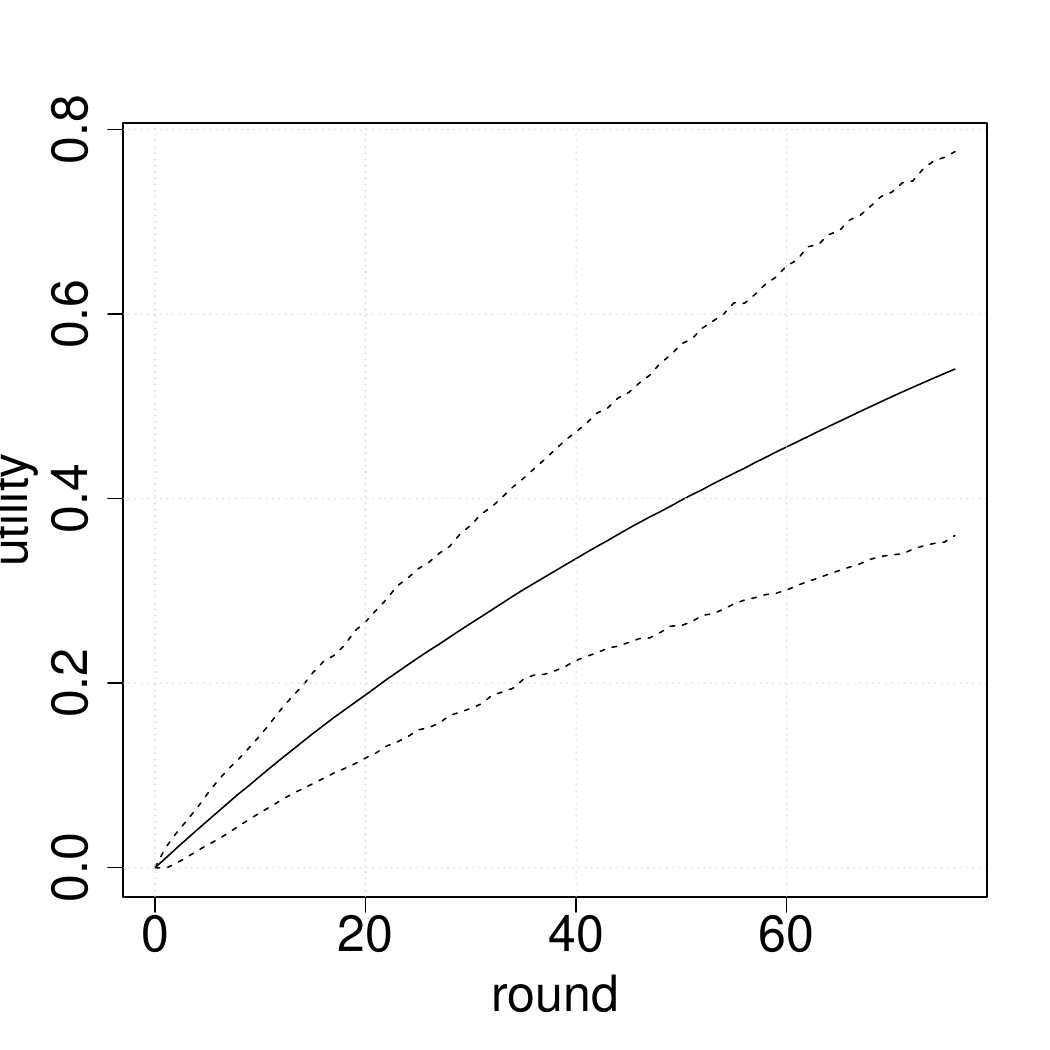}
		\caption{Mutual}
	\end{subfigure}
	\begin{subfigure}[b]{0.35\textwidth}
		\centering
		\includegraphics[width=\textwidth]{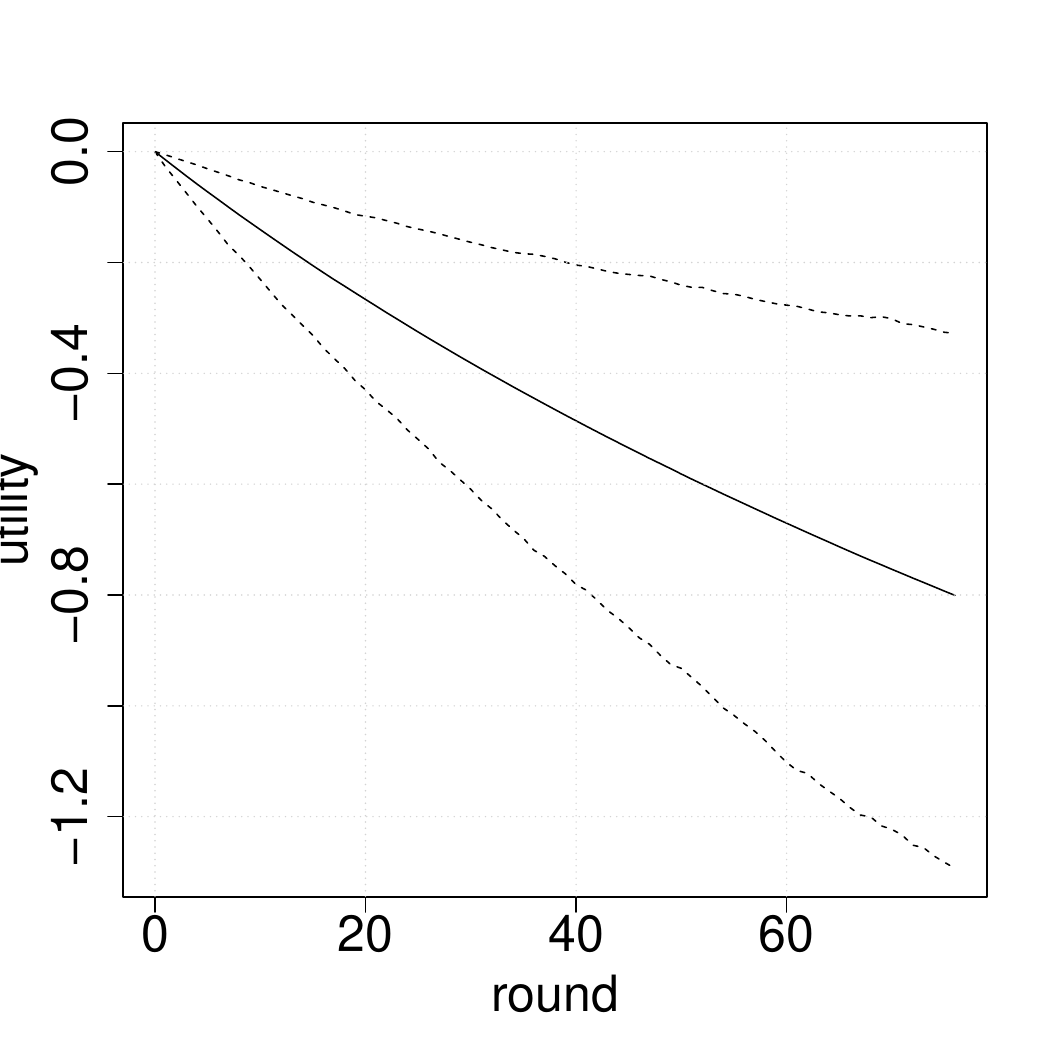}
		\caption{Indirect/Popularity}
	\end{subfigure}
\end{figure}

\section{Application: sensitivity of main estimates when doubling the estimate of the number of rounds}
\label{Supplement-double_tau}
In this Appendix, we analyze the sensitivity of our results to doubling the estimated number of rounds used in estimation. \Cref{Supplement-tab_ep_double} reports posterior quantities of structural parameters when we use our baseline Expectation Propagation algorithm but double our estimate $\hat \tau_0$ in the first step. \Cref{Supplement-tab_ols_ep_double} reports the measures of homophily in the four counterfactuals. Results in the counterfactual analyses remain qualitatively mostly unchanged compared to the estimates reported in the main text.

% latex table generated in R 4.3.0 by xtable 1.8-4 package
% Sun Apr 23 21:47:02 2023
\begin{table}[H]
\centering
\caption{Posterior estimates - Expectation propagation (doubling $\hat \tau_0$)} 
\label{Supplement-tab_ep_double}
\begin{tabular}{lllll}
  \hline
  & Mean & Q 0.025 & Q 0.975 & Prob $<$ 0 \\ 
  \hline \multicolumn{5}{c}{ Meeting process } \\ \hline  \hline
distance in age &  0.5287 &  0.3185 &  0.7388 &  0.0000 \\ 
  distance in gender &  1.0789 &  0.5879 &  1.5700 &  0.0000 \\ 
  distance in cognitive skills &  0.8302 & -1.0042 &  2.6647 &  0.1875 \\ 
  distance in conscientiousness & -0.8902 & -1.3269 & -0.4534 &  1.0000 \\ 
  distance in neuroticism & -0.1250 & -0.5521 &  0.3021 &  0.7170 \\ 
  g\_ij &  4.8507 &  4.1908 &  5.5105 &  0.0000 \\ 
  (1-g\_ij)*distance in classlist & -0.7495 & -1.1449 & -0.3542 &  0.9999 \\ 
   \hline \multicolumn{5}{c}{ Utility -- Direct Links } \\ \hline intercept &  4.5526 &  4.0571 &  5.0480 &  0.0000 \\ 
  distance in age & -0.1654 & -0.4586 &  0.1278 &  0.8656 \\ 
  distance in gender & -4.3272 & -5.1442 & -3.5103 &  1.0000 \\ 
  distance in cognitive skills &  0.3276 & -1.6268 &  2.2819 &  0.3713 \\ 
  distance in conscientiousness & -2.3413 & -2.9935 & -1.6890 &  1.0000 \\ 
  distance in neuroticism &  0.4318 & -0.2097 &  1.0733 &  0.0935 \\ 
   \hline \multicolumn{5}{c}{ Utility -- Reciprocity Links } \\ \hline intercept &  1.0623 &  0.1576 &  1.9669 &  0.0107 \\ 
  distance in age &  0.6437 & -0.3031 &  1.5905 &  0.0913 \\ 
  distance in gender & -2.0621 & -3.9248 & -0.1994 &  0.9850 \\ 
  distance in cognitive skills & -1.9591 & -4.8656 &  0.9474 &  0.9068 \\ 
  distance in conscientiousness &  3.1624 &  2.2032 &  4.1216 &  0.0000 \\ 
  distance in neuroticism &  0.2732 & -0.9808 &  1.5272 &  0.3347 \\ 
   \hline \multicolumn{5}{c}{ Utility -- Indirect/Popularity } \\ \hline intercept & -0.0370 & -0.4356 &  0.3617 &  0.5721 \\ 
  distance in age &  0.0733 & -0.1822 &  0.3288 &  0.2870 \\ 
  distance in gender & -4.5280 & -5.6134 & -3.4425 &  1.0000 \\ 
  distance in cognitive skills & -0.5451 & -2.1256 &  1.0353 &  0.7505 \\ 
  distance in conscientiousness & -0.7154 & -1.2041 & -0.2267 &  0.9979 \\ 
  distance in neuroticism & -0.6307 & -1.0832 & -0.1783 &  0.9969 \\ 
   \hline
\end{tabular}
\end{table}

% latex table generated in R 4.3.0 by xtable 1.8-4 package
% Sun Apr 23 21:47:14 2023
\begin{table}[H]
\centering
\caption{Projection coefficients and edge statistics - Expectation propagation (doubling $\hat \tau_0$)} 
\label{Supplement-tab_ols_ep_double}
\begin{adjustbox}{width=\textwidth}
\begin{tabular}{llllll}
  \hline
  & Data & Base case & Random matching & Tracking & Random friendship \\ 
  \hline \multicolumn{4}{c}{ Regression coefficients } \\ \hline  \hline
distance in classlist & -0.0016 & -0.0013 & -0.0020 & -0.0034 & -0.0009 \\ 
    & [-0.0028;-0.0004] & [-0.0018;-0.0008] & [-0.0028;-0.0012] & [-0.0042;-0.0025] & [-0.0012;-0.0007] \\ 
  distance in age &  0.0018 & -0.0093 &  0.0011 & -0.0041 &  0.0010 \\ 
    & [-0.0048; 0.0083] & [-0.0152;-0.0038] & [-0.0078; 0.0088] & [-0.0080;-0.0006] & [-0.0006; 0.0025] \\ 
  distance in gender & -0.1864 & -0.1402 & -0.3526 & -0.0836 &  0.0011 \\ 
    & [-0.2072;-0.1657] & [-0.1498;-0.1307] & [-0.3784;-0.3292] & [-0.0905;-0.0770] & [-0.0015; 0.0040] \\ 
  distance in cognitive skills & -0.1275 & -0.0834 & -0.1884 & -0.0122 & -0.0056 \\ 
    & [-0.2172;-0.0378] & [-0.1358;-0.0324] & [-0.2712;-0.1128] & [-0.0668; 0.0452] & [-0.0194; 0.0086] \\ 
  distance in conscientiousness & -0.0193 & -0.0268 & -0.0621 & -0.0214 & -0.0004 \\ 
    & [-0.0332;-0.0053] & [-0.0350;-0.0176] & [-0.0791;-0.0443] & [-0.0273;-0.0155] & [-0.0026; 0.0019] \\ 
  distance in neuroticism & -0.0231 &  0.0006 & -0.0109 & -0.0033 & -0.0006 \\ 
    & [-0.0356;-0.0106] & [-0.0079; 0.0092] & [-0.0301; 0.0084] & [-0.0107; 0.0040] & [-0.0026; 0.0016] \\ 
   \hline \multicolumn{4}{c}{ Edge summary statistics } \\ \hline $\mathbb{E}[g_{ij}]$ & 0.1788 &  0.0711 &  0.2149 &  0.0440 &  0.0062 \\ 
    & [0.168;0.1896] & [0.0663;0.0758] & [0.2039;0.2266] & [0.0407;0.0475] & [0.0050;0.0077] \\ 
  $\operatorname{sd}[g_{ij}]$ & 0.3832 &  0.2569 &  0.4107 &  0.2050 &  0.0787 \\ 
    & [0.374;0.3921] & [0.2488;0.2647] & [0.4029;0.4186] & [0.1975;0.2127] & [0.0703;0.0872] \\ 
   \hline
\end{tabular}
\end{adjustbox}
\end{table}

\section{Application: auxiliary tables}
\label{Supplement-application_tables}
\begin{figure}[H]
	\centering
	\caption{Example: a 3rd-grade baseline classroom}
	\includegraphics[scale = 0.9]{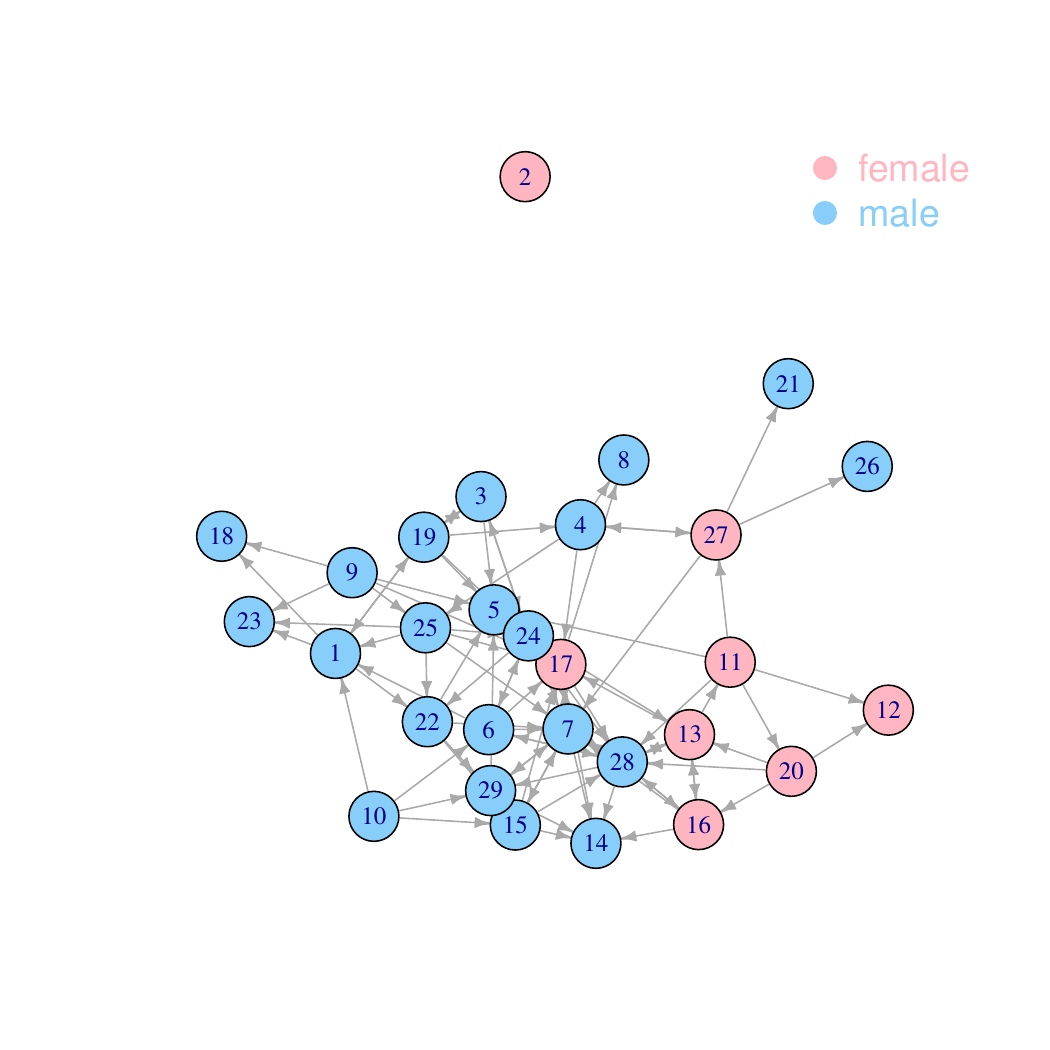}
	\begin{minipage}{0.9\textwidth} % choose width suitably
		{\centering  \textit{Note:} The figure presents a 3rd-grade classroom network from our baseline data. Numbered circles represent students. An arrow stemming from the circle ``x''  to ``y'' denotes student ``x'' nominated ``y'' as a friend in mid 2014.  }
	\end{minipage}
	\label{Supplement-fig:network}
\end{figure}

% Table created by stargazer v.5.2.2 by Marek Hlavac, Harvard University. E-mail: hlavac at fas.harvard.edu
% Date and time: Sun, Jul 14, 2019 - 23:16:23
\begin{table}[H] \centering
  \caption{Pairwise distance in covariates -- Summary statistics}
  \label{Supplement-tab_summary_dyad}
    \begin{adjustbox}{width = \textwidth}
\begin{tabular}{@{\extracolsep{5pt}}lccccccc}
\\[-1.8ex]\hline
\hline \\[-1.8ex]
Statistic & \multicolumn{1}{c}{N} & \multicolumn{1}{c}{Mean} & \multicolumn{1}{c}{St. Dev.} & \multicolumn{1}{c}{Min} & \multicolumn{1}{c}{Pctl(25)} & \multicolumn{1}{c}{Pctl(75)} & \multicolumn{1}{c}{Max} \\
\hline \\[-1.8ex]
edge (baseline) & 17,736 & 0.172 & 0.377 & 0 & 0 & 0 & 1 \\
edge (follow up) & 17,736 & 0.179 & 0.383 & 0 & 0 & 0 & 1 \\
distance in classlist (baseline) & 17,736 & 8.009 & 5.553 & 1 & 3 & 12 & 30 \\
distance in age (years) (baseline) & 17,736 & 0.827 & 0.908 & 0.000 & 0.249 & 1.041 & 6.633 \\
distance in gender (baseline) & 17,736 & 0.502 & 0.500 & 0 & 0 & 1 & 1 \\
distance in cognitive skills (baseline)& 17,736 & 0.099 & 0.081 & 0.000 & 0.035 & 0.144 & 0.530 \\
distance in conscientiousness (baseline) & 17,736 & 0.568 & 0.485 & 0.000 & 0.207 & 0.798 & 3.275 \\
distance in neuroticism (baseline) & 17,736 & 0.676 & 0.538 & 0.000 & 0.249 & 0.976 & 3.491 \\
\hline \\[-1.8ex]
\end{tabular}
\end{adjustbox}
\end{table}

\section{Counterfactuals in a model of peer effects}
\label{Supplement-model_peer}
In this Appendix, we couple a peer effects model of cognitive skills to our network formation algorithm and show how the counterfactuals in \Cref{application} of the main text translate into measures of productivity and inequality in cognitive skills.

We consider the following structural model for cognitive skills in classroom $j$:

\begin{equation}
	\label{eq_model}
	\boldsymbol{c}_{j,1}= {\alpha}_j \iota_{N_j \times 1} + \boldsymbol{X}_{j,0}\beta + \gamma  \boldsymbol{G}_{j,0} \boldsymbol{c}_{j,1} +  \boldsymbol{G}_{j,0} \boldsymbol{X}_{j,0} \delta + \boldsymbol{\epsilon}_{j,1} \, ,
\end{equation}
where $\boldsymbol{c}_{j,1}$ is the vector of classroom cognitive skills in the\textbf{ followup period}; $\boldsymbol{G}_{j,0}$ is the (row-sum normalised) adjacency matrix \textbf{at the baseline}; and $\boldsymbol{X}_{j,0}$ is a $N_j \times d$ matrix of traits \textbf{at the baseline}. The $\alpha_j$ is a classroom effect, and $\boldsymbol{\epsilon_{j,1}}$ are individual unobservables. This is a standard version of the linear-in-means model of peer effects \citep{dePaula2016}. Following the notation used by the literature, $\gamma \boldsymbol{G}_{j,0} \boldsymbol{c}_{j,1}$ is known as an endogenous effect, whereas $\boldsymbol{G}_{j,0}\boldsymbol{X}_{j,0}\delta$ are contextual effects. We assume exogeneity:

\begin{equation}
	\label{eq_exog_model}
	\mathbb{E}[\boldsymbol{\epsilon}_{j,1}| \boldsymbol{G}_{j,0}, \boldsymbol{X}_{j,0}] = 0 \, .
\end{equation}

Recall our network formation algorithm postulates that, for each $t$:

$$\boldsymbol{G}_{j,t} = \phi_j(\boldsymbol{G}_{j,t-1}, \boldsymbol{X}_{j,0},\boldsymbol{u}_{j,t})\, ,$$
where $\boldsymbol{u}_{j,t}$ are the unobservables driving the network formation process between  $t-1$ and $t$. Therefore, for \eqref{eq_exog_model} to hold, it is sufficient to assume:

\begin{equation}
	\mathbb{E}[\boldsymbol{\epsilon}_{j,1}|\boldsymbol{X}_{j,0}, \boldsymbol{G}_{j,-1}, \boldsymbol{u}_{j,0}] = 0 \, ,
\end{equation}
i.e., taste shocks driving network formation until the baseline period are mean independent of the unobservable determinants of cognitive skills at the follow-up, conditional on observable characteristics at the baseline.

Under the above assumption, the strategy in \cite{Bramoulle2009} can be adopted. First, we remove the group effects by working with a transformed model:

\begin{equation*}
	(I-\boldsymbol{H}_j)\boldsymbol{c}_{j,1}=   	(I-\boldsymbol{H}_j)\boldsymbol{X}_{j,0}\beta + \gamma 	(I-\boldsymbol{H}_j) \boldsymbol{G}_{j,0} \boldsymbol{c}_{j,1} +  	(I-\boldsymbol{H}_j) \boldsymbol{G}_{j,0} \boldsymbol{X}_{j,0} \delta + 	(I-\boldsymbol{H}_j)\boldsymbol{\epsilon}_{j,1},
\end{equation*}
where $\boldsymbol{H}_j$ is such that $\boldsymbol{H}_j\iota_{j} = \iota_{j}$. We put $\boldsymbol{H}_j = \frac{1}{N_j}\iota_{j} \iota_{j}'$.\footnote{As discussed in \cite{Bramoulle2009}, an alternative transformation would be to choose $\boldsymbol{H}_j = \boldsymbol{G}_{j,0}$, provided that every student nominates at least one friend (so that all rows are normalized to one). This is not our case.} We then estimate the model by 2SLS, instrumenting $(I-\boldsymbol{H}_j) \boldsymbol{G}_{j,0} \boldsymbol{c}_{j,1}$ with $(I-\boldsymbol{H}_j) \boldsymbol{G}_{j,0}^2 \boldsymbol{X}_{j,0}$.

Table \ref{Supplement-tab_peers} reports the estimates of our model. Again, standard errors are clustered at the classroom level.

% latex table generated in R 4.2.1 by xtable 1.8-4 package
% Mon Oct 31 10:59:08 2022
\begin{table}[ht]
\centering
\caption{Estimates from peer effects model} 
\label{Supplement-tab_peers}
\begin{tabular}{ll}
  \hline
  & Estimate \\ \hline
Endogenous & -0.2407  \\ 
   & (0.40436) \\ 
  age in years -- self & -0.0065 *** \\ 
   & (0.00247) \\ 
  girl dummy -- self &  0.0105 ** \\ 
   & (0.00458) \\ 
  cognitive factor (baseline) -- self &  0.7282 *** \\ 
   & (0.03443) \\ 
  noncognitive factor 1 (baseline) -- self &  0.0315 *** \\ 
   & (0.00483) \\ 
  noncognitive factor 2 (baseline)  -- self &  0.0184 *** \\ 
   & (0.00351) \\ 
  age in years -- contextual & -0.0068  \\ 
   & (0.00802) \\ 
 girl dummy -- contextual & -0.0235 *** \\ 
   & (0.00899) \\ 
  cognitive factor (baseline) -- contextual &  0.1674  \\ 
   & (0.25391) \\ 
  noncognitive factor 1 (baseline) -- contextual &  0.0197  \\ 
   & (0.01222) \\ 
  noncognitive factor 2 (baseline) -- contextual &  0.0149 ** \\ 
   & (0.00640) \\ 
  Intercept  -- contextual &  0.1074  \\ 
   & (0.11769) \\ 
   \hline
\end{tabular}
\begin{minipage}{0.8\textwidth}
	\vspace{1em}
	\footnotesize
	\textit{Notes:} $***$: significant at the 1\% level;
	$**$: significant at the 5\% level; $*$: significant at the 10\% level.
\end{minipage}
\end{table}

After estimating the peer effect model, we evaluate its implications at an eventual ``next year''. Specifically, we compute:

$$\tilde{\phi}_j = (I-\hat{\gamma}\tilde{\mathbf{G}}_{j,1})^{-1} (\boldsymbol{X}_{j,1}\hat{\beta}+\tilde{\mathbf{G}}_{j,1}\boldsymbol{X}_{j,1}\hat{\delta}),$$
where $\boldsymbol{X}_{j,1}$ is the observed matrix of traits at the \textbf{followup period}, and $\tilde{\mathbf{G}}_{j,1}$ is the adjacency matrix at the followup period, computed under each counterfactual set of parameters of the network formation process. Measure $\tilde{\phi}_j$ will have a structural interpretation as the expected value of cognitive skills, net of classroom effects, in an eventual ``next year'' under different network policies, provided that changes in the parameters of the network formation process between the baseline and followup do not affect the process determining $\boldsymbol{X}_{j,1}$ nor the model \eqref{eq_model}.\footnote{This is related to the notion of structural invariance in Structural Econometrics \citep{Engle1983}. } The first of these requirements will be satisfied if a model like \eqref{eq_model} determines the elements in $\boldsymbol{X}_{j,1}$, and if the change in the network formation process does not affect $\boldsymbol{G}_{j,0}$ nor the parameters of the model determining $\boldsymbol{X}_{j,1}$. For this reason, we do not compute $\tilde{\phi}_j$ for the tracking counterfactual.\footnote{If a model like \eqref{eq_model} determines each column of  $\boldsymbol{X}_{j,1}$, then we could construct an alternative measure that nets out from  $\boldsymbol{X}_{j,1}$ the effects of classroom effects. This alternative measure could then assess the effects of tracking policies.}

Given $\tilde{\phi}_j$, we compute two measures a school planner might be interested in. The first one is the overall average of $\tilde{\phi}_j$:

$$A =\frac{ \sum_{j=1}^C \tilde{\phi}_j'\iota_{N_j}}{\sum_{j=1}^C N_j} .$$

The second measure is a weighted average of the within-classroom distance between the expected cognitive skills of boys and girls, which we compute as:

$$D =\sum_{j=1}^C \frac{N_j}{\sum_{c=1}^C N_c } | \tilde{\phi}_{j,\text{boys}} - \tilde{\phi}_{j,\text{girls} }|,  $$
where $\tilde{\phi}_{j,\text{boys}} $ is the average of the entries of vector $\tilde{\phi}_j$ among those corresponding to boys.

Table \ref{Supplement-tab_peer_counter} reports the posterior mean and 95\% credible intervals of measures $A$ and $D$ under the base, random matching, and random friendship scenarios. We notice that both random matching and random friendship appear to raise cognitive skills, with random matching producing larger effects. There is also evidence that random friendships may lead to larger differences between boys' and girls' scores in the same classroom, though credible sets in the random friendships and base case for the $D$ measure overlap. Such an increase would be expected since preferences exhibit homophily in gender, the contextual effect of having a girl friend on cognitive skills' is negative, and girls have higher cognitive skills on average than boys. Since random friendships would reduce homophily in gender, this would increase average cognitive skills among women and decrease it among men, leading to a higher gap between both groups.

Overall, the results in this section suggest that policies designed at changing the determinants of network formation may produce nontrivial impacts on students' outcomes.

% latex table generated in R 4.2.1 by xtable 1.8-4 package
% Mon Oct 31 12:16:02 2022
\begin{table}[H]
\centering
\caption{Cognitive skills model: counterfactuals}
\label{Supplement-tab_peer_counter} 
\begin{tabular}{rlll}
  \hline
 & Base case & Random matching & Random friendships \\ 
  \hline
A & 0.02343 & 0.03111 & 0.02713 \\ 
   & [0.02246;0.02435] & [0.03038;0.03178] & [0.02521;0.02899] \\ 
  D & 0.0515 & 0.05097 & 0.05269 \\ 
   & [0.05017;0.05289] & [0.04966;0.05236] & [0.05098;0.05449] \\ 
   \hline
\end{tabular}
\end{table}

\end{document}